\documentclass[useAMS,usenatbib]{mn2e}

\usepackage{epsfig}
\usepackage{longtable}
\usepackage{times}

\usepackage{amsmath}
\usepackage{amsmath,amssymb}

\usepackage{txfonts}

\usepackage[T1]{fontenc}
\usepackage{ae,aecompl}

\usepackage{graphicx}	
\usepackage{amssymb}	

\usepackage{amssymb}

\usepackage{graphicx}

\def \inte {$INTEGRAL$}
\def \xmm {$XMM$-$Newton$}

\def \hcm {\hbox {\ifmmode $ atom cm$^{-2}\else atom cm$^{-2}$\fi}}

\def \apj {ApJ}

\def \aap {A\&A}

\def \mnras {MNRAS}

\def \ssr {Space Science Reviews}

\def \aapr {A\&A Rev.}

\def \extras {EXTraS}

\newcommand{\be}{\begin{equation}}

\newcommand{\ee}{\end{equation}}

\def\beq#1{\begin{equation}\label{#1}}
\def\eeq{\end{equation}}
\def\beqa#1{\begin{eqnarray}\label{#1}}
\def\eeqa{\end{eqnarray}}

\def\Eq#1{Eq.~(\ref{#1})} 

\def\myfrac#1#2{\left(\frac{#1}{#2}\right)}

\def\comment#1{\relax}

\newcommand{\ergs}{\:\mbox{erg\,s}$^{-1}$ }
\newcommand{\gs}{\:\mbox{g\,s}$^{-1}$ }

\title[Magnetospheric instability in SFXTs]{Supergiant Fast X-ray Transients uncovered by the \extras\ project: flares reveal the development of magnetospheric instability in accreting neutron stars 
}
\author[Sidoli et al.]{Lara~Sidoli,$^{1}$\thanks{E-mail: lara.sidoli@inaf.it} Konstantin A. Postnov,$^{2,3}$ Andrea Belfiore,$^{1}$ Martino Marelli,$^{1}$ 
\newauthor David Salvetti,$^{1}$ Ruben Salvaterra,$^{1}$ Andrea De~Luca,$^{1}$ and Paolo Esposito$^{1}$ 
\smallskip\\
$^{1}$INAF, Istituto di Astrofisica Spaziale e Fisica Cosmica, Via Alfonso Corti 12,   I-20133 Milano,  Italy   \\
$^{2}$ Sternberg Astronomical Institute, M.V. Lomonosov Moscow State University, 13, Universitetskij pr., 119234 Moscow, Russia\\
$^{3}$ Kazan Federal University, Kremlevskaya 18, 420008 Kazan, Russia
} 

\begin{document}

\date{Accepted 2019 May 01. Received 2019 April 30; in original form 2019 February 25}

\pagerange{\pageref{firstpage}--\pageref{lastpage}} \pubyear{2019}

\maketitle

\label{firstpage}

\begin{abstract}
The low luminosity, X--ray flaring activity, of the sub-class of high mass X--ray binaries called Supergiant Fast X-ray Transients, 
has been investigated using \xmm\ public observations, taking advantage of the products made publicly available by the \extras\ project. 
One of the goals of  \extras\  was to extract from the \xmm\ public archive 
information on the aperiodic variability of all sources observed in the soft X--ray range with EPIC  (0.2--12 keV). 
Adopting a Bayesian block decomposition of the X--ray light curves of a sample of SFXTs, we picked out 144 X--ray flares, 
covering a large range of soft X--ray luminosities (10$^{32}$-10$^{36}$~erg~s$^{-1}$).
We measured temporal quantities, like the rise time to and the decay time from
the peak of the flares, 
their duration and the time interval between adjacent flares. 
We also estimated the peak luminosity, average accretion rate  and energy release in the flares. The observed soft X-ray properties of low-luminosity flaring activity from SFXTs is in qualitative agreement with what is expected by the application of the Rayleigh-Taylor instability model in accreting plasma near the neutron star magnetosphere. In the case of rapidly rotating neutron stars, sporadic accretion from temporary discs cannot be excluded.
\end{abstract}


        \section{Introduction\label{intro}}

Supergiant Fast X--ray Transients (SFXTs) are a kind of high mass X--ray binaries (HMXBs) where a neutron star (NS) 
accretes a fraction of the wind of an early-type supergiant donor 
(see \citealt{Sidoli2017review}, \citealt{Martinez-Nunez2017}, \citealt{Walter2015}, for the most recent reviews). 
They were recognized as a new class of massive binaries
thanks to  rare, short and 
bright flares, reaching a peak luminosity L$_X$$\sim$10$^{36}$--10$^{37}$~erg~s$^{-1}$, caught during
\inte\ observations (\citealt{Sguera2005, Sguera2006}; \citealt{Negueruela2006}).

Low level X--ray flaring activity characterizes also their emission outside outbursts,  
down to L$_X$$\sim$10$^{32}$~erg~s$^{-1}$.
Since SFXT flares usually display a complex morphology,
it is somehow difficult to 
disentangle multiple flares (or structured flares) from the quiescent level
and measure interesting quantities, such as, e.g., 
their duration, rise and decay times, time interval between flares. 
The investigation of SFXT flares requires a twofold approach: 
firstly,  high  throughput and uninterrupted observations are needed, both
to detect flares even at very low X--ray fluxes and to measure the 
flare timescales without data gaps (within each observation) that might bias them; 
secondly, an efficient, systematic procedure to pick out flares 
and determine their observational properties, to be compared with the theory.
Both conditions are met by the database of products made available to the community by the \extras\ project.

\extras\
(acronym of ``Exploring the X--ray Transient and variable Sky") is a project funded within the EU/FP7 framework \citep{Deluca2017},  aimed at extracting from the \xmm\ public archive 
the temporal information (periodic and aperiodic variability) of all sources observed by the EPIC cameras in the 0.2--12 keV energy range.

In the first part of the paper we report  on the behaviour of some essential flare quantities that 
we have extracted from the \extras\ database, 
by means of a Bayesian blocks analysis of the light curves of  a sample of SFXTs. 

In the second part of the paper, we discuss the behaviour of the flare properties and discovered dependences in terms of the interchange instability of accreting matter near NS magnetosphere.  We show that for slowly rotating NSs, the development of Rayleigh-Taylor instability in a quasi-spherical shell above NS magnetosphere can qualitatively describe the observed properties of the SFXT flares. For rapidly rotating NSs accreting from the stellar wind, the propeller mechanism could lead to the formation of an equatorial dead disc that may trigger the magnetospheric instability once the centrifugal barrier at its inner edge is overcome.

This {\em paper} is organized as follows:
Sect.~\ref{extras_red} introduces the \extras\ project, for what is relevant here (for more details, we refer the 
reader to \citealt{Deluca2017}); Sect.~\ref{sec:obs} reports on the \xmm\ observations we have considered in our study; 
Sect.~\ref{results} reports on the automatic procedure used to pick out flares and to measure the parameters of the flares,
 together with the observational results. 
 Sect.~\ref{discussion} discusses the results in the context of the interchange instability near the NS magnetosphere. 
In Sect.~\ref{conclusion}, we summarize our findings and conclusions.


\section{\extras}
 	 \label{extras_red}

The \extras\ project dived into the public soft X--ray data archive of \xmm,
building on top of the 3XMM-DR4 catalog \citep{Rosen2016}, and aimed at 
characterising the variability of as many EPIC point sources as possible.
It provided the community with tools, high-level data products, catalogues,
documentation, all available through the EXTraS web-site\footnote{http://www.extras-fp7.eu}
and the LEDAS astronomical data archive\footnote{https://www88.lamp.le.ac.uk/extras/archive}.
One of these products is a set of adaptively binned light curves,
one for each exposure (observation segment specific to a single EPIC
camera), particularly suited for the systematic analysis carried out
in this paper. In particular, these light curves make it possible to identify easily 
flares and provide some parameters (like the slope between two blocks, see below)
useful to their characterisation.

EXTraS addressed the characterisation of variability of point-like
X-ray sources under various aspects, each one dealing with specific
problems and dedicated techniques \citep{deluca16}. In particular,
aperiodic variability has been treated separately on the short term
(within a single observation segment or exposure) and on the long
term (combining different pointings to the same source, including
slew data). We will focus in this paper on short-term variability
products. Dealing with separated exposures avoids the problem of combining
data collected with different EPIC cameras, with different filters and operating modes.
Although EXTraS provides also a characterisation of aperiodic
variability in the frequency domain, we use here only light curves, in
the form of count rates versus time. All the light curves considered
in this paper always remain in the Poisson regime, and are not affected
by significant instrumental pile-up.

\begin{figure}
\hspace{-0.3cm}
\includegraphics[scale=0.31,angle=0]{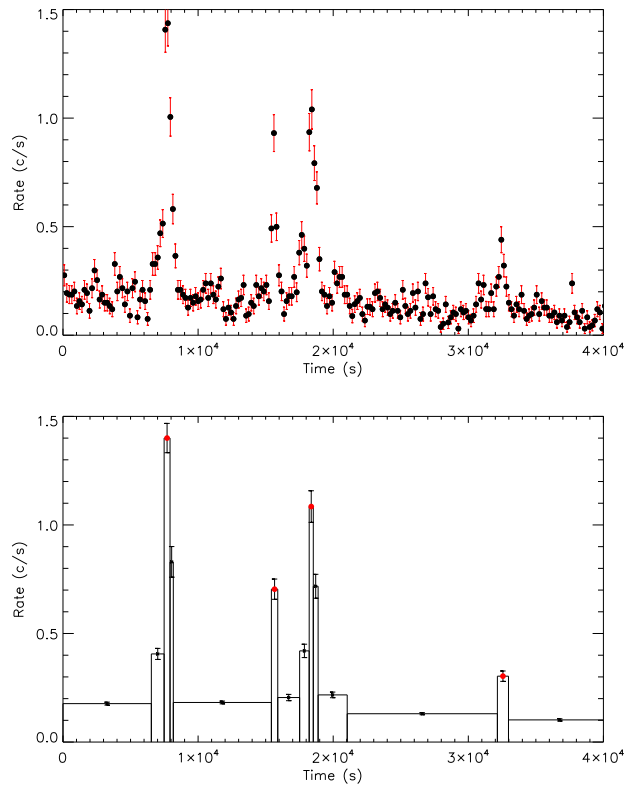} 
\caption{IGR~J08408-4503: comparison between the EPIC pn light curve in its original form (uniform bin time = 187 s; upper panel) and segmented in B.b. (lower panel).
}
\label{fig:orig_bb}
\end{figure}

An important aspect of the EXTraS approach is the technique of background
modelling and subtraction. EPIC data can be affected for a large
fraction of the observing time by strong and rapid background flares
due to soft protons (up to 35\%, \citealt{marelli17}). Standard recommendations
for data screening lead to discard 21\% of the data \citep{Rosen2016}.

EXTraS instead takes care of modelling the
background, by disentangling its steady and variable components,
and evaluating separately their distribution on the
detector. Background maps for both components are built 
in order to take into account the background distribution
on the detector. In this way it is possible to effectively subtract
the variable EPIC background from source regions and use 
all the exposure time  (see \citealt{marelli17} for more details).
The source region has a circular shape, while the background region 
covers most of the detector, excluding circular regions around contaminating sources. 
All radii are chosen to maximise the signal
to noise ratio while minimising the contamination from
other sources (the region files are provided through the
EXTraS archive).

One of the algorithms implemented within EXTraS for the characterisation
of short term (within a single, uninterrupted exposure) variability
is based on Bayesian blocks (hereafter B.b., \citealt{scargle98, scargle13}). 
This is a segmentation
technique, often applied to astronomical time series, which aims to
split the data into the maximum number of adjacent blocks, in such
a way that each block is statistically different from the next. The
B.b. algorithm starts from a fine segmentation of a time
series in cells, that are subsequently merged in a statistically optimal
way. We define a time cell as containing at least 50 source photons or
50 background photons expected in the source region. This algorithm for defining
the initial segmentation allows us to identify features both in the
source and background light curves, which can be considered constant
within each cell. As a figure of merit we consider the likelihood
of an average net source rate within each block, reduced by a fixed
amount. This represents a cost for each added block, implementing
an Occam's razor: the higher the cost, the more significant the difference
between two neighboring blocks. EXTraS provides two sets of Bayesian
blocks light curves (nominally at 3$\sigma$ and 4$\sigma$, respectively): 
one with a lower cost, more sensitive to small
variations in rate; the other with an higher cost, more robust in
its segmentation. In this work we use the latter.

If we consider two neighboring blocks, we can be confident that the
rate of the source has changed between blocks, while it is consistent
with a constant within each block. However, this is not sufficient for us to tell whether
the rate of the source has changed sharply or smoothly and to this
extend we introduce a parameter that we call slope (S). This is the minimum
rate of change in the counts rate of the source between two neighboring
blocks. To find S, we shrink each of the blocks until their associated
rates, $R_{1}$ and $R_{2}$, are compatible within 3$\sigma$, assuming
that the uncertainty in the rates, $\delta R_{1}$ and $\delta R_{2}$,
decreases with time as $T^{-\frac{1}{2}}$, as expected for Poisson
events. Then, we assume that the rate of the source has changed linearly
for the duration of the 2 blocks, $T_{1}+T_{2}$, compatibly with
the 2 rates, and obtain:
\[
S=\frac{2}{9}\frac{\left(R_{2}-R_{1}\right)^{3}}{\left(\delta R_{1}\times\sqrt{T_{1}}+\delta R_{2}\times\sqrt{T_{2}}\right)^{2}}
\]
For similar blocks that are $n\sigma$ apart (as expected from a source
that undergoes a linear trend in flux, with no background flares),
this relation reduces to: 
\[
S\cong2\left(\frac{n}{3}\right)^{2}\frac{R_{2}-R_{1}}{T_{1}+T_{2}}
\]
.

\begin{figure*}
\includegraphics[scale=0.32,angle=0]{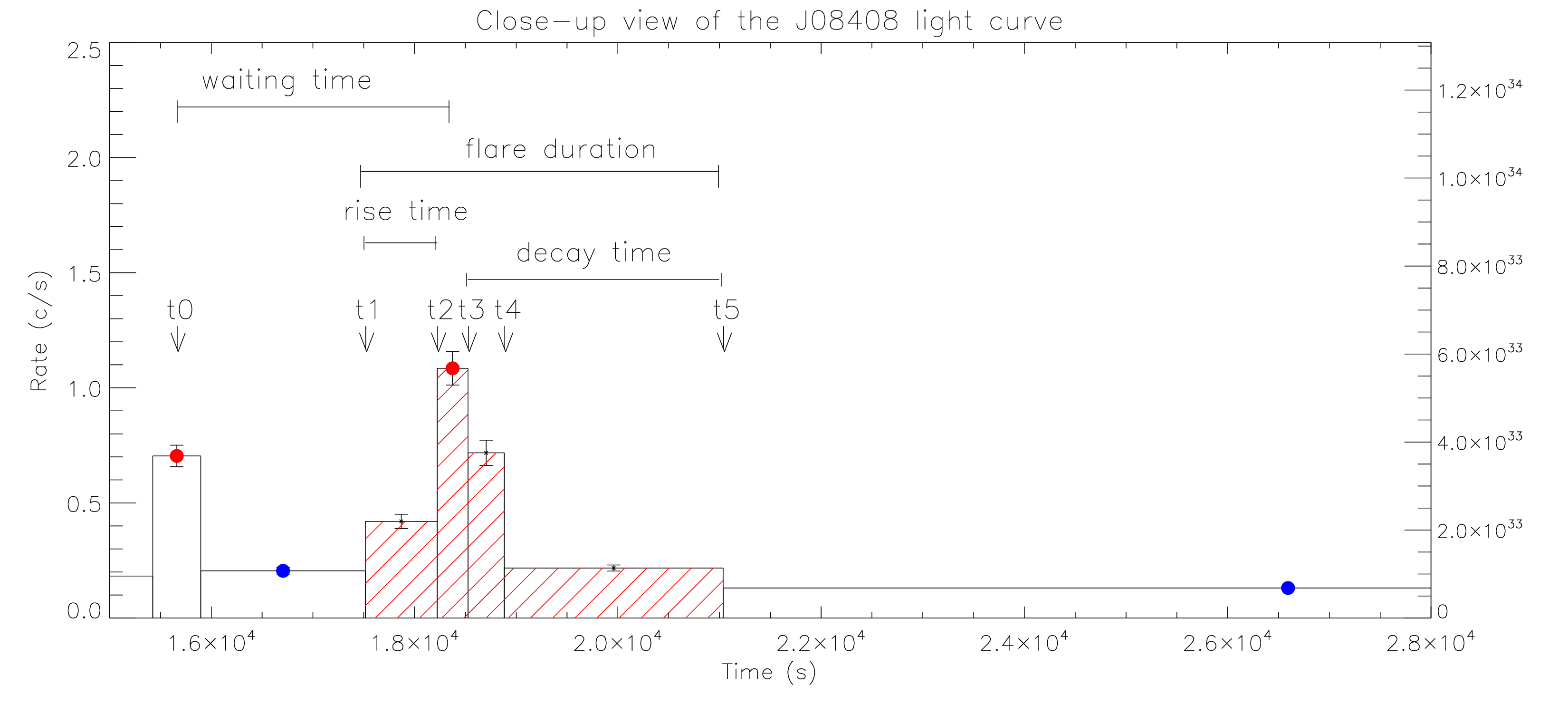} \\
\caption{A close-up view of a flare from IGR~J08408-4503 is shown to display a typical resolved event 
and how the observational quantities have been defined. 
The red dots mark the flare peaks (local maxima). 
The blue dots indicate the valleys (local minima). 
The dashed red area indicates the B.b. covered by this particular flare. 
This area has been used to calculate the energy released in the flare. 
The five timescales (t$_i$) marked in the upper region of this plot have  the following meaning: 
t$_0$ marks the midtime of the block containing the first flare. 
The times  t$_1$, t$_2$, t$_3$, t$_4$ and t$_5$ mark the start and/or  stop times
of the four B.b. composing the second flare. From them, we have defined the following timescales: 
waiting time between two consecutive flares, $\Delta$T, as $\Delta$T=[t$_2$ + 0.5(t$_3$ -- t$_2$) -- t$_0$]. 
The rise time to the peak of the flare is defined as $\delta t_{rise}$ = t$_2$--t$_1$.
The decay time from the peak of the flare is $\delta t_{decay}$ = t$_5$--t$_3$. 
The duration of the flare is: $\Delta$t$_{f}$ = t$_5$--t$_1$.
All the timescales are explicitely marked in the upper region of the plot.
}
\label{fig:explain}
\end{figure*}

\section{\xmm\ observations}
\label{sec:obs}

We searched for the SFXTs in the \extras\ database, 
selecting the  B.b. EPIC light curves from observations pointed on the following 
members of the class: 
IGR~J08408-4503, IGR~J11215-5952,  IGR~J16328-4726, IGR~J16418-4532,   
XTE~J1739-302, IGR~J17544-2619,   IGR~J18410-0535,  IGR~J18450-0435 and IGR~J18483-0311. 
Multiple pointings were available for three sources, as reported in Table~\ref{tab:obslog},
where we list the observations considered in our study. 

The B.b. light curves extracted from observations performed after 2012 are not present in the public \extras\ database,
but were produced for this work by the team, using the same techniques explained in Sect.~\ref{extras_red}.
For the sake of completness, note that we excluded from this investigation two observations present in the \extras\ database and
targeted on IGR~J16479-4514 and on IGR~J18483-0311, because no flares were present (according to the definition assumed below).

All pointings are Guest Observer observations, 
except two Target of Opportunity ones (targeted on IGR~J11215-5952 and the 2011 observation targeted on IGR~J16418-4532)
which were triggered at the occurrence of an outburst.

We have considered only the EPIC pn light curves (all in Full Frame mode), except for IGR~J11215-5952
and the second observation of IGR~J18450-0435, where the EPIC MOS light curves were considered.
The reason is because in these observations EPIC pn 
operated in Small Window mode, where the background  cannot be appropriately treated using the \extras\ techniques.

A systematic analysis of the \xmm\ observations considered in this work, but focussed on the X-ray spectroscopy, can be found in 
\citet{Gimenez2015}, \citet{Bozzo2017}, \citet{Pradhan2018} (and references therein).

We list the source properties in Table~\ref{tab:prop}, adopting the same values published by \citet{SP2018}.
For sources with no published range of variability of the distance 
(the ones with no uncertainty present in Table~\ref{tab:prop}), 
an error of $\pm{1}$~kpc has been assumed, when needed.

\begin{table}
 \centering
  \caption{Logbook of the \xmm\ observations used here.
}
 \begin{tabular}{llcc}
\hline
Target              & Obs. ID    &     Start Date     &    Duration      \\
                    &            &                    &      (ks)        \\
\hline
IGR J08408-4503 &    0506490101     &  2007-05-29        &        45.7      \\     
IGR J11215-5952 &    0405181901     &  2007-02-09        &        22.2      \\     
IGR J16328-4726 &    0654190201 (a) &  2011-02-20        &        21.9      \\  
IGR J16328-4726 &    0728560201 (b) &  2014-08-24        &        36.2      \\  
IGR J16328-4726 &    0728560301 (c) &  2014-08-26        &        23.0      \\  
IGR J16418-4532 &    0206380301 (a) &  2004-08-19        &        23.2      \\  
IGR J16418-4532 &    0405180501 (b) &  2011-02-23        &        39.6      \\  
XTE J1739-302   &    0554720101     &  2008-10-01        &        43.1      \\  
IGR J17544-2619 &    0148090501     &  2003-09-11        &        11.2      \\     
IGR J18410-0535 &    0604820301     &  2010-03-15        &        45.9      \\  
IGR J18450-0435 &    0306170401 (a) &  2006-04-03        &        19.2      \\          
IGR J18450-0435 &    0728370801 (b) &  2014-10-13        &        22.9      \\   
IGR J18483-0311 &    0694070101     &  2013-04-18        &        57.6      \\       
\hline
\end{tabular}
\label{tab:obslog}
\end{table}

 	 \section{Results}
 	 \label{results}

In Fig.~\ref{fig:orig_bb} we show the comparison between a SFXT light curve 
with a uniform binning and the one obtained with a B.b. segmentation.
In Appendix (Figs.~\ref{fig:lc1} and~\ref{fig:lc2}) the B.b. light curves of the other SFXTs in our sample are reported.

The adoption of the B.b. segmentation of the light curves offers 
an efficient and systematic way to select a flare without the need of
assuming any specific model for its profile. 
Indeed, we have considered a ``flare" as a statistically significant peak (i.e., a B.b. containing a local maximum), 
with respect to the surrounding, adjacent emission.
We also used the local minima (``valleys") to calculate the duration of each flare (see below).
This automated procedure picked out 144 SFXT flares.

\begin{table}
 \centering
  \caption{Source properties (see \citealt{SP2018} and referencees therein).
}
 \begin{tabular}{lcccc}
\hline
Name                   & Dist        & $P_{\mathrm{orb}}$    &  orbital      &     $P_{\mathrm{spin}}$   \\
                       & (kpc)       &  (d)                  & eccentricity  &        (s)          \\
\hline
IGR J08408-4503 &       2.7          &       9.54            &  0.63    &    $-$          \\     
IGR J11215-5952 &      7.0$\pm{1.0}$ &     164.6             &  $>$0.8  &   187          \\   
IGR J16328-4726 &      7.2$\pm{0.3}$ &      10.07            &  $-$     &    $-$             \\     
IGR J16418-4532 &      13            &       3.75            &  0.0     &    1212          \\    
XTE J1739-302   &      2.7           &      51.47            &  $-$     &    $-$          \\    
IGR J17544-2619 &      3.0$\pm{0.2}$ &       4.93            &  $<$0.4  &   71.49($^a$)     \\   
IGR J18410-0535 &       3$\pm{2}$    &       6.45            &  $-$     &    $-$       \\     
IGR J18450-0435 &      6.4           &       5.7             &  $-$     &    $-$       \\          
IGR J18483-0311 &      3.5$\pm{0.5}$ &      18.52            &$\sim$0.4 &  21.05        \\       
\hline
\end{tabular}
\flushleft{
 $^{a}$This spin period is still uncertain.}
\label{tab:prop}
\end{table}

\subsection{Measuring observational quantities}
\label{sect:obsdef}

After the selection of the local maxima and minima in the light curves, 
we have estimated  the following temporal quantities 
(which are also explained graphically in Fig.~\ref{fig:explain}, for clarity):

\begin{itemize}

\item {\bf Waiting time ($\Delta T$):} we define it as the time interval between the peaks of subsequent flares. 
Given our B.b. segmentation of the SFXT light curve, for each B.b. containing a flare peak (a local maximum), 
the waiting time  is calculated as the difference between the midtime of the B.b. containing the peak of the flare,
and the midtime of the B.b. containing the peak of the previous flare. 
For all the first flares of the light curves the waiting time could not be calculated.
The observations are uninterrupted so that, within each EPIC observation, $\Delta T$s do not suffer any bias, as well as other parameters in Fig.~\ref{fig:explain}.
In Fig.~\ref{fig:explain}, $\Delta T$=[t$_2$ + 0.5(t$_3$ -- t$_2$) -- t$_0$].

\item  {\bf Rise time ($\delta t_{rise}$)}
to the peak of the flare: for unresolved flares (those made of a single B.b.), the rise time
is calculated as $\delta t_{rise}$= $\Delta$R/S, 
where  $\Delta$R is the difference in count rate between the two adjacent B.b.  
(the one containing the flare, and the B.b. located immediately before it), 
and ``S" is the positive slope measured before the peak (in units of counts~s$^{-1}$~ks$^{-1}$). 
For the resolved flares (those spanning more than one B.b.), 
it is calculated as the time interval between the stop time of the B.b. containing the valley before the flare, 
and the start time of the block containing the peak of the flare. In Fig.~\ref{fig:explain}, $\delta t_{rise}$=t$_2$-t$_1$.
These definitions return the best estimates for the flare rise time.
However note that  in case of unresolved flares, the true rise time to the peak might be formally (although unphysically) zero.
On the other hand, the rise time defined above for resolved flares can be considered a minimum value, by definition. 
The same is valid for the decay time, as defined below.

\item  {\bf Decay time ($\delta t_{decay}$)} from the peak of the flare: for the unresolved flares, 
it is calculated as $\delta t_{decay}$= $\Delta$R/S, where  $\Delta$R is the difference in count rate 
between the two adjacent B.b. (the one containing the peak, and the B.b. just after it), 
and ``S'' is the (negative) slope measured at the peak (in units of counts~s$^{-1}$~ks$^{-1}$). 
Note that we have considered the absolute value of the decay times.   
For the resolved flares (those spanning more than one B.b.), it is calculated as the time interval between 
the stop time of the  block containing the peak of the flare and the start time of the block containing the valley next to it.
In Fig.~\ref{fig:explain}, $\delta t_{decay}$=t$_5$-t$_3$.

\item  {\bf Flare Duration ($\Delta$t$_{f}$):} 
it is defined as the time interval comprised between the two local minima surrounding the flare,
subtracting the time interval covered by the B.b. containing these same minima. 
Therefore, it can be calculated as the sum of the time intervals covered by the B.b. in-between two local minima. 
For flare peaks  which do {\em not} lie 
between two local minima (this might occur at the beginning and/or at the end of an observation), the total 
duration cannot be  calculated. 
In Fig.~\ref{fig:explain}, $\Delta$t$_{f}$=t$_5$-t$_1$.
The only exception to this rule is the first peak in the light curve of the SFXT IGRJ18410--0535: since its profile is very 
well defined (fast rise and exponential decay, hereafter FRED), we could measure its duration, although formally a valley is not present before it.

\end{itemize}

Besides the above timescales, we have calculated the following quantities, for each flare $j$:

\begin{itemize}

\item {\bf Flare peak Luminosity (L$_j$)}: the EPIC count rate of the B.b. containing the peak of the 
flare has been converted to unabsorbed flux (1--10 keV) using \textsc{WebPIMMS}\footnote{https://heasarc.gsfc.nasa.gov/cgi-bin/Tools/w3pimms/w3pimms.pl} 
assuming a power law spectrum with a photon index $\Gamma$=1 
and a column density N$_H$=1.5$\times$10$^{22}$~cm$^{-2}$. \footnote{This spectral shape represents an average between somehow 
harder X-ray emission observed in SFXT bright flares (e.g. \citealt{Sidoli2007}), 
and softer emission from  
outside outbursts \citep{Sidoli2008:sfxts_paperI}.  
The energy range 1--10 keV has been assumed to better compare with the literature.  
}
This resulted in a conversion factor of 10$^{-11}$~erg~cm$^{-2}$~count$^{-1}$ (EPIC pn).
Then, the flare peak  X-ray luminosities have been calculated assuming the source distances 
reported by \citet{SP2018} and listed  in Table~\ref{tab:prop}, for clarity. 
The luminosity is also reported in the y-axis, on the right side of the graphs in Figs.~\ref{fig:lc1} and~\ref{fig:lc2}, 
to enable a proper comparison between different sources.

\item {\bf Energy (E$_j$) released in a flare $j$:} it is calculated summing the products (L (B.b.$_i$)$\times$ dur(B.b.$_i$), where L(B.b.$_i$) is the luminosity
reached by a single block $i$, while dur(B.b.$_i$) is the time duration of the block $i$) 
over the blocks  covered by a single flare:

\begin{equation}
E_j = \sum\nolimits_{B.b._i} L (B.b._i) \times dur  (B.b._i)  
\end{equation}

\item {\bf Average Luminosity ($\langle{L}_j\rangle$) during a flare $j$: }

For each flare $j$, we calculate the average X-ray luminosity as: 
 
\begin{equation}
\langle{L}_j\rangle = E_j / \Delta t_{f_j}  
\end{equation}

where  $\Delta~t_{f_j}$ is the total duration of the flare $j$ (as defined before).

\end{itemize}

The last quantity relevant for this investigation is the  pre-flare X--ray luminosity (or accretion rate; L$_{X_q}$).
With this term we mean  the X--ray luminosity level displayed by the valley just before each flare.

We report in Appendix 
the Table~\ref{tab:flares} with the values of the flare quantities defined above.
We collected 144 SFXT flares, which  can be ``unresolved" or ``resolved", depending on whether they extend on just one or more B.b., respectively. 
About one third of the flares are unresolved. We have clearly marked them with an asterisk in Table~\ref{tab:flares}.

\begin{figure*}
\includegraphics[scale=0.31,angle=0]{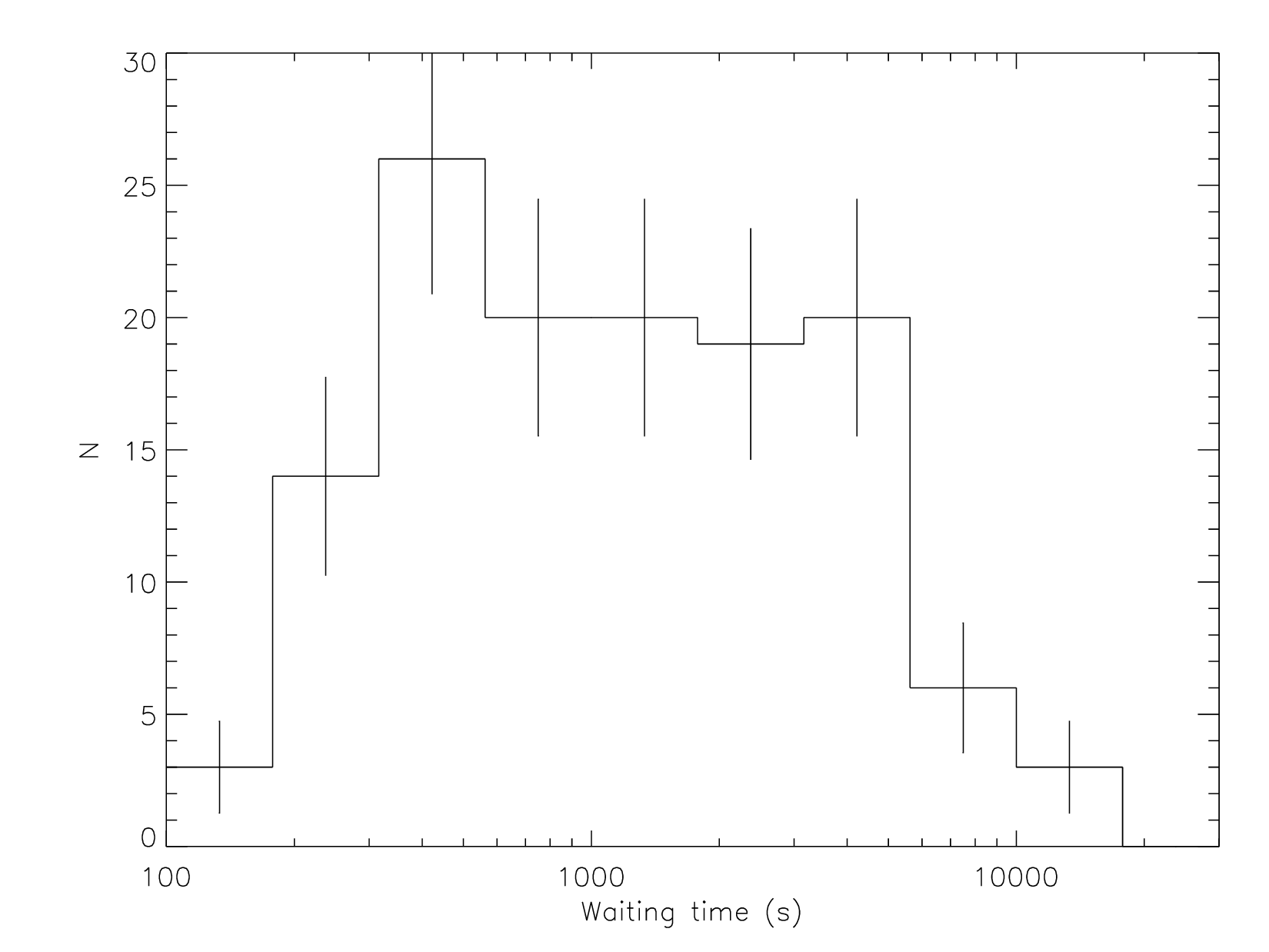} 
\includegraphics[scale=0.31,angle=0]{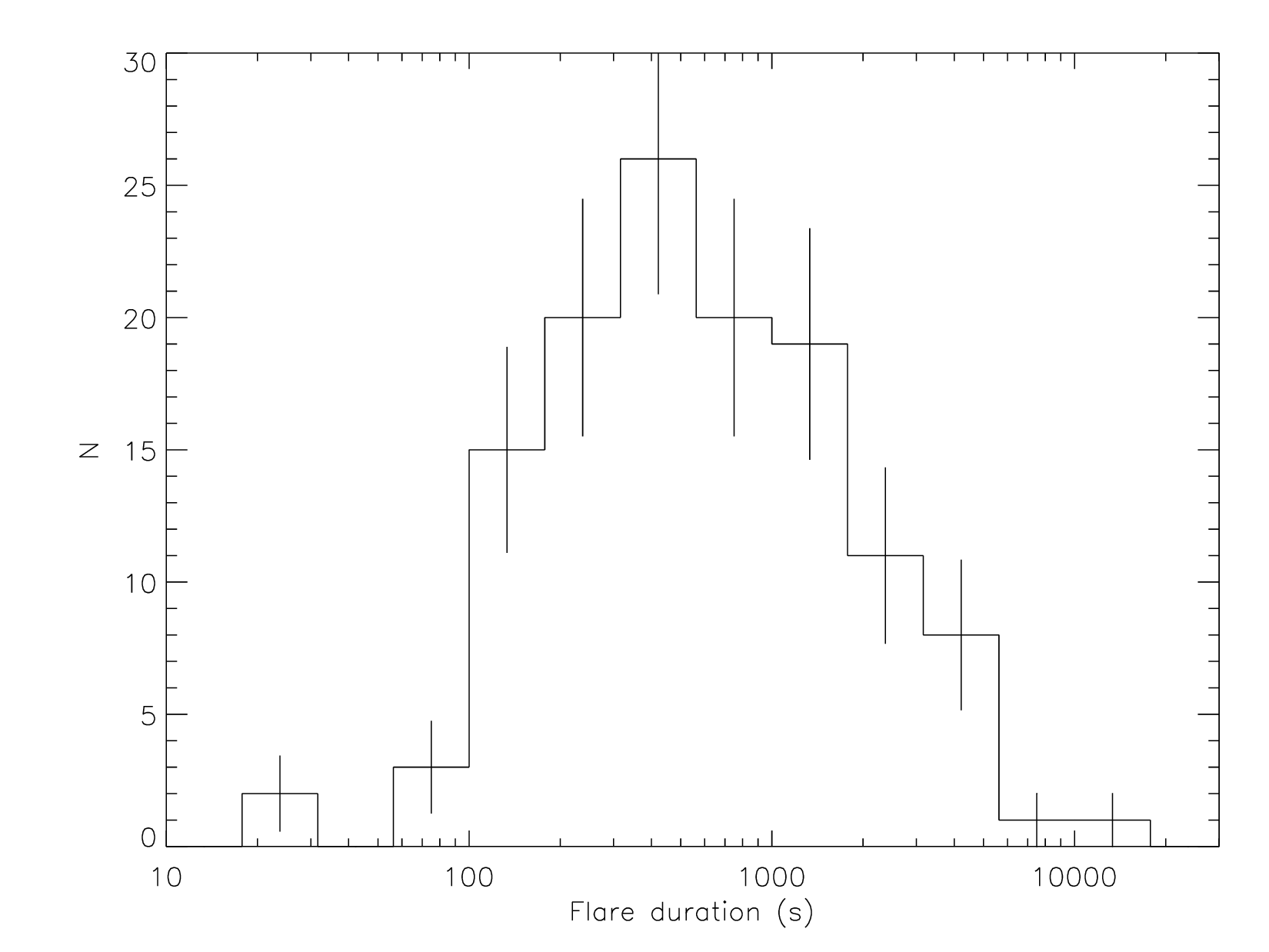} \\
\includegraphics[scale=0.31,angle=0]{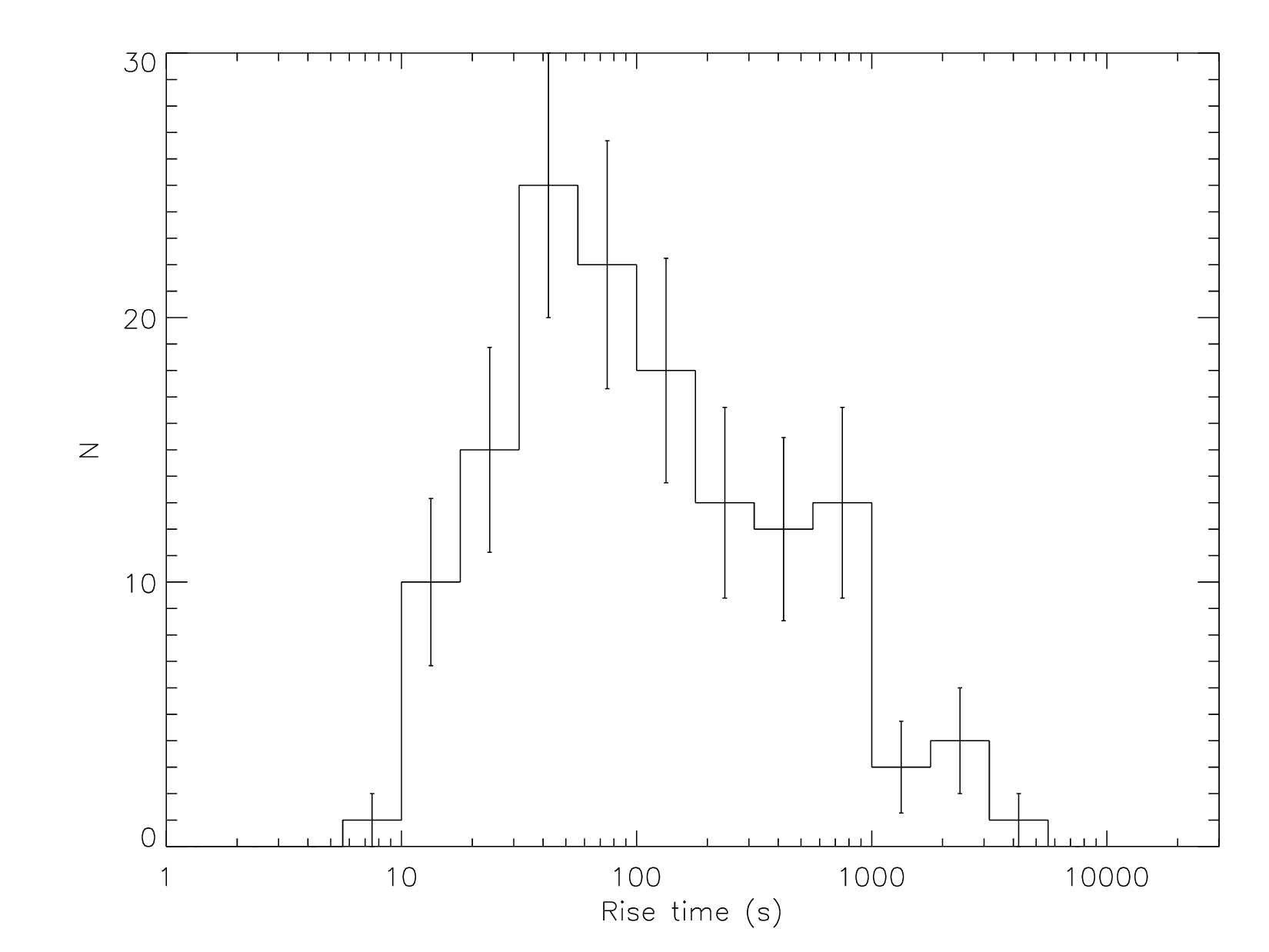} 
\includegraphics[scale=0.31,angle=0]{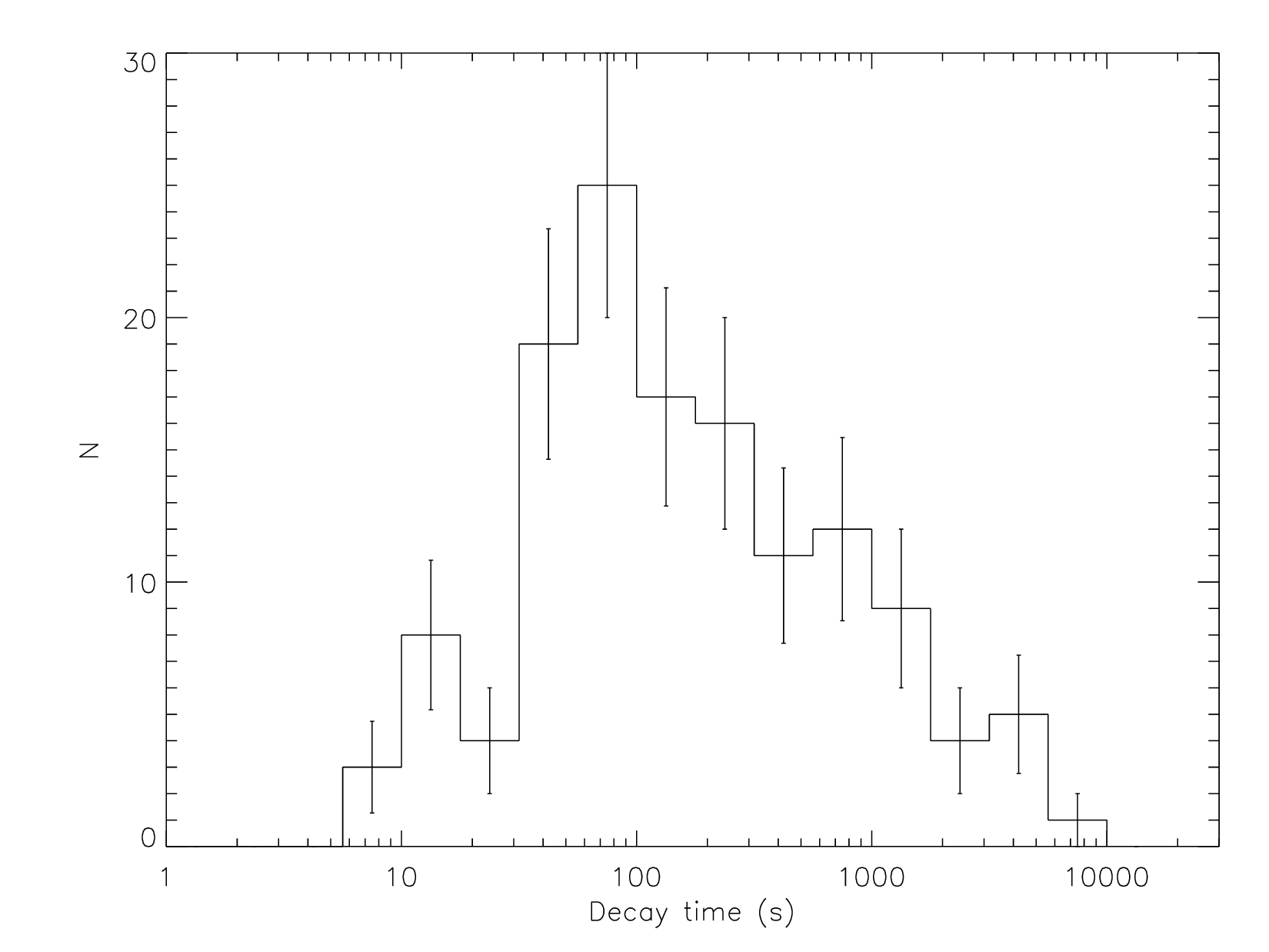} \\
\caption{Histogram of the flare timescales adopting a logarithmic binning: from top to bottom, from left to right: 
waiting times ($\Delta T$), rise times ($\delta t_{rise}$), flare durations ($\Delta t_f$) and decay times ($\delta t_{decay}$). 
The error bars are $\pm{\sqrt{N}}$.
}
\label{fig:histotemp}
\end{figure*}

\begin{figure*}   
\includegraphics[scale=0.31,angle=0]{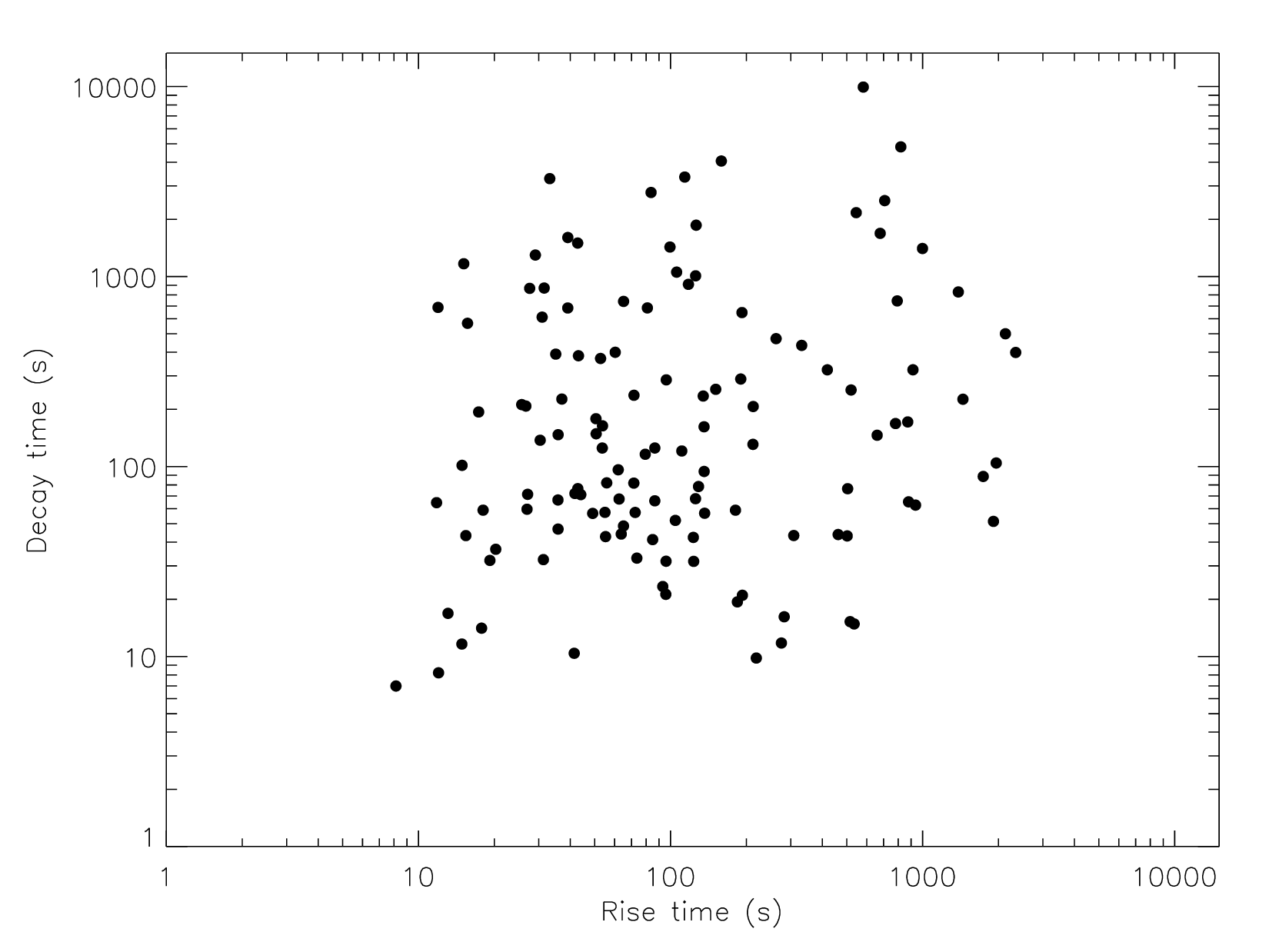} 
\includegraphics[scale=0.32,angle=0]{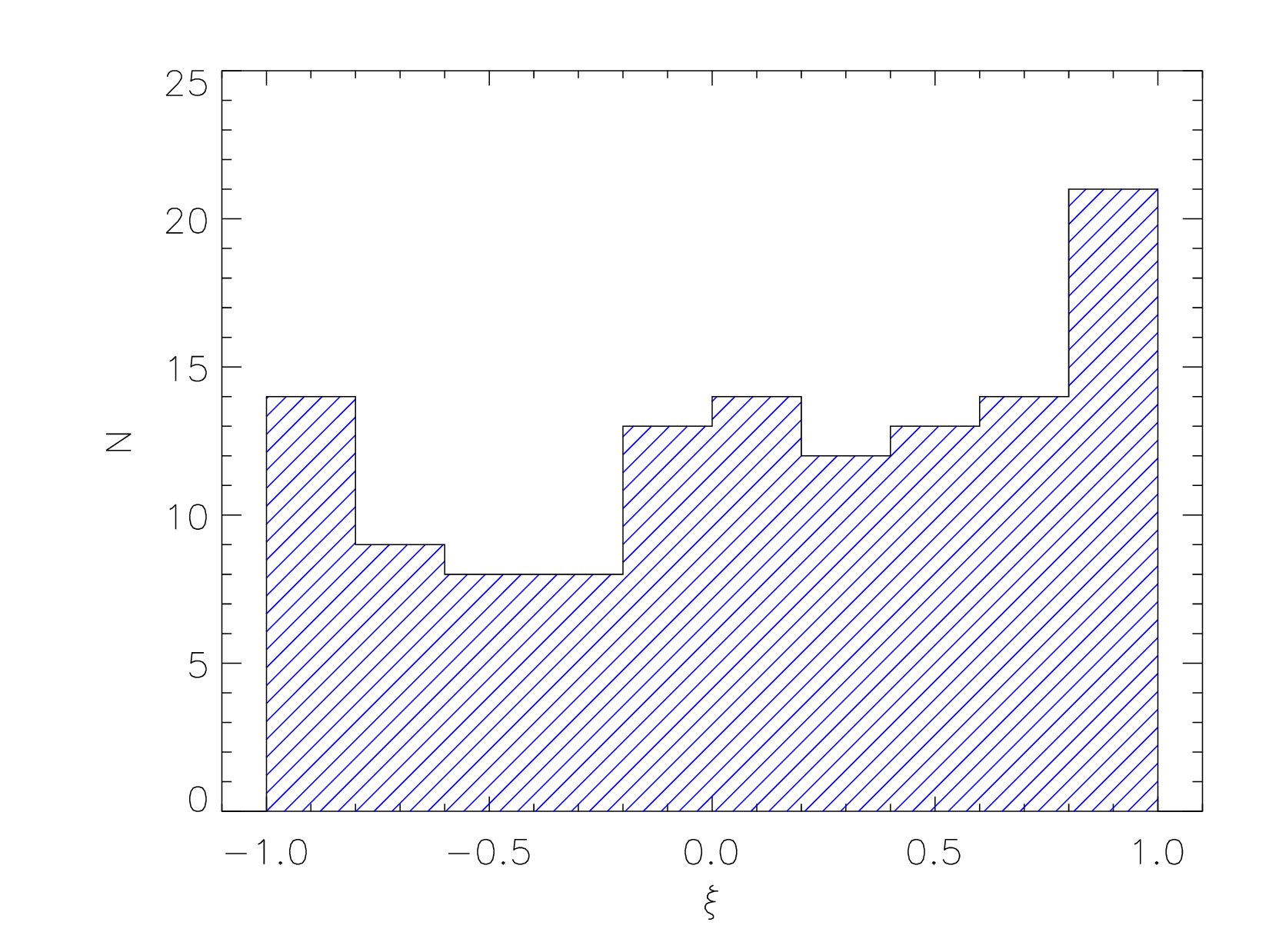} 
\caption{Flare rise and decay times. Their dependence is shown on the left, while 
the histogram of the flare asymmetry parameter, $\xi$,  is displayed on the right (see the text for its definition). 
}
\label{fig:csi}  
\end{figure*}

\begin{figure*}
\includegraphics[scale=0.33,angle=-90]{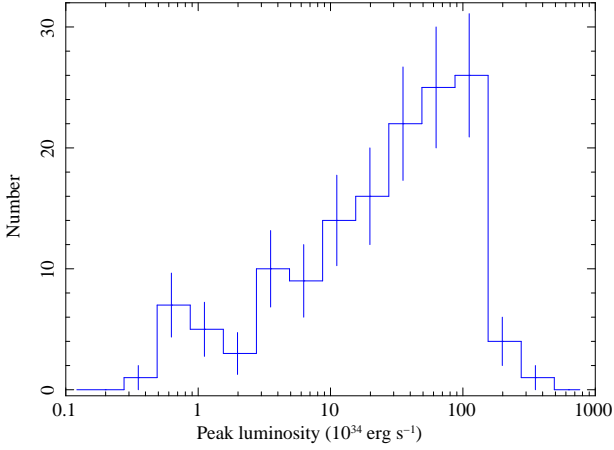}
\includegraphics[scale=0.33,angle=-90]{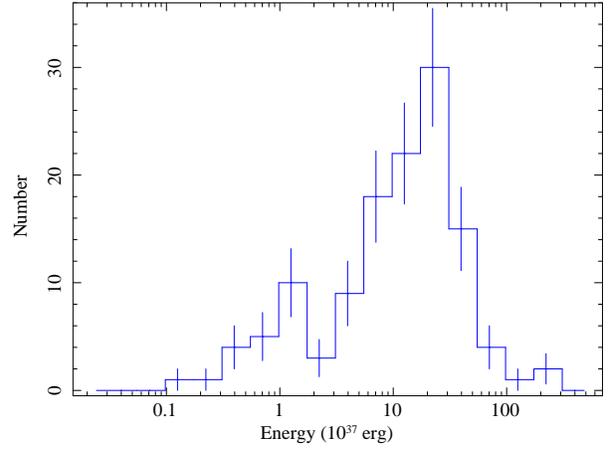} \\
\caption{Histogram of the luminosity at the flare peak (on the left) and of the energy released during flares (on the right), 
in logarithmic binning.
}
\label{fig:histo_ene_lx}
\end{figure*}
 
\begin{figure}
\includegraphics[scale=0.31,angle=0]{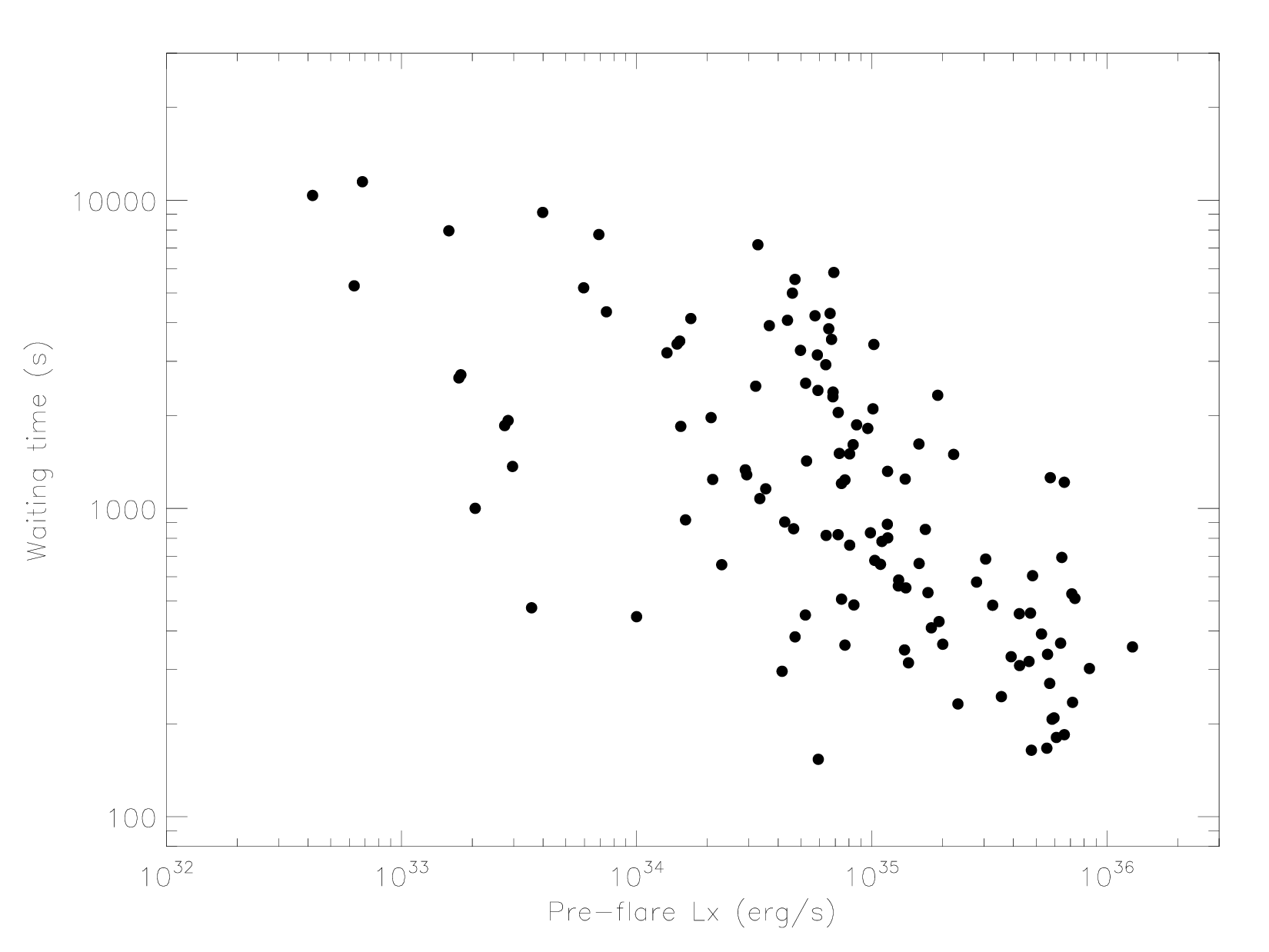} 
\caption{Flare waiting time against the pre-flare X-ray luminosity. For individual sources, see Fig.~\ref{fig:wait_lx_quiesc_sources}.
}
\label{fig:wait_lx_quiesc} 
\end{figure}

\begin{figure}
\includegraphics[scale=0.31,angle=0]{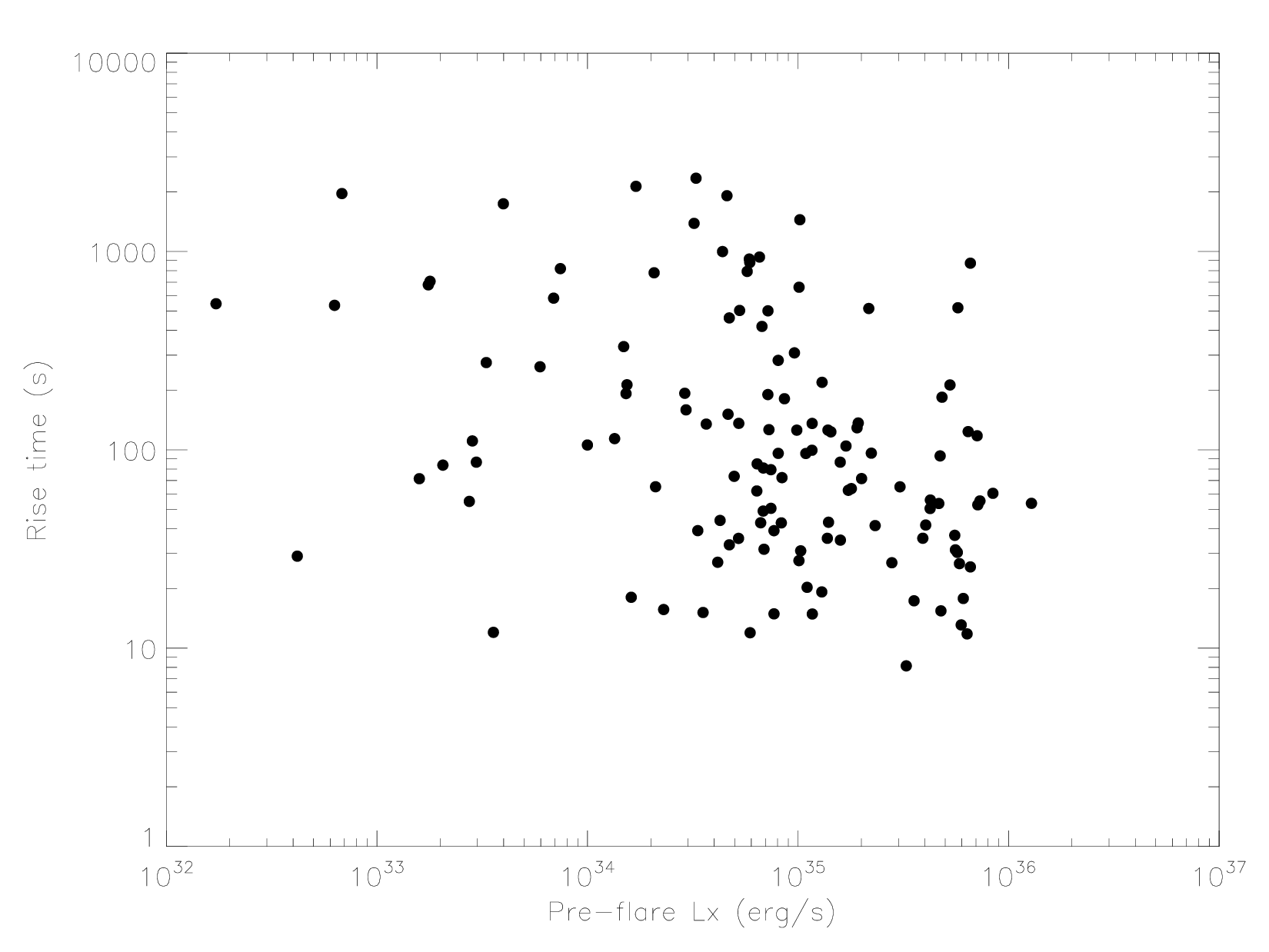} 
\caption{Rise time to the flare peak vs pre-flare X-ray luminosity (defined as the luminosity before the flare). 
For individual sources, see Fig.~\ref{fig:riset_vs_lx_quiesc_sources}.
}
\label{fig:riset_vs_lx_quiesc}
\end{figure}

\begin{figure}
\includegraphics[scale=0.31,angle=0]{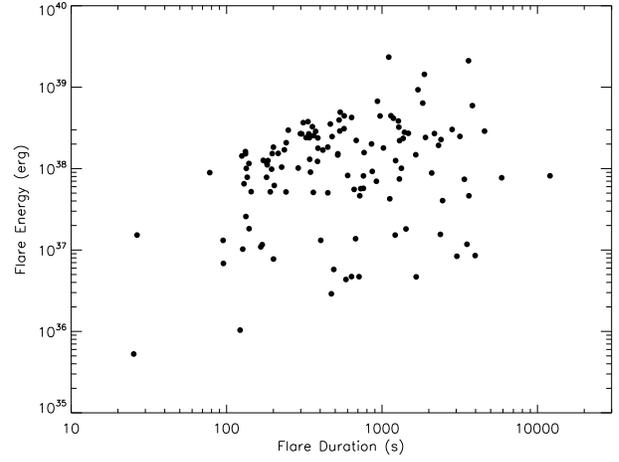} 
\caption{Energy released in flares vs  flare duration. For individual sources, see Fig.~\ref{fig:ene_dur_sources}. 
}
\label{fig:ene_dur}
\end{figure}

\begin{figure}
\includegraphics[scale=0.31,angle=0]{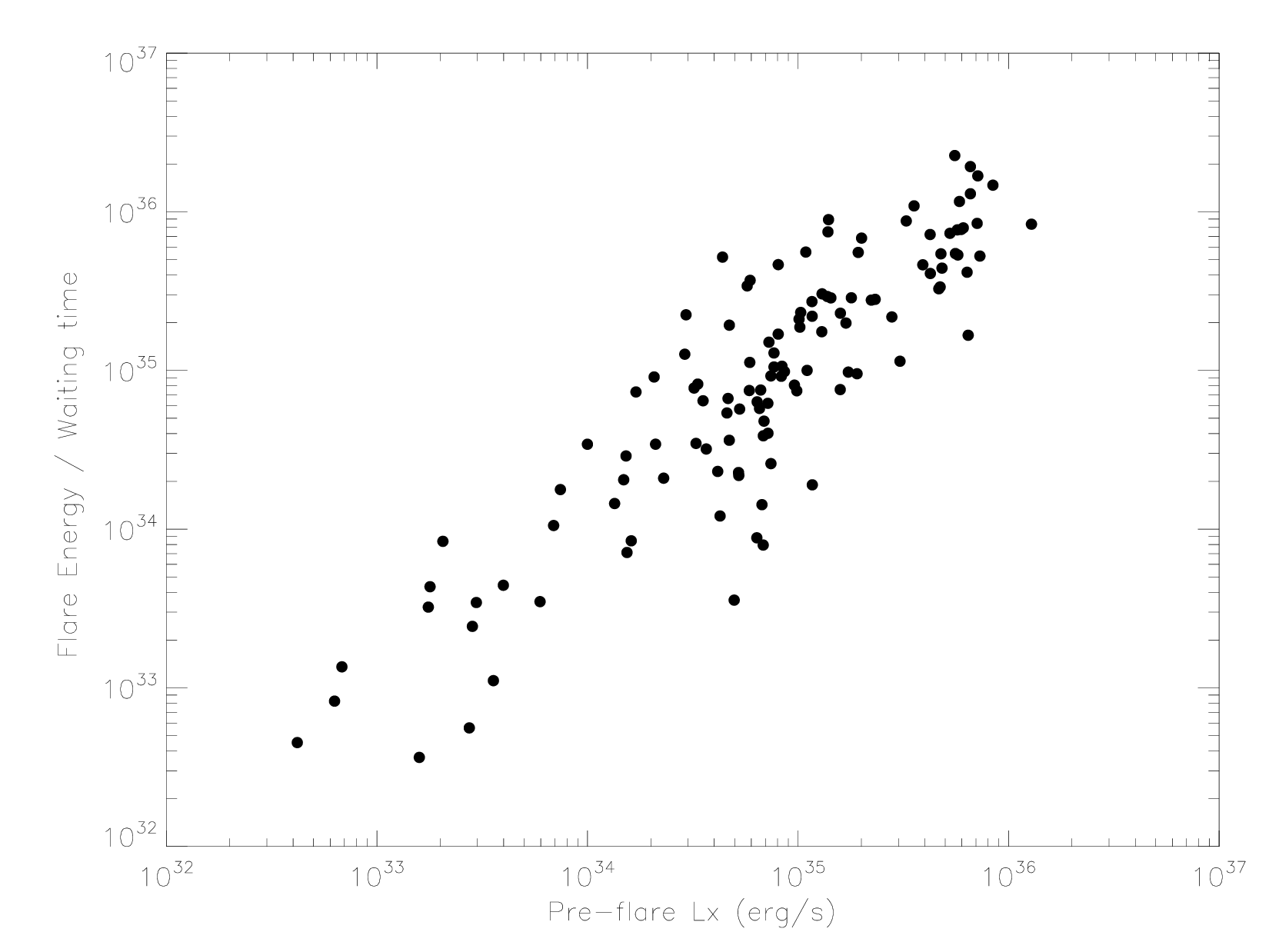} 
\caption{Ratio of the energy released in flares to the waiting times between consecutive flares, 
plotted against the pre-flare luminosity, defined as the
luminosity level at the local minimum just before the flare.  
For individual sources, see Fig.~\ref{fig:ene_wt_vs_lx_quies_sources}. 
}
\label{fig:ene_wt_vs_lx_quies}
\end{figure}

\section{Discussion} \label{discussion}

The automated procedure we have adopted to select the SFXT flares from \xmm\ light curves 
has lead to the measurement of observational quantities which can be now compared with the theory.
In Sect.~\ref{disc:rti}
we will discuss the phenomenology of the flares in the framework of 
the development of the Rayleigh-Taylor instability (RTI) in accreting plasma trying 
to enter the magnetosphere of  slowly rotating neutron stars.

In order to perform this comparison, we will highlight the behaviour of flares from single sources, 
within the global behaviour of SFXT flares taken as a whole. 
In this way it is somehow possible to identify
trends which are  not found in flares occurred in a single source, but are
due to the superposition of flares from different sources, lying in different regions of the parameter space, 
and {\em vice versa}: trends in single sources might be in principle 
mixed up when all SFXT flares are considered together.

In Fig.~\ref{fig:histotemp} we show the overall distributions of flare temporal properties 
(flare duration, waiting times, rise and decay times), while 
in Fig.~\ref{fig:csi} (left panel), the decay time is plotted against the rise time,
to investigate the flare shape. 
In order to quantify the degree of asymmetry in the flare profiles, we calculated the parameter
$\xi$ = ($\delta t_{decay}$ - $\delta t_{rise}$) / ($\delta t_{decay}$ + $\delta t_{rise}$).
This implies that flares where the rise to the peak is much faster than the decay have $\xi$$\sim$1 (as in FRED-like profiles),
while flares with a much slower rise than the decay show $\xi$$\sim-$1.

The result is reported in the right panel of
Fig.~\ref{fig:csi}, where it is evident that all range of values is covered by the SFXT flares analysed here.

We show in Fig.~\ref{fig:histo_ene_lx} the histograms of the peak luminosity and of the energy emitted during flares.
From the latter histogram, a bimodality in the energy released in flares might be present, 
above and below $\sim$2$\times$10$^{37}$~erg, but the relatively low statistics 
do not permit to draw a firm conclusion. 
Note that most of the flares with energies below 10$^{37}$~erg are contributed by the source XTE~J1739--302.

In Figs.~\ref{fig:wait_lx_quiesc}, ~\ref{fig:riset_vs_lx_quiesc}, ~\ref{fig:ene_dur} and \ref{fig:ene_wt_vs_lx_quies}
we show flare global behaviours  which will be compared with the theory in the next subsections. 
From Fig.~\ref{fig:wait_lx_quiesc} an apparent anticorrelation is present between 
the waiting time between two adjacent flares and the X-ray luminosity in-between flares. 
The plot of energy released in flares versus the flare duration might indicate a positive trend, 
while the plot of the rise time of all flares, versus their pre-flare luminosities does not apparently show any correlation.
A correlation is shown over several orders of magnitude
by the ratio between the flare energy and the waiting time versus the pre-flare luminosity.
Similar plots are reported in  
Figs.~\ref{fig:wait_lx_quiesc_sources}, ~\ref{fig:riset_vs_lx_quiesc_sources}, ~\ref{fig:ene_dur_sources} and \ref{fig:ene_wt_vs_lx_quies_sources}, where the flares from single sources are highlighted. 
They are discussed in Sect.~\ref{disc:rtigrowth}.

\begin{figure}  
\includegraphics[scale=0.31,angle=0]{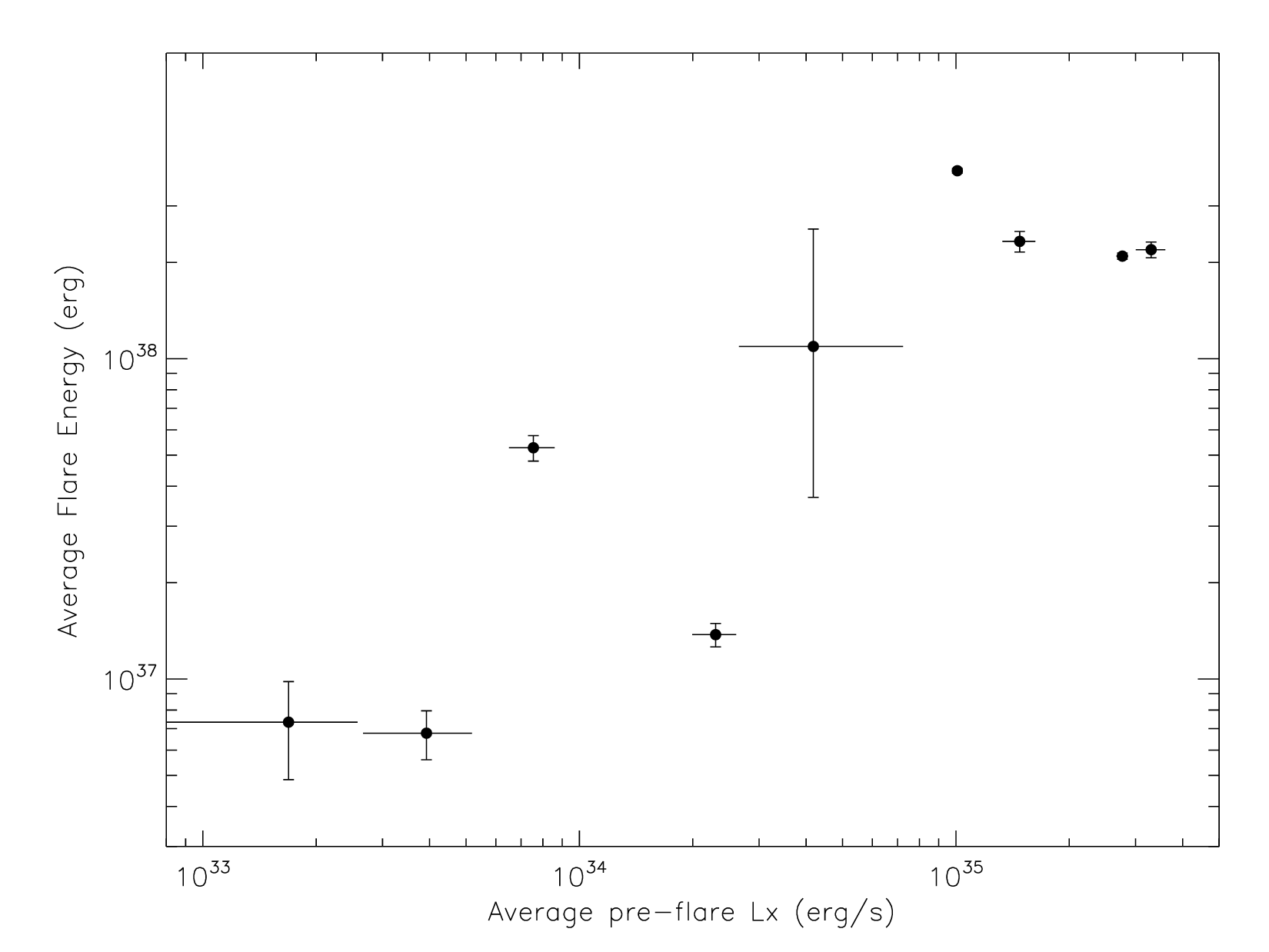}
\caption{Average flare energy vs average pre-flare luminosity. Each solid circle indicates a source.
}
\label{fig:av_ene_lx_quiesc}
\end{figure}

We note that the same \xmm\ observations reported here have been  analyzed by \citet{Gimenez2015}, \citet{Bozzo2017}, \citet{Pradhan2018} (and references therein). Although their temporal-selected spectra of flares were extracted from intervals much longer than the B.b.s adopted here (so that they cannot be directly compared), some spectra of SFXT flares showed a column density larger than 1.5$\times$10$^{22}$~cm$^{-2}$ (the value we assume for all flares), implying, in principle, a larger conversion factor from count rate to flux. However, even a very high absorption of $\sim$2$\times$10$^{23}$~cm$^{-2}$ would imply only a $\sim$3 times larger unabsorbed flux (1-10 keV), with no impact on the conclusions of our work, where observational facts and theory are compared over four orders of magnitude in X-ray luminosity (and emitted energy).

\subsection{Rayleigh-Taylor instability}
\label{disc:rti}

It has long been recognized that plasma entry in NS magnetosphere in accreting X-ray binaries occurs via interchange instability -- Rayleigh-Taylor (RTI) in the case of slowly rotating NSs \citep{1976ApJ...207..914A,1977ApJ...215..897E}  or Kelvin-Helmholtz (KH) in rapidly rotating NSs  \citep{1983ApJ...266..175B}. 
In the case of disc accretion, the plasma penetration into magnetosphere via RTI was compellingly demonstrated by multi-dimensional numerical MHD simulations \citep{2008MNRAS.386..673K}. However, global MHD simulations of large NS magnetospheres ($\sim 10^9$ cm) have not been performed yet, and information about physical processes near NS magnetospheres should be inferred from observations. 

During quasi-spherical wind accretion onto slowly rotating NSs, there is a characteristic X-ray luminosity $L^*\simeq 4\times 10^{36}$~\ergs that separates two physically distinct accretion regimes: the free-fall Bondi-Hoyle supersonic accretion occurring at higher X-ray luminosity, when the effective Compton cooling time of infalling plasma is shorter than the dynamical free-fall time \citep{1984ApJ...278..326E}, and subsonic settling accretion at lower luminosities, during which a hot convective shell forms above the NS magnetosphere \citep{Shakura2012,2018ASSL..454..331S}. In the latter case, a steady plasma entry rate is controlled by plasma cooling (Compton or radiative) and is reduced 
compared to the maximum possible value determined by the Bondi-Hoyle gravitational capture rate $\dot M_B$ from the stellar wind by a factor $f(u)^{-1}\approx (t_\mathrm{cool}/t_\mathrm{ff})^{1/3}>2$. 
 
The necessary conditions for settling accretion are met at low-luminosity stage in SFXTs.
Low X-ray luminosities make it difficult 
to detect X-ray pulsations and therefore to answer the question of where the observed X-rays are actually produced. They can be either generated near the NS surface (if the inefficient plasma entry rate into the magnetosphere is provided by diffusion, cusp instabilities, etc., as discussed e.g. by \citealt{1984ApJ...278..326E}), or be a thermal emission from magnetospheric accretion, like in the model developed for $\gamma$ Cas stars \citep{2017MNRAS.465L.119P}. It is quite possible that at low-luminosity states of SFXTs, no RT-mediated plasma penetration into the NS magnetosphere occurs at all. This may be the case if the plasma cooling time is longer than the time a plasma parcel spends near the magnetosphere because of convection: $t_\mathrm{cool}>t_\mathrm{conv}\sim t_\mathrm{ff}(R_\mathrm{B})\sim 300-1000$~s. Once this inequality is violated, 
RTI can start to develop. 

At this stage, the magnetospheric instability can occur for different reasons. For example, it was conjectured \citep{Shakura2014} that bright flares in SFXTs are due to sudden break of the magnetospheric boundary caused by the magnetic field reconnection with the field carried along with stellar wind blobs. This can give rise to short strong outbursts occurring in the dynamical (free-fall) time scale during which accretion rate onto NS reaches the maximum possible Bondi value from the surrounding stellar wind.   

Another reason for the instability can be due to stellar wind inhomogeneities which can disturb the settling accretion regime and even lead to free-fall Bondi accretion episodes.

At typical X-ray luminosities $L_q\simeq 10^{33}-10^{34}$ erg s$^{-1}$, SFXTs occasionally demonstrate 
less pronounced flares with phenomenology as described in previous Sections. 
Below we present a possible scenario of development of such flares based on the consideration of non-linear growth of RTI. 
During these flares, an RTI-mixed layer is advected to the NS magnetosphere, but X-ray power generated is still below $\sim 10^{36}$~\ergs 
to provide effective plasma cooling for steady RTI. In a sense, the observed short flares during low-luminosity SFXT state are due to `failed' RTI.

These will enable us to explain, without making additional assumptions, the main observed properties of SFXT flares inferred from the statistical analysis presented in this paper.
In all numerical estimates below, we assume the NS mass $M_x=1.5 M_\odot$ and normalize the NS magnetospheric radius as $R_\mathrm{m}=10^9 [\mathrm{cm}] R_9$, the mass accretion rate onto NS as $\dot M_x=10^{16}[\mathrm{g\,s}^{-1}]\dot M_{16}$
and the NS magnetic moment as $\mu=10^{30}[\mathrm{G\,cm}^3] \mu_{30}$. 

\subsection{Non-linear RTI growth}
\label{disc:rtigrowth}

At the settling accretion stage, in a subsonic convective shell around the NS magnetosphere, external wind perturbations gravitationally captured from stellar wind of the optical components at the Bondi radius $R_\mathrm{B}=2GM_x/v_w^2$ travel down to the NS magnetosphere $R_\mathrm{m}$ with the convective motions. Therefore, the response time of the magnetosphere to the external perturbations is not shorter than about free-fall time from the Bondi radius, $t_\mathrm{ff}(R_\mathrm{B})=\sqrt{R_\mathrm{B}^3/2GM_x}$, ($R_\mathrm{B}=2GM_x/v^2\simeq 4\times 10^{10}[\mathrm{cm}]v_8^{-2}$~ cm is the Bondi radius for the relative wind  velocity $v=10^8[\mathrm{cm\,s}^{-1}]v_8$), typically a few hundred seconds for the stellar wind velocity from OB-supergiant $v_w\sim 1000$~ km s$^{-1}$. 

In an ideal case with constant boundary conditions, the development of RTI occurs via production of a collection of bubbles (or, rather, flutes) with different size \citep{1976ApJ...207..914A}, and the mean plasma entry rate $u$ into magnetosphere is determined by the slowest linear stage of RTI in a changing effective gravity acceleration determined by plasma cooling \citep{Shakura2012,2018ASSL..454..331S}: $u\approx f(u)v_\mathrm{ff}(R_\mathrm{m})$, 
where $v_\mathrm{ff}(R_\mathrm{m})=\sqrt{2GM_x/R_\mathrm{m}}$ is the free-fall velocity at the magnetospheric radius.

In the quiescent state, the mass accretion rate is determined by the average plasma velocity $u$ at the magnetospheric boundary, $\dot M_z=4\pi R_\mathrm{m}^2\rho u$. At low X-ray luminosity $\lesssim 10^{35}$ erg s$^{-1}$, radiative plasma cooling dominates the Compton cooling. The characteristic plasma cooling time is 
\beq{e:radcool}
t_\mathrm{rad}\approx \frac{3k_\mathrm{B}T}{n_e\Lambda(T)}   
\eeq
where $n_e$ is the electron number density, $\Lambda(T)\approx 2.5\times 10^{-27}\sqrt{T}$ [erg cm$^3$~s$^{-1}$] is thermal cooling function dominated by bremsstrahlung at the characteristic temperatures of the problem (1-10 keV). Taking into account that the plasma temperature at the basement of the shell near the magnetosphere is about the adiabatic value, $T\approx 2/5(GM_x)/{\cal R}R_\mathrm{m}\simeq 10^{10}R_9$~K, and by 
expressing $n_e$ from the mass continuity equation,  we obtain the radiative cooling time
\beq{e:radcool1}
t_\mathrm{rad}\approx 7\times 10^3 [\mathrm{s}] \frac{R_9}{\dot M_{16}} f(u)_\mathrm{rad}\sim 700 [\mathrm{s}]\zeta^{2/9}\mu_{30}^{2/3}\dot M_{16}^{-1}\,,
\eeq{}
In the last equation, the magnetospheric radius and factor $f(u)_\mathrm{rad}$ are derived as \citep{2013MNRAS.428..670S,2018ASSL..454..331S}: 
\beq{e:Rm}
R_\mathrm{m}\approx 10^9 [\mathrm{cm}] \zeta^{4/81}\mu_{30}^{16/27}\dot M_{q,16}^{-6/27}
\eeq
\beq{e:fu}
f(u)_\mathrm{rad}\approx 0.1 \zeta^{14/81}\mu_{30}^{2/27}\dot M_{16}^{6/27}
\eeq
(the dimensionless parameter $\zeta\lesssim 1$ characterizes the size of the RTI region in units of the magnetospheric radius $R_ m$).

Increase in the density $\rho_m$ near the magnetosphere would shorten the plasma cooling time and lead to an increase in the X-ray photon production from the NS surface, which in turn would enhance the Compton  plasma cooling and increase the plasma entry rate $f(u)$. This would result in an X-ray flare (or a collection of flares) on top of the quiescent X-ray luminosity level.

\subsubsection{Flare waiting time}

We start with the estimate of the flare waiting time $\Delta T$. 
Consider the spreading of the RTI layer during the instability development.
If there is no plasma penetration into the NS magnetosphere, the thickness of RTI mixing layer at the late non-linear stage grows as   
\beq{}
Z\sim \alpha A gt^2\,,
\eeq
where $g=GM_x/R_\mathrm{m}^2$ is the gravity acceleration, the dimensionless factor $\alpha\sim 0.03$, $A\lesssim 1$ is the effective Atwood number (see e.g. \cite{2017A&A...605A.101C} for a recent discussion of numerical calculations of the non-linear growth of magnetic RTI).

However, in our problem the layer struggles against the mean plasma flow with velocity $u(t)$ determined by the plasma cooling that further slows down the RTI, and therefore the net distance the RTI layer extends above the magnetosphere is
\beq{e:Z'}
Z'=Z-\int u(t)dt 
\eeq
As long as the time is shorter than the cooling time, $t<t_\mathrm{rad}$, during 
the linear stage of RTI development in the unstable region we can write \citep{Shakura2012}:
\beq{}
u(t)=\frac{gt^2}{2t_\mathrm{rad}}\cos\chi
\eeq
(here $\chi$ is the latitude from the magnetosphere equator where RT modes are the most unstable; below we set $\cos\chi=1$). Therefore, \Eq{e:Z'} takes the form:
\beq{e:Z''}
Z'=\alpha A gt^2- g\frac{t^3}{6t_\mathrm{rad}}\,.
\eeq

With time, the second (negative) term in \Eq{e:Z''} overtakes the first (positive) one. The negative value of the RTI layer height above magnetosphere $Z'<0$ would inhibit instability growth because there will be no room for plasma flutes to interchange with magnetic field above the magnetospheric boundary (more precisely, above the layer in which the mean plasma entry rate is sustained for a given plasma cooling rate). Therefore, the growth of the RTI mixing layer size 
at the nonlinear stage should be restricted by the time for the net travel distance of rising blobs  above the magnetosphere to become zero.
 Thus, the time it takes for the RTI layer to grow is:
\beq{e:flaret}
\Delta T\approx 6\alpha A t_\mathrm{rad}\approx 0.18 \myfrac{\alpha}{0.03}A t_\mathrm{rad}\,.
\eeq

We can identify this time with intervals between
consecutive flares (the `waiting time'). That this time turned out to be of the order of the plasma cooling time is intuitively clear: the next portion of RT-unstable plasma is accumulated during the characteristic time needed for plasma to cool down to enable RTI. 

Substituting $t_\mathrm{rad}$ from \Eq{e:radcool} and the expression for the magnetospheric radius $R_\mathrm{m}$ for radiation cooling, \Eq{e:Rm}, into \Eq{e:flaret}, we find
\beq{e:tfrad}
\Delta T \approx 130 [\mathrm{s}] \myfrac{\alpha}{0.03} A \zeta^{2/9}\mu_{30}^{2/3}\dot M_{16}^{-1}\,.
\eeq
This estimate shows that flare waiting time can be as long
as a few thousand seconds.
In this model, the comparison of flare waiting times in a particular source enables us to evaluate the dimensionless combination of parameters $A\zeta^{2/9}<1$, which is impossible to obtain from theory.  

Fig.~\ref{fig:wait_lx_quiesc} displays the flare waiting time against the pre-flare luminosity.
Flares from single sources are overlaid  in Fig.~\ref{fig:wait_lx_quiesc_sources}.
The straight line indicates the dependence 
$\Delta T = 130  [\mathrm{s}] \dot M_{16}^{-1}$ (where $\dot M_{16}$ is, in this context, 
the accretion rate measured before each flare). 
Most flares from single sources follow this relation, with some scatter. 
Flares from XTE~J1739-302 are notable in following
this anticorrelation but with a significantly lower normalization. 
In this source,  we can derive $\myfrac{\alpha}{0.03} A \zeta^{2/9}\mu_{30}^{2/3}$$\sim$0.03 (the total range covered is 0.01-0.1). A similar situation may be valid for IGR~J08408-4503 (although with only two flares from this target it is impossible to draw any conclusion). Also flares from IGR~J18483-0311 appear sistematically shifted to 
a lower value of $\myfrac{\alpha}{0.03} A \zeta^{2/9}\mu_{30}^{2/3}$$\sim$0.3.
On the other hand, no trend is apparent from IGR~J18410-0535, probably because all these flares 
come from the decaying part of a FRED-like flare, a unique behaviour among SFXTs studied here.

\subsubsection{Flare rising time $\delta t$, duration $\Delta t$ and energy $\Delta E$}

\textbf{Flare rising time.} 
Consider a plasma blob rising due to RTI. When the instability starts, the rising blob struggles against the flow velocity towards the magnetosphere, so the net blob velocity is
\beq{e:vb}
v_b=a\sqrt{Ag{\cal R}}-gt^2/2t_\mathrm{rad}
\eeq
where $g=GM_x/R_\mathrm{m}^2$ is the gravity acceleration, the dimensionless factor $a\sim 0.1$, $A\lesssim 1$ is the effective Atwood number, ${\cal R}=2\piup/k$ is the blob curvature radius that we will associate with the instability wavelength $\lambda=2\pi/k$, $k$ is the wavenumber. 

In the convective shell above the magnetosphere, plasma is likely to be turbulent \citep{Shakura2012}.
In this case, the effective viscosity in the plasma is, according to the Prandtl rule, 
\beq{e:nut}
\nu_t=\frac{1}{3}v_tl_t
\eeq
where $v_t$ and $l_t$ is the characteristic turbulent velocity and scale, respectively. Below we shall scale these quantities with the free-fall velocity and magnetospheric radius, respectively: $v_t=\alpha_v v_\mathrm{ff}$, $l_t=\alpha_l R_\mathrm{m}$, so that the turbulent viscosity can be written in the form $\nu_t=(\alpha_t/3) v_\mathrm{ff}R_\mathrm{m}$,
where $\alpha_t=\alpha_v\alpha_l\lesssim 1$ is the effective turbulent viscosity parameter \citep{1973SvA....16..756S}.

One of the viscosity effect on RTI is the appearance of the fastest growing mode \citep{1974PhFl...17....1P}:
\beq{e:lmax}
\lambda_{max}=4\pi\myfrac{\nu^2}{Ag}
\eeq
Substituting \Eq{e:lmax} into \Eq{e:vb} using \Eq{e:nut}, from the condition $v_b=0$  we find the time of the most rapidly growing mode:
\beq{e:rt}
t_{k_{max}}
\simeq 1.5(A\alpha_t^2)^{1/6}\sqrt{t_\mathrm{ff}t_\mathrm{rad}}
\simeq 30[\mathrm{s}]\,\zeta^{4/27}\mu_{30}^{7/9}\dot M_{16}^{-2/3}\,.
\eeq
We assume that this is the time after which the entire RTI-mixed layer falls onto the NS in the dynamical time producing a flare. Therefore, we identify this time with the flare rising time, $\delta t_{rise}=t_{k_{max}}$ 
(see Figs.~\ref{fig:riset_vs_lx_quiesc} and \ref{fig:riset_vs_lx_quiesc_sources}).

\textbf{Flare duration}. 
During a flare triggered by an external perturbation, 
the mass $\Delta M$ accumulated in the mixing layer is assumed to fall onto the NS surface over the characteristic dynamical time of the entire shell
$t_\mathrm{ff}(R_\mathrm{B})= R_\mathrm{B}^{3/2}/\sqrt{2GM_x}$:
\beq{e:tf}
\Delta t \sim t_\mathrm{ff}(R_\mathrm{B})= \frac{R_\mathrm{B}^{3/2}}{\sqrt{2GM_x}}\approx 400 [\mathrm{s}]\myfrac{v_w}{1000\,[\mathrm{km\,s}^{-1}]}^{-3}
\eeq
As this time is most sensitive to the wind velocity, the observed dispersion in the flare duration should reflect the stellar wind velocity fluctuations, $\delta \Delta t/\Delta t=-3\delta v_w/v_w$, and cannot vary more than by a factor of 2-3. Apparently, the long duration of some flares may be a result of `gluing' of several shorter flares into a longer one.

See Fig.~\ref{fig:ene_dur_sources} for the range of flare duration in individual sources.

\textbf{Flare energy}. 
The mass accumulated in the RTI mixing layer can be estimated as $\Delta M=\dot M_q\Delta T$:
\beq{e:DM1}
\Delta M \approx 1.3\times 10^{18} [\mathrm{g}] \myfrac{\alpha}{0.03} A \zeta^{2/9}\mu_{30}^{2/3}\,.
\eeq
 In this approach, the characteristic flare energy due to accretion of the mass $\Delta M$ onto NS, 
$\Delta E = 0.1 \Delta M c^2$, turns out to be:
\beq{e:DE1}
\Delta E\approx 1.3\times 10^{38} [\mathrm{erg}] \myfrac{\alpha}{0.03} A \zeta^{2/9}\mu_{30}^{2/3}\,,
\eeq
which is very close to what is observed (see Fig.~\ref{fig:histo_ene_lx}). 

The mean mass accretion rate and hence mean X-ray luminosity during the flare is 
\beq{e:dMdt1}
\langle L \rangle \equiv \frac{\Delta E}{\Delta t}\approx 3\times 10^{35}[\mathrm{erg\,s}^{-1}]\myfrac{\alpha}{0.03} A \zeta^{2/9}\mu_{30}^{2/3}v_8^3\,.
\eeq

Remarkably, in this model the waiting time between flares $\Delta T$ is independent on the mass accretion rate $\dot M_x$ between flares (to within the possible dependence of the Atwood number $A$ and parameter $\zeta$ on the mass accretion rate).
Therefore, these qualitative considerations
suggest that the power of flares and mean accretion luminosity in flares in particular sources should be of the same value during individual \xmm\ observations. Moreover, the ratio of the flare energy to the waiting time
\beq{}
\frac{\Delta E}{\Delta T}=10^{36}[\mathrm{erg\,s}^{-1}]\dot M_{16}
\eeq
does not depend on unknown RTI parameters $\zeta, 
\alpha, A$ (which can vary in individual sources) 
and is proportional only to the mean mass accretion rate between the flares 
(Figs.~\ref{fig:ene_wt_vs_lx_quies} and ~\ref{fig:ene_wt_vs_lx_quies_sources}, for individual sources).  
This explains the tight correlation seen in these plots.
 
Clearly, the actual mass accretion rate during the flares in individual sources is determined by the RTI details (e.g., the fraction of the magnetospheric surface subject to the instability, the effective Atwood number etc.), which cannot be calculated 
theoretically. However, we note good agreement of the expected flare duration and mean flare energy obtained from this qualitative considerations with observations (Fig.~\ref{fig:ene_dur_sources}).
In these estimates, one should also keep in mind the inevitable dispersion, from source to source, in the NS magnetic field (see the dependence on
$\mu_{30}$ in above formulas).

If the external mass accretion rate from the stellar wind increases, however, the higher average mass accretion rate can be reached automatically to enable Compton cooling to control plasma entry. We remind that once $\dot M_x\gtrsim 4\times 10^{16}$ \gs, the settling regime itself ceases altogether, and free-fall supersonic flow occurs until the magnetospheric boundary with the subsequent formation of a shock above the magnetosphere, as was described and studied in more detail in earlier papers \citep{1976ApJ...207..914A,1983ApJ...266..175B}.

\subsection{Could the SFXT flares at their low-luminosity state be 
due to propeller mechanism?}

The propeller mechanism \citep{1975A&A....39..185I} has been also suggested for the SFXT phenomenon \citep{2007AstL...33..149G,2008ApJ...683.1031B} as a mechanism for gating accretion onto a rapidly rotating magnetized NS. It is feasible for disc accretion and is likely observed in luminous transient X-ray pulsars \citep{2016A&A...593A..16T}. In the case of quasi-spherical accretion, the propeller mechanism can be involved to explain major observational features of enigmatic $\gamma$~Cas stars \citep{2017MNRAS.465L.119P}. For low-states of SFXTs, the propeller mechanism, which requires centrifugal barrier for accretion by the condition that the Alfv\'en radius $R_\mathrm{m}\sim \dot M^{-2/7}$ be larger than the corotation radius,
$R_\mathrm{c}=(GM_\mathrm{x}P_x^2/4\piup^2)^{1/3}$, would need either a fast NS rotation or a large NS magnetic field: $P_x\le P_{cr}\simeq 9[\mathrm{s}]\mu_{30}^{6/7}\dot M_{16}^{-3/7}$. Clearly, with the quiescence X-ray luminosity $L_x\sim 10^{34} $ erg~s$^{-1}$,  the fastest SFXT from Table~2, IGR~J18483-0311, with a pulse period of $P_x\sim 21$ s could be at the propeller stage. If so, the low X-ray luminosity can be due to the leakage of matter across the  magnetospheric surface (for example, close to the rotational axis). 

In the case of quasi-spherical turbulent shell above magnetosphere, the propeller regime should correspond to 
the so-called `strong coupling' between the magnetic field and surrounding matter, when the toroidal field component is approximately equal to the poloidal one, $B_t\sim B_p$
\citep{Shakura2012,2018ASSL..454..331S}. In this regime, the NS spins down at a rate
\begin{equation}
    \frac{\dot P_x}{P_x}=K_2\frac{\mu^2P_x}{4\piup^2IR_\mathrm{m}^3}\simeq 2\times 10^{-12} (P_x/10s)\mu_{30}^2 R_\mathrm{m,9}^{-3}
    \label{e:prop}
\end{equation}
($I\approx 10^{45}$ g cm$^2$ is the NS moment of inertia, $K_2\simeq 7.6$ is the numerical coefficient accounting for the structure of quasi-spherical NS magnetosphere, \citealt{1976ApJ...207..914A}),
corresponding to a spin-down time of less than $10^5$ yrs. 
This short time suggests that a fast spinning magnetized NS rapidly approaches the critical period to become accretor, $R_\mathrm{m}\sim R_\mathrm{c}$, and this fact was stressed already in the original paper by \cite{1975A&A....39..185I}.

There is a  difference between the propeller stage for disc accretion and quasi-spherical accretion. In the former case, the disc is produced by accreting matter and the material is expelled by the rotating NS magnetosphere along open magnetic field lines \citep{2014MNRAS.441...86L}. In the disc case, sporadic accretion episodes during the transition to the accretion stage were found once $R_\mathrm{m}\simeq R_\mathrm{c}$  (e.g, for pre-outburst flares in A~0535+26, \citealt{2008A&A...480L..21P}). However, the numerical simulations by \cite{2014MNRAS.441...86L} were carried out for small magnetospheres, and in the case of large magnetospheres the situation remains unclear.

In the quasi-spherical case onto large magnetospheres (low accretion rates or high magnetic fields), the matter acquires the (super-Keplerian) specific angular momentum of the magnetosphere $\sim \omega_xR_\mathrm{m}^2$. If the cooling time of this matter (e.g., if it is expelled in the form of dense blobs) is short compared to the dynamic (convection) time in the shell, an equatorial ring with radius $R_\mathrm{p}\simeq R_\mathrm{m}(R_\mathrm{m}/R_\mathrm{c})^3$ and some thickness $h\ll R_\mathrm{p}$ \citep{Shakura2012} is likely to form. This ring spreads over in the viscous diffusion time scale, $t_d\sim t_\mathrm{K}(R_\mathrm{p})(R_\mathrm{p}/h)^2$ ($t_\mathrm{K}$ is the Keplerian time), and may end up with an accretion episode once the inner disc radius overcomes the centrifugal barrier. Therefore, it is possible to characterize the time between flares by the ring diffusion time scale, over which the disc replenishes mass by freshly propelled matter.
 
In so far as accretion through such a disc is centrifugally prohibited, its structure should be described by equations of `dead' discs \citep{1977SvAL....3..138S}, with the characteristic thickness $h\sim t_\mathrm{K}(R_\mathrm{p})^{6/7}\Sigma_0^{3/14}$, where $\Sigma_0$ is the surface density at its outer edge. Assume that the mass stored in this disc in a time interval $\Delta T$ be $M_\mathrm{d}\sim \dot M_\mathrm{p} \times \Delta T$, where $\dot M_\mathrm{p}$ is the 
fraction of the mass accretion rate propelled from the magnetosphere, $\dot M_\mathrm{p}=\dot M-\dot M_\mathrm{x}$ ($\dot M_\mathrm{x}$ -- the fraction of the mass accretion rate that reaches the NS surface and produces the inter-flare X-ray luminosity). In the simplest case,  $\dot M_\mathrm{x}$ is geometrically determined by the centrifugally free fraction of the magnetosphere surface, $\dot M_\mathrm{x}=\dot M(1-\sqrt{1-(R_\mathrm{c}/R_\mathrm{m})^2}$. 

Next, we use the relation $M_\mathrm{d}\sim R_\mathrm{p}^2\Sigma_0$ and note that the corotation radius $R_\mathrm{c}$ remains pretty much constant over short time intervals. Then by identifying the time between flares $\Delta T$ with the viscous time of such a dead disc, we arrive at the relation $\Delta T\sim R_\mathrm{p}^{5/4}\dot M_\mathrm{p}^{-3/10}$. As the ring radius $R_\mathrm{p}$ scales with mass accretion rate as  $R_\mathrm{m}^4\sim \dot M^{-8/9}$ and the propelled mass rate $\dot M_\mathrm{p}$ scales as $\dot M$, we arrive at $\Delta T\sim \dot M^{-127/90}\approx \dot M^{-1.4}$. The mass accumulated in the dead disc between the flares turns out to be inversely dependent on the mass accretion rate between the flares, $M_\mathrm{d}\sim \dot M^{-0.4}$.

In a quite different setup, the magnetospheric instability could be related to perturbations in the magnetosphere. These perturbations propagates with the Alfv\'en velocity, $v_A=B_p/\sqrt{4\piup\rho}\sim v_{ff}(R_\mathrm{m})$. Therefore, the characteristic time between accretion episodes due to these perturbations would be $\Delta T\sim t_A\sim R_\mathrm{m}/v_A\propto R_\mathrm{m}^{3/2}$. For any regime (disc or quasi-spherical), $R_\mathrm{m}$ scales with $\dot M$ not stronger than $R_\mathrm{m}\sim \dot M^{-2/7}$, and thus $\Delta T\sim \dot M^{-3/7}$ (disc) or $\sim \dot M^{-1/3}$ (quasi-spherical, radiation cooling). 
In the last plot of Fig.~\ref{fig:wait_lx_quiesc_sources}, which outlines the case of the the fastest pulsar in a SFXT known to date (IGR~J18483--0311), we show the dependences $\Delta T \propto \dot M^{-1/3}$ and $\Delta T \propto \dot M^{-1.4}$.  

Thus we conclude that the propeller model for SFXT flares can also provide the qualitative inverse dependence of the flare waiting time on the pre-flare luminosity.
A more detailed analysis of the propeller mechanism at low accretion rates onto large NS magnetospheres definitely deserves further investigation, which is far beyond the scope of the present paper.

\section{Summary}
   \label{conclusion}

To summarize, here we propose the following model explaining the flaring behaviour of SFXTs at their low luminosity 
state which is based on the statistical analysis of properties of the \xmm\ B.b. light curves:

\begin{itemize}
    \item 
    
    At the quiescent states of SFXTs with low X-ray luminosity $\sim 10^{33}$~\ergs, the RTI is ineffective to enable rapid plasma penetration into the NS magnetosphere. Instead, either a steady settling accretion regime controlled by the radiative plasma cooling occurs and  the mass accretion rate onto the NS is reduced by  factor $f(u)\ll 1$ compared to maximum available Bondi-Hoyle-Littleton value $\dot M_B\simeq \rho_w R_\mathrm{B}^2/v_w^3$, $\dot M_x=f(u)\dot M_B\approx \dot M_B(t_\mathrm{ff}/t_\mathrm{rad})^{1/3}$ \citep{Shakura2012,2018ASSL..454..331S}, or plasma enters the magnetosphere via ineffective processes (e.g., diffusion or magnetospheric cusp dripping, see \citealt{1984ApJ...278..326E}). In the last case, a low X-ray luminosity of 
    $10^{32}-10^{33}$~\ergs can be sustained by the thermal X-ray emission of the hot magnetospheric shell (see \cite{2017MNRAS.465L.119P} for a more detailed discussion and possible applications to the  $\gamma$~Cas phenomenon).
    
    \item
    
    A series of flares can be triggered by an external fluctuation of the stellar wind properties (density $\rho_w$ and/or velocity $v_w$).  Individual \xmm\ X-ray light  curves of SFXTs (see Figs.~\ref{fig:lc1} and ~\ref{fig:lc2}) suggest that in most cases a series of flares is initiated by a small- or moderate-amplitude flare, with subsequent development of more powerful flares and gradual decrease in flare amplitudes. Typically, such flare series last for about $\sim 1000$~s, a few dynamical time scales of the problem.

    \item
    
    The analysis of individual flares shows (see, e.g., Table~\ref{tab:flares}, sixth column) 
    that the mean X-ray luminosity during the flare 
    very rarely 
    exceeds 
    $\sim 10^{36}$~erg~s$^{-1}$. This can explain why these flares cannot switch-on the development of Compton-cooling controlled RTI and thus does not turn the source into a steady-state wind accreting state like Vela X-1. Instead, a series of flare terminates when all matter stored in the magnetopsheric shell is exhausted by the small RTI-flares. This is to be contrasted with bright SFXT X-ray flares during which the entire magnetospheric shell can accrete onto the NS because of the magnetosphere breakage due to, for example, reconnection of the magnetic field carried out by stellar wind plasma \citep{Shakura2014}. 
    
    \item
    
    On average, the flare energy in individual sources should be proportional to the fraction of the shell subject to RTI, $\delta M/M_{sh} \sim Z/R_\mathrm{m}$,
where $Z$ is the size of the RTI layer.
In the convective/turbulent shell, there is a turbulent viscosity that singles out a specific wavelength growing most rapidly, $Z\sim 4\pi [2\alpha^2/(45 A)]^{1/3}$,
with $\alpha<1$ being the turbulent viscosity coefficient (a la Shakura-Sunyaev in discs), $A\lesssim 1$  the Atwood number. The shell mass is  (Shakura et al 2014) $M_{sh}\sim \dot M_x t_\mathrm{ff}(R_\mathrm{B}) \sim \dot M_x v_w^{-3}$.
Therefore, the mean energy of flares in individual source $<\Delta E_f>\sim \delta M \sim <L_{x,q}> v_w^{-3}$, i.e. on average \textit{linearly} grows with the mean pre-flare X-ray luminosity $<L_{x,q}>$ (Fig.~\ref{fig:av_ene_lx_quiesc}).

\item 

In each individual source, $E_{f} \sim \dot M_x \times \Delta T$, where $\Delta T$ is the `waiting time' between individual flares, which is $\Delta T\sim t_\mathrm{rad} \sim 1/\dot M_x$. 
Thus, in each source, $E_f$ must be independent on variations of $\dot M_x$ between the flares.

\item 
The spread of the mean  X-ray luminosities $<L_{x,q}>$ between flares in individual sources is determined by the fractional change in the mass accretion rate onto the NS due to variations in the Bondi mass accretion rate captured from the stellar wind. For example, in the settling accretion theory with radiative plasma cooling 
$\dot M_x = f(u)\dot M_B \sim \dot M_B \dot M_x^{2/9}$, hence $\dot M_x\sim \dot M_B^{9/7}$.
Therefore, the fractional change in $L_{x,q}$ in the source between flares is $
\delta L_x/L_x = 9/7 \delta \dot M_B/\dot M_B = (9/7) \delta \rho_w/\rho_w -(27/7) \delta v_w/v_w$,
where $\delta \rho_w/\rho_w$ and $\delta v_w/v_w$ is the stellar wind density and velocity fluctuations, respectively.
These variations in $L_x$ up to one order of magnitude can be produced 
during  the active RTI stage.

\item 

The rising time of a flare can correspond to the  fastest growing RTI mode in the turbulent shell, $\delta t_r \sim 30 \mbox{s}\, \dot M_{16}^{-2/3}$. 
The inverse dependence of the flare rising time on the X-ray luminosity between flares 
can be seen for some individual sources (Fig.~\ref{fig:riset_vs_lx_quiesc_sources}).

\item

For the fastest pulsar in IGR~J18483--0311, the centrifugal barrier at the magnetospheric boundary may lead to the  formation of an equatorial dead cold disc which could trigger a flaring activity of the source once the centrifugal barrier at its inner edge is overcome. The waiting time between the flares in this case can be characterized by the viscous time scale of the disc evolution and is also inversely proportional  to the pre-flare X-ray luminosity as $\Delta T\sim \dot M_x^{-1.4}$.

\end{itemize}

We conclude that SFXT flares observed during low-luminosity states could be qualitatively compatible with the development of the Rayleigh-Taylor instability in plasma accreted from the stellar wind of the companion which tries to enter the NS magnetosphere. 

However, the full development of the RTI fails because the radiative plasma cooling 
during the flares turns out to be insufficient for Compton cooling 
to enable steady-state magnetospheric plasma penetration, 
as in the case of persistent wind accreting X-ray pulsars like Vela X-1.  

Thus, SFXT flares offer unique possibility to probe complicated processes of plasma entering into magnetopsheres of magnetic NSs through interchange instabilities under natural conditions.

\section*{Acknowledgments}

This work is based on observations obtained with \xmm, an ESA science mission with instruments and contributions directly funded by ESA Member States and NASA. We have made use of data produced by the \extras\ project, funded by the European Union's Seventh Framework Programme under grant agreement no 607452. We acknowledge financial support from the Italian Space Agency (ASI) through the ASI-INAF agreement 2017-14-H.0. The \extras\ project acknowledges the usage of computing facilities at INAF - Astronomical Observatory of Catania. The \extras\ project acknowledges the CINECA award under the ISCRA initiative, for the availability of high performance computing resources and support. 
The work of KAP is supported by RFBR grant 18-502-12025,  and by the grant of Leading Scientific Schools of Moscow University `Physics of stars, relativistic objects and galaxies'.
The authors thank the anonymous referee for the useful and constructive comments, and suggestions for alternative explanations for the SFXT flaring activity.

\clearpage
\begin{appendix}

\section{Flare properties}
\label{app:flares}

\begin{figure*}
\includegraphics[scale=0.31,angle=0]{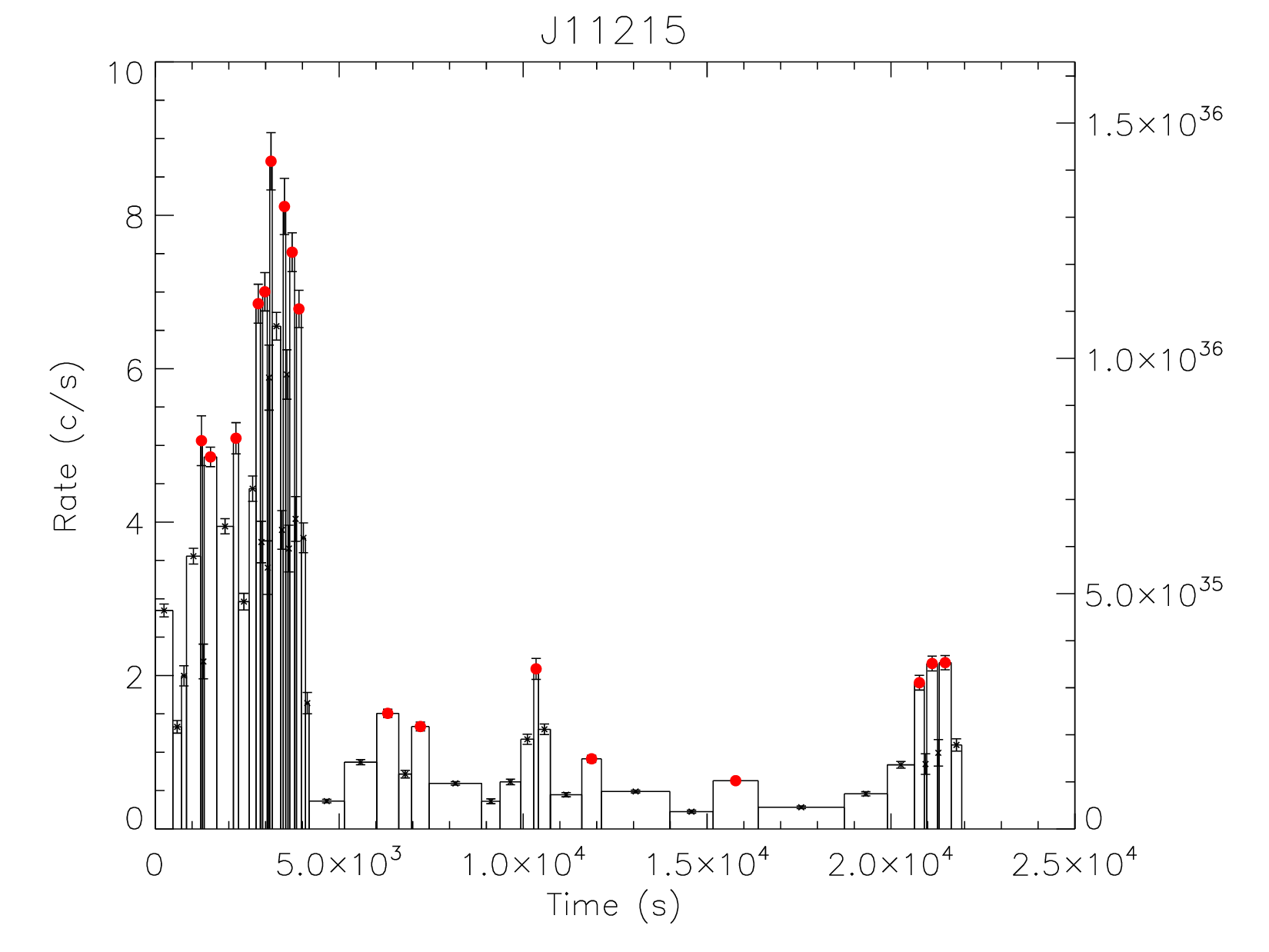} 
\includegraphics[scale=0.31,angle=0]{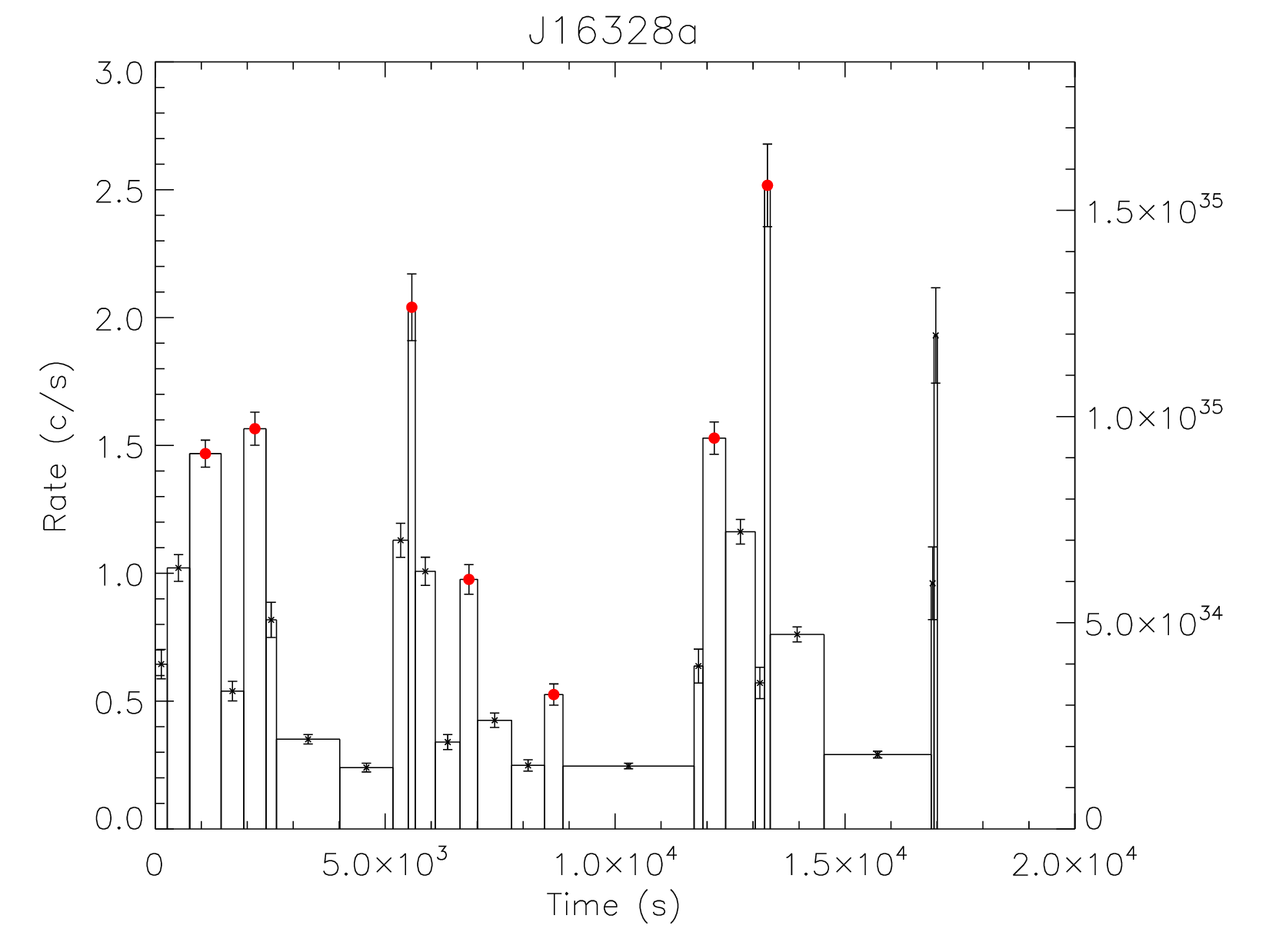} \\
\includegraphics[scale=0.31,angle=0]{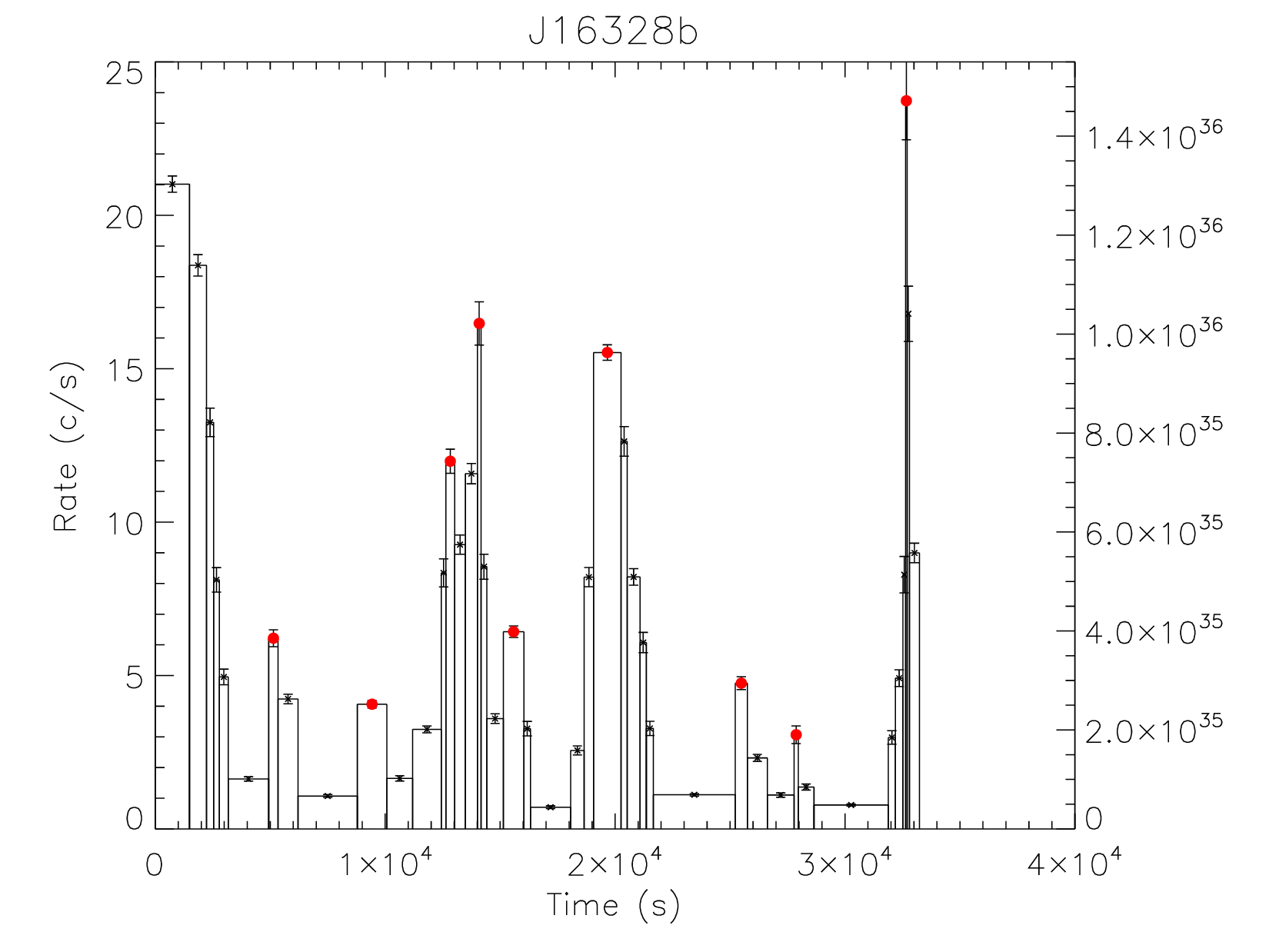}
\includegraphics[scale=0.31,angle=0]{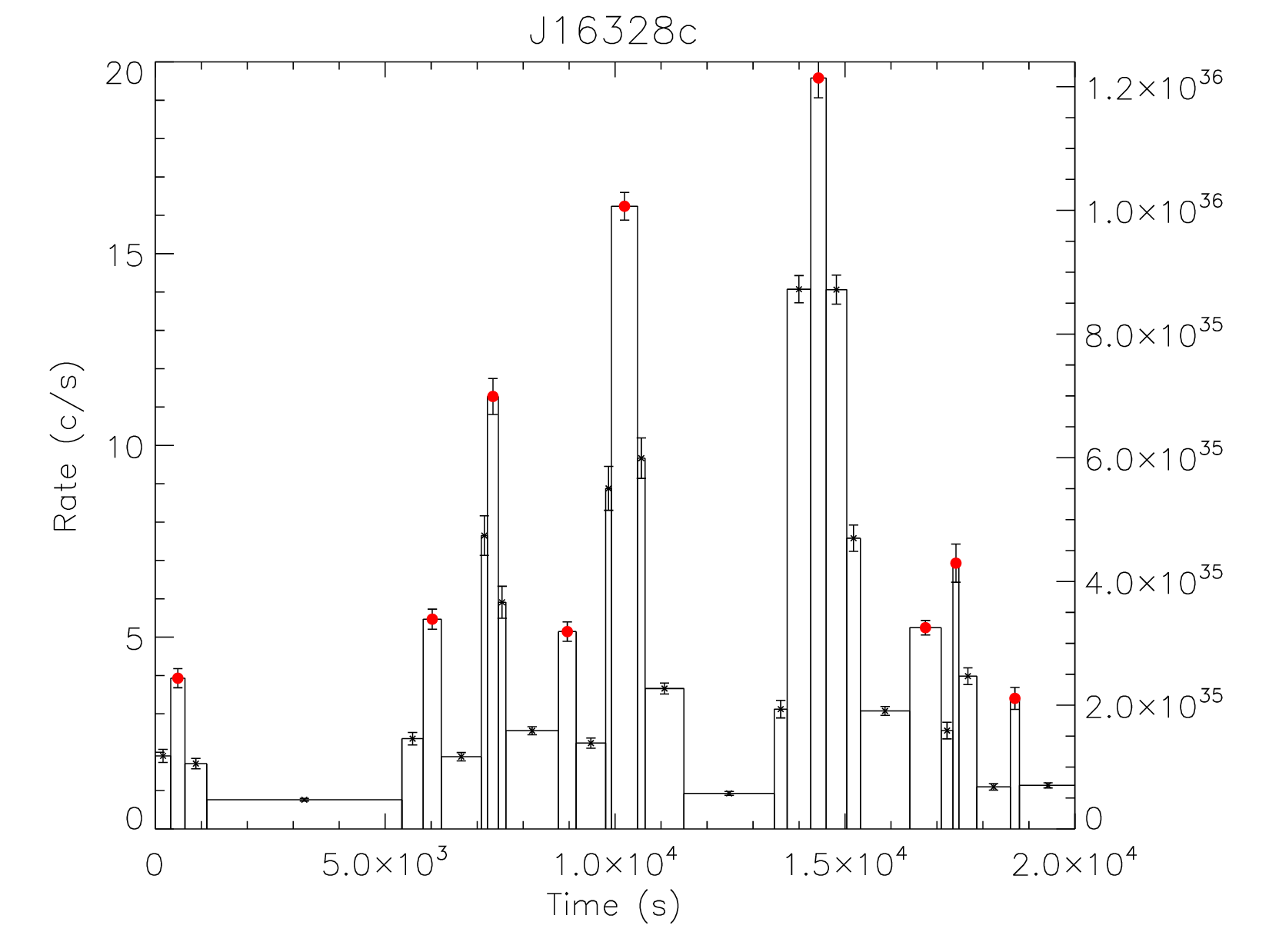} \\
\includegraphics[scale=0.31,angle=0]{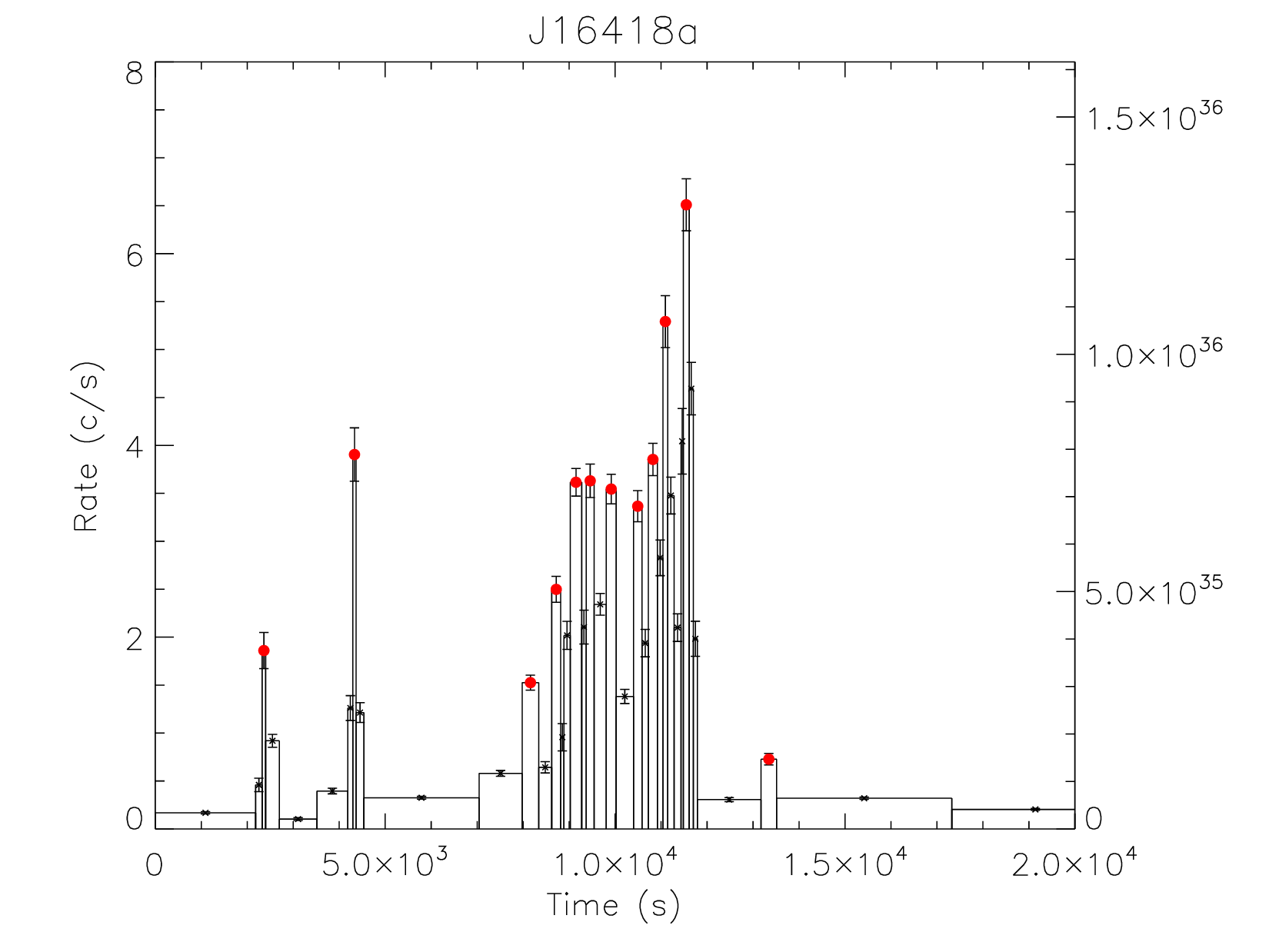} 
\includegraphics[scale=0.31,angle=0]{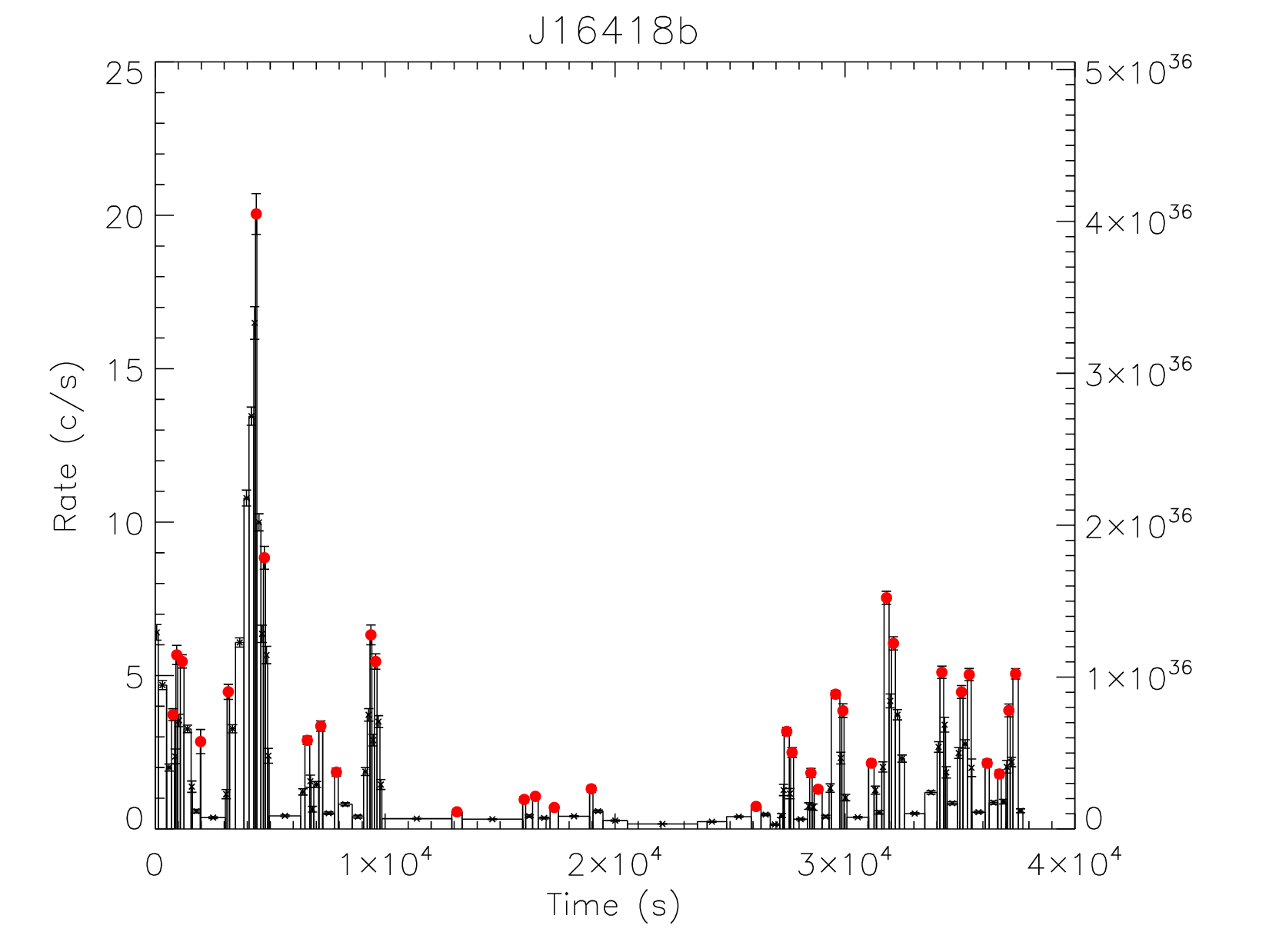} \\
\caption{B.b. light curves of the SFXTs analysed here. 
Red dots mark the B.b. including the  flare peaks. 
On the y-axis, both count rates (on the left) and the estimated luminosity (on the right) are reported (1--10 keV).
}
\label{fig:lc1}
\end{figure*}

\begin{figure*}
\includegraphics[scale=0.31,angle=0]{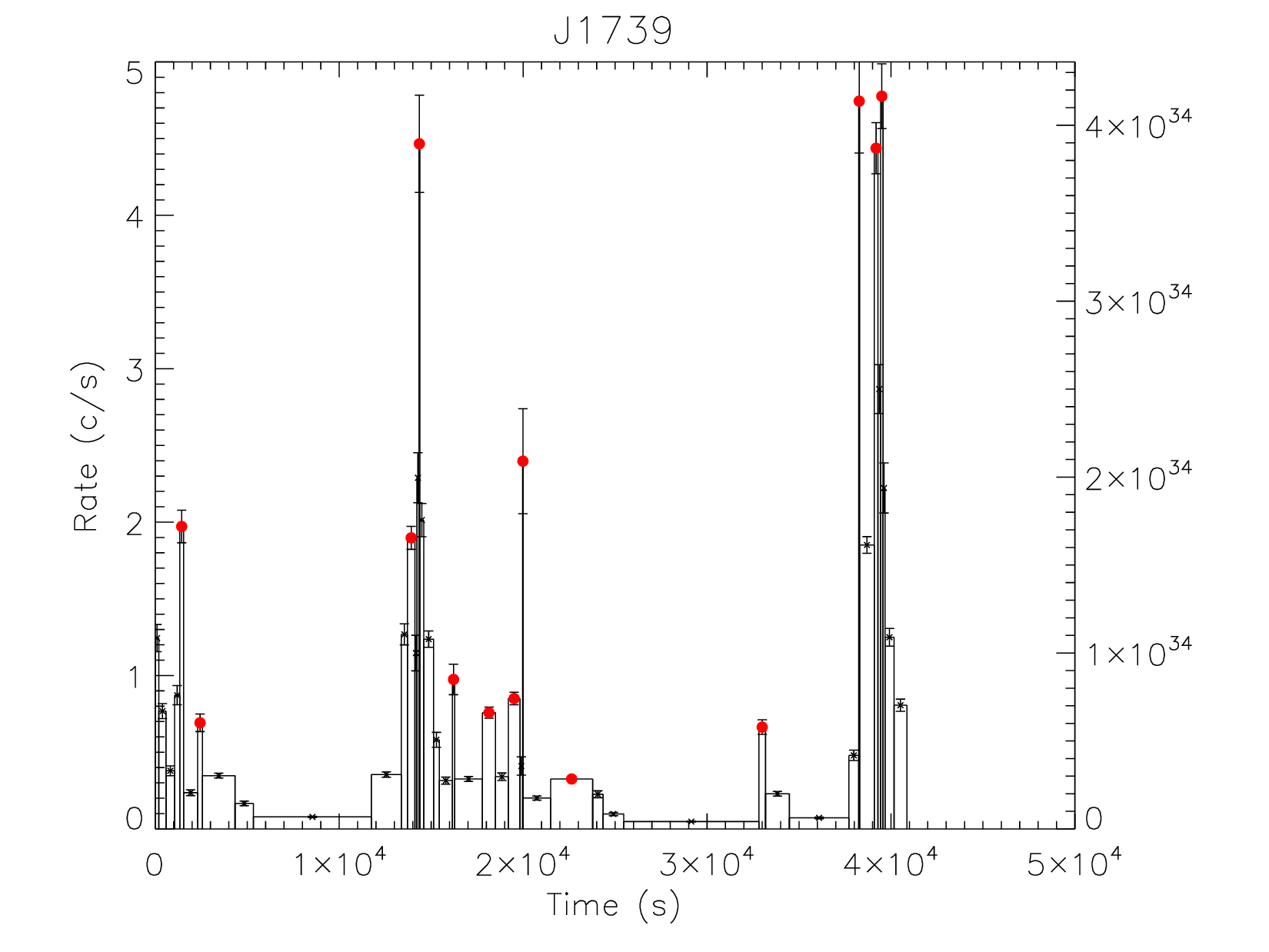}
\includegraphics[scale=0.31,angle=0]{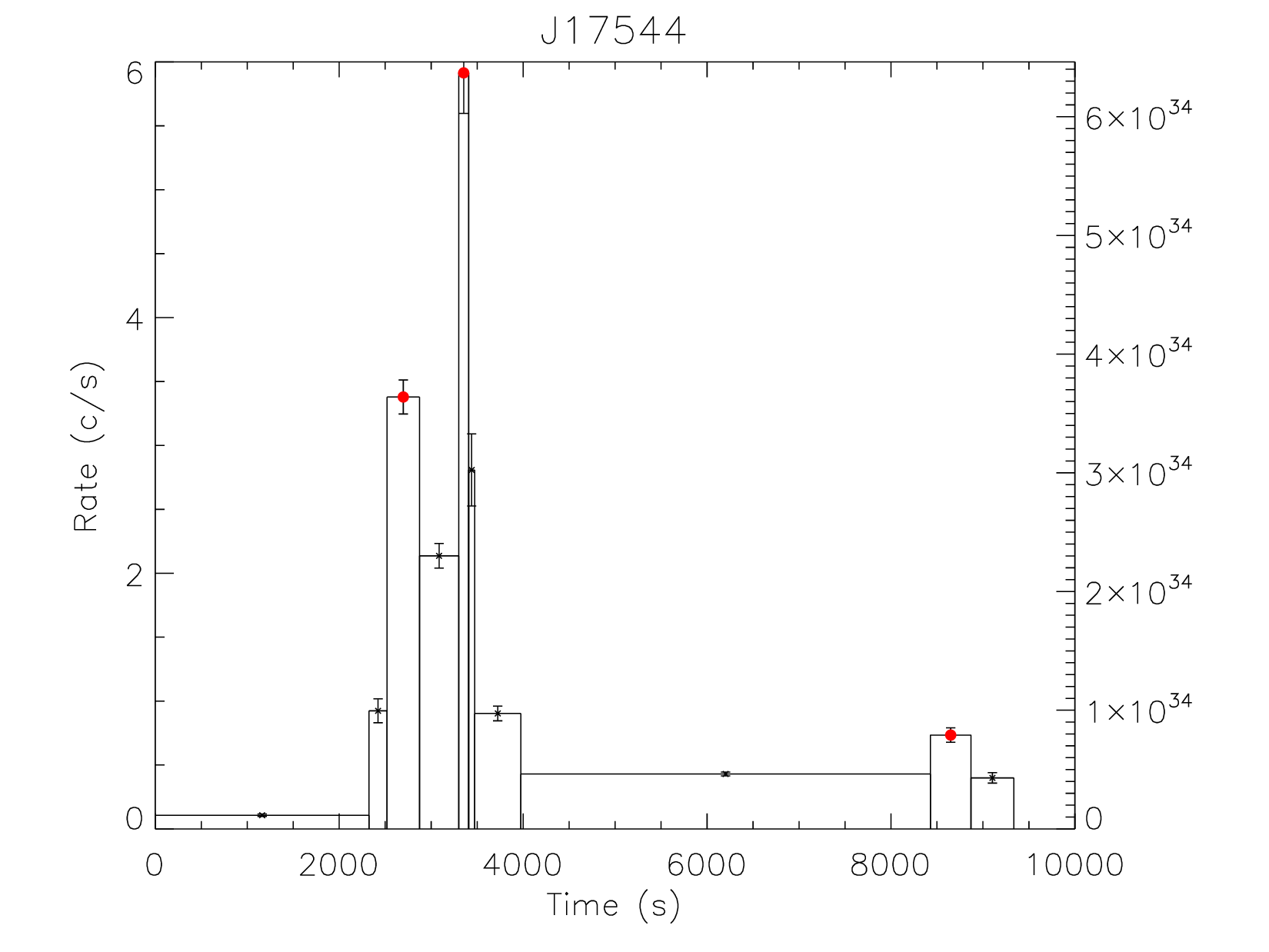} \\
\includegraphics[scale=0.31,angle=0]{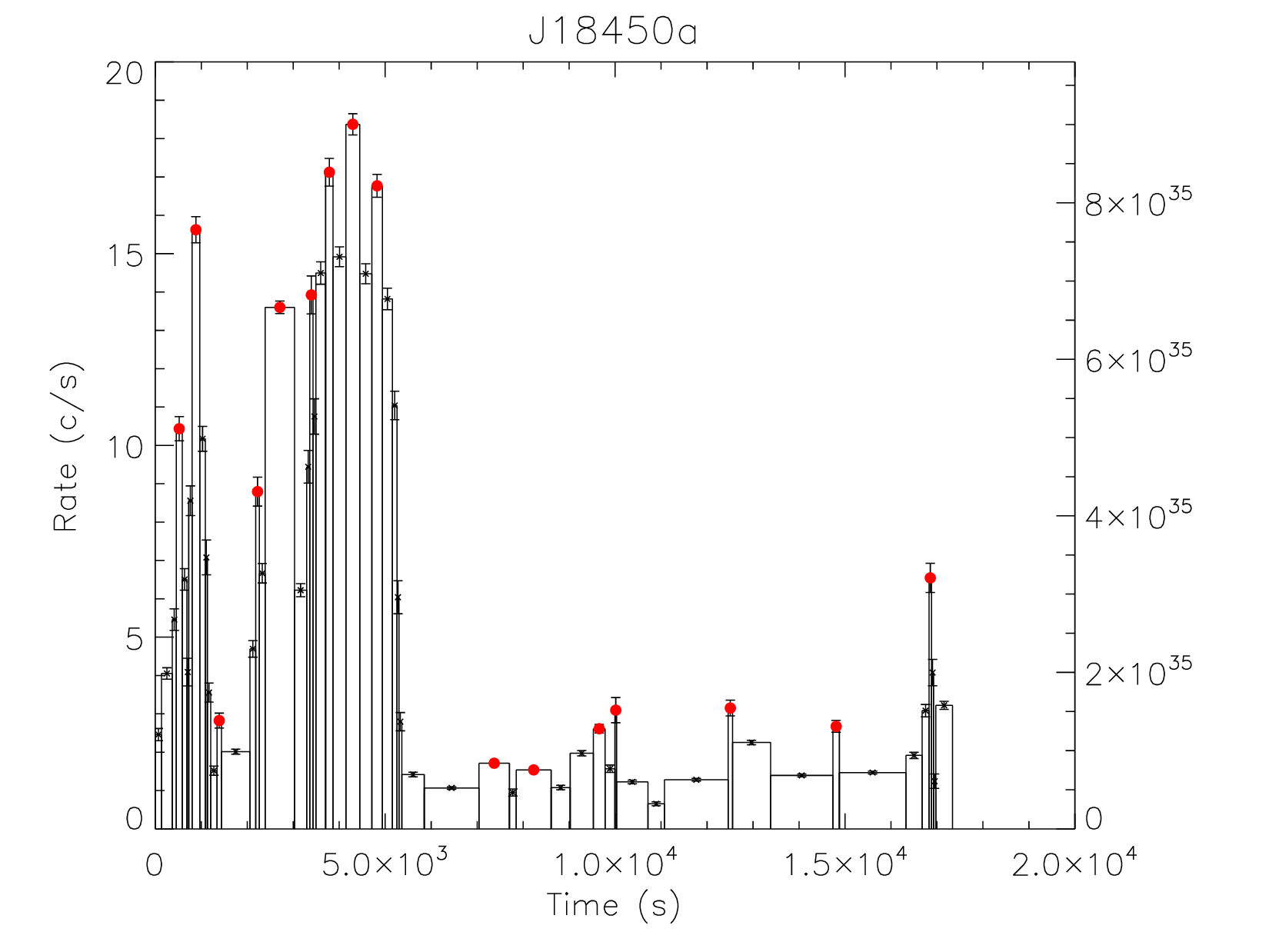}
\includegraphics[scale=0.31,angle=0]{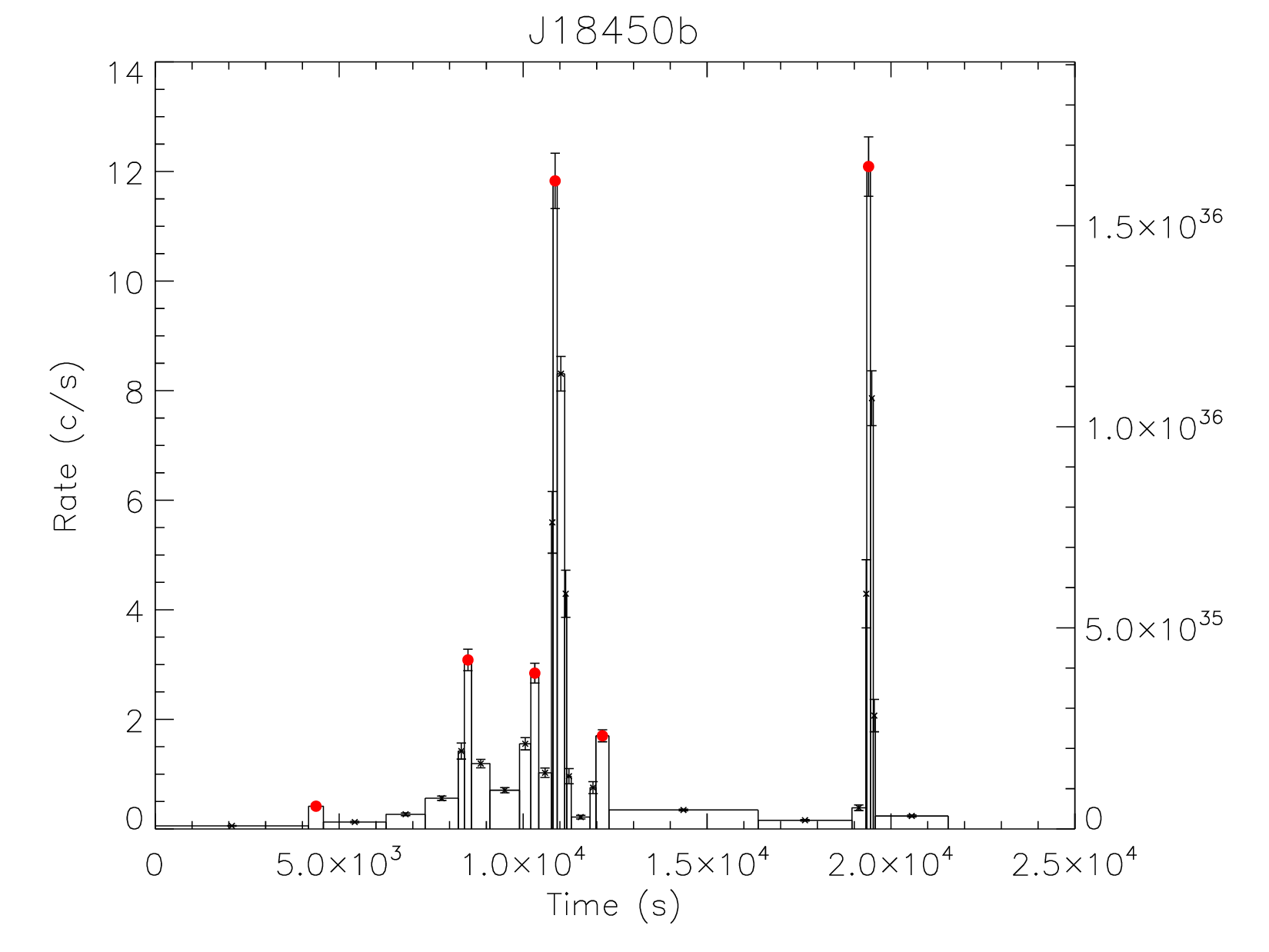} \\
\includegraphics[scale=0.31,angle=0]{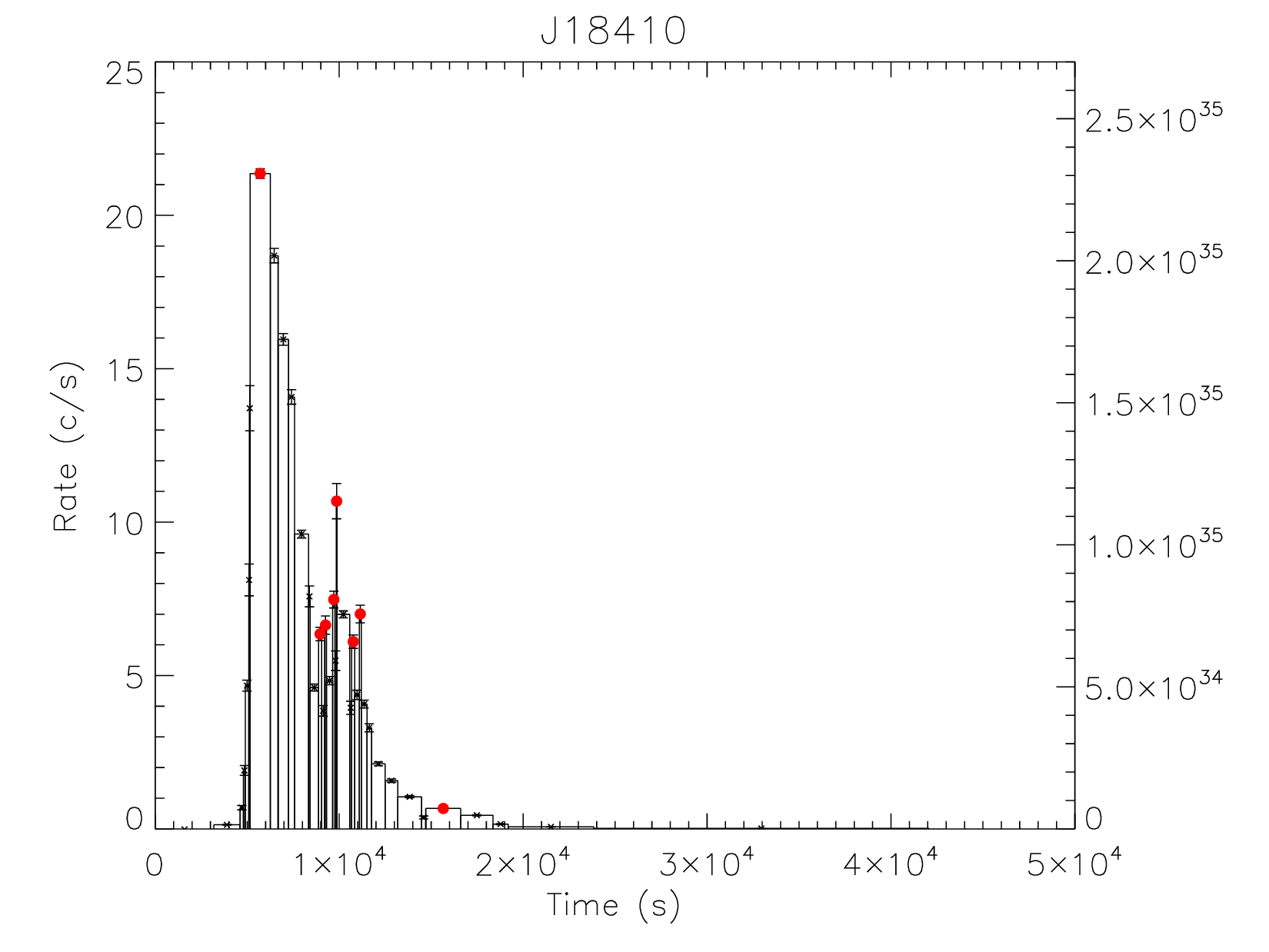}
\includegraphics[scale=0.31,angle=0]{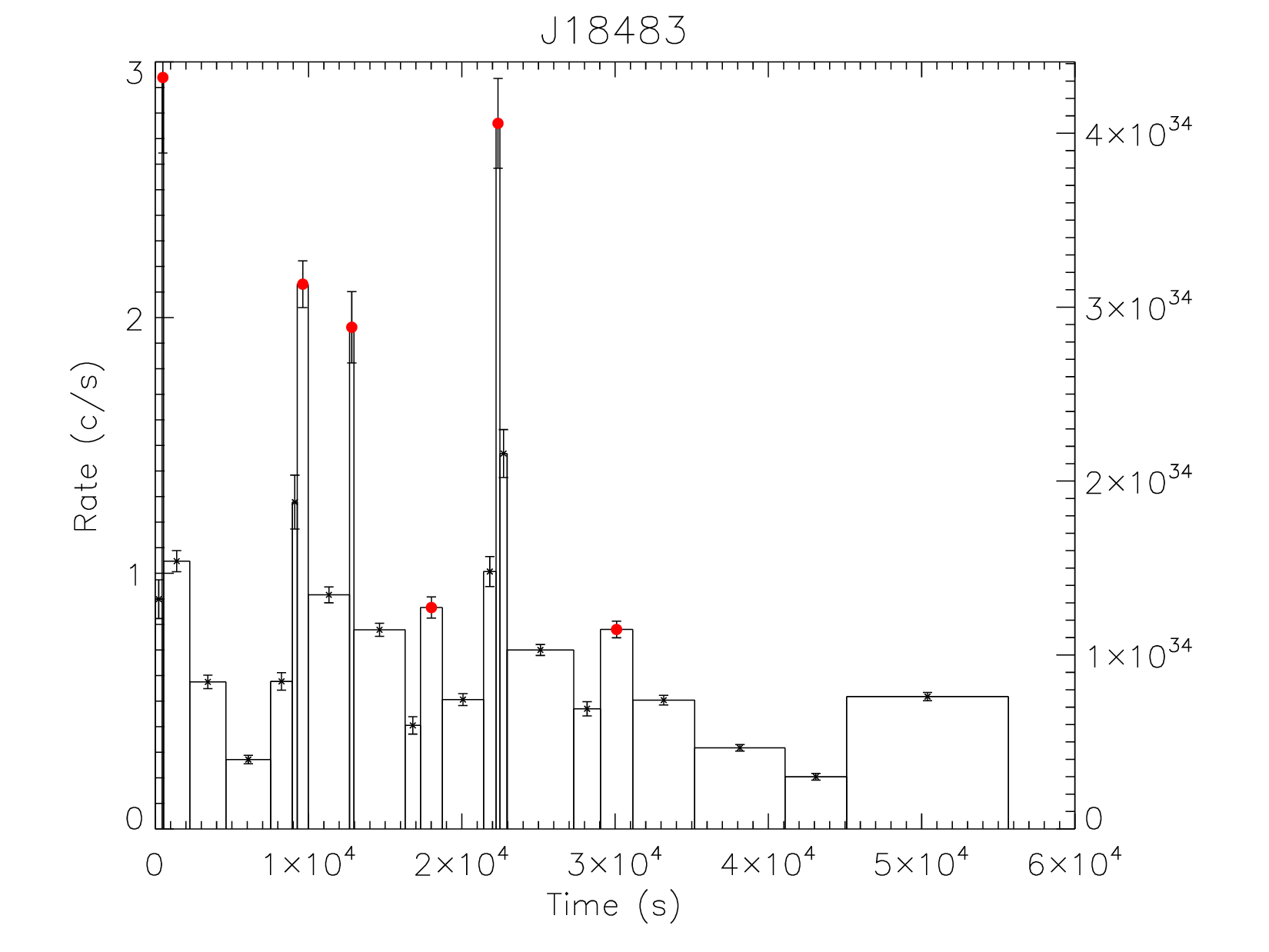} \\
\caption{B.b. light curves of SFXTs. The symbols have the same meaning as in Fig.~\ref{fig:lc1}.
}
\label{fig:lc2}
\end{figure*}

\begin{figure*}
\includegraphics[scale=0.29,angle=0]{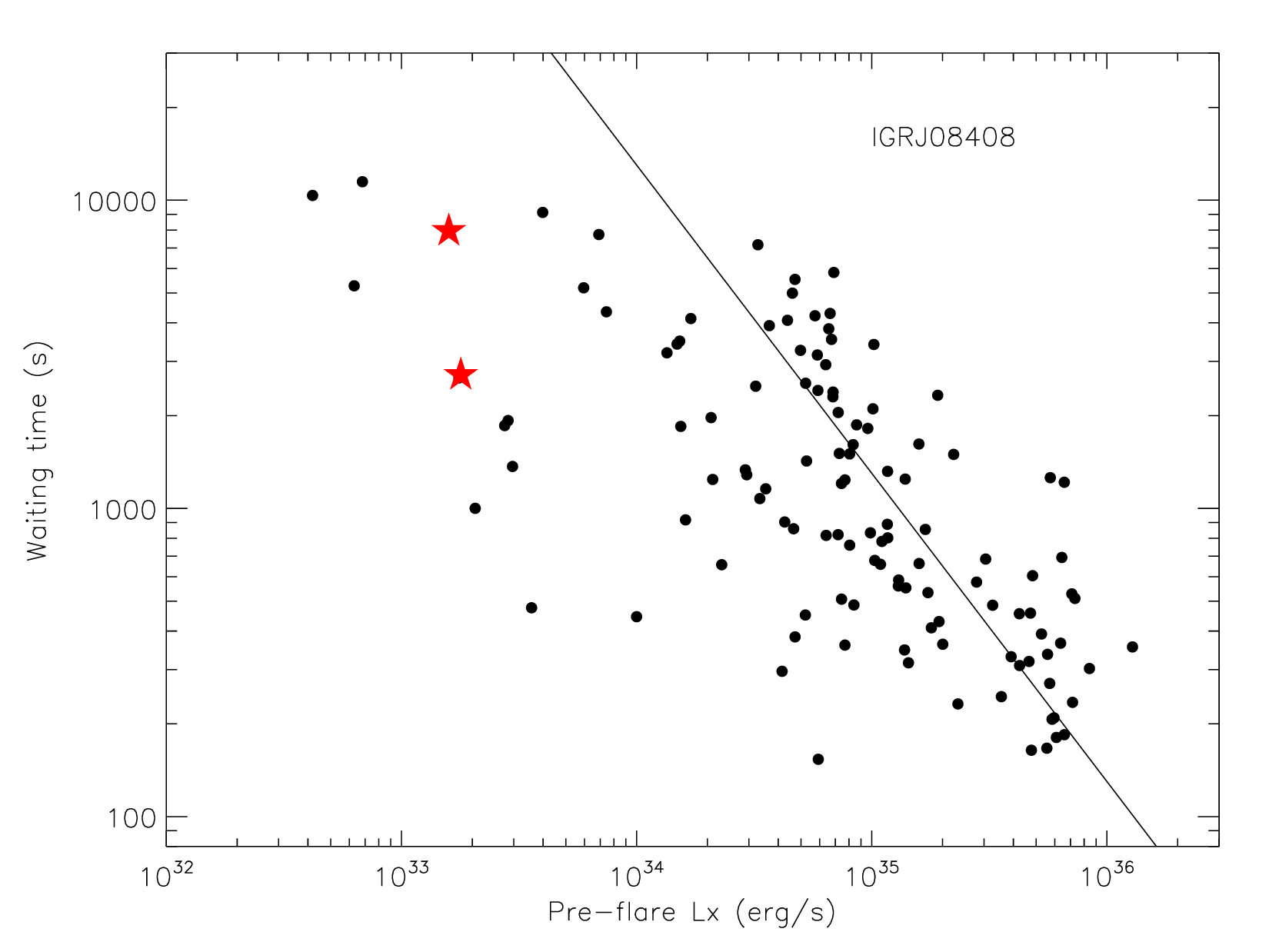} 
\includegraphics[scale=0.29,angle=0]{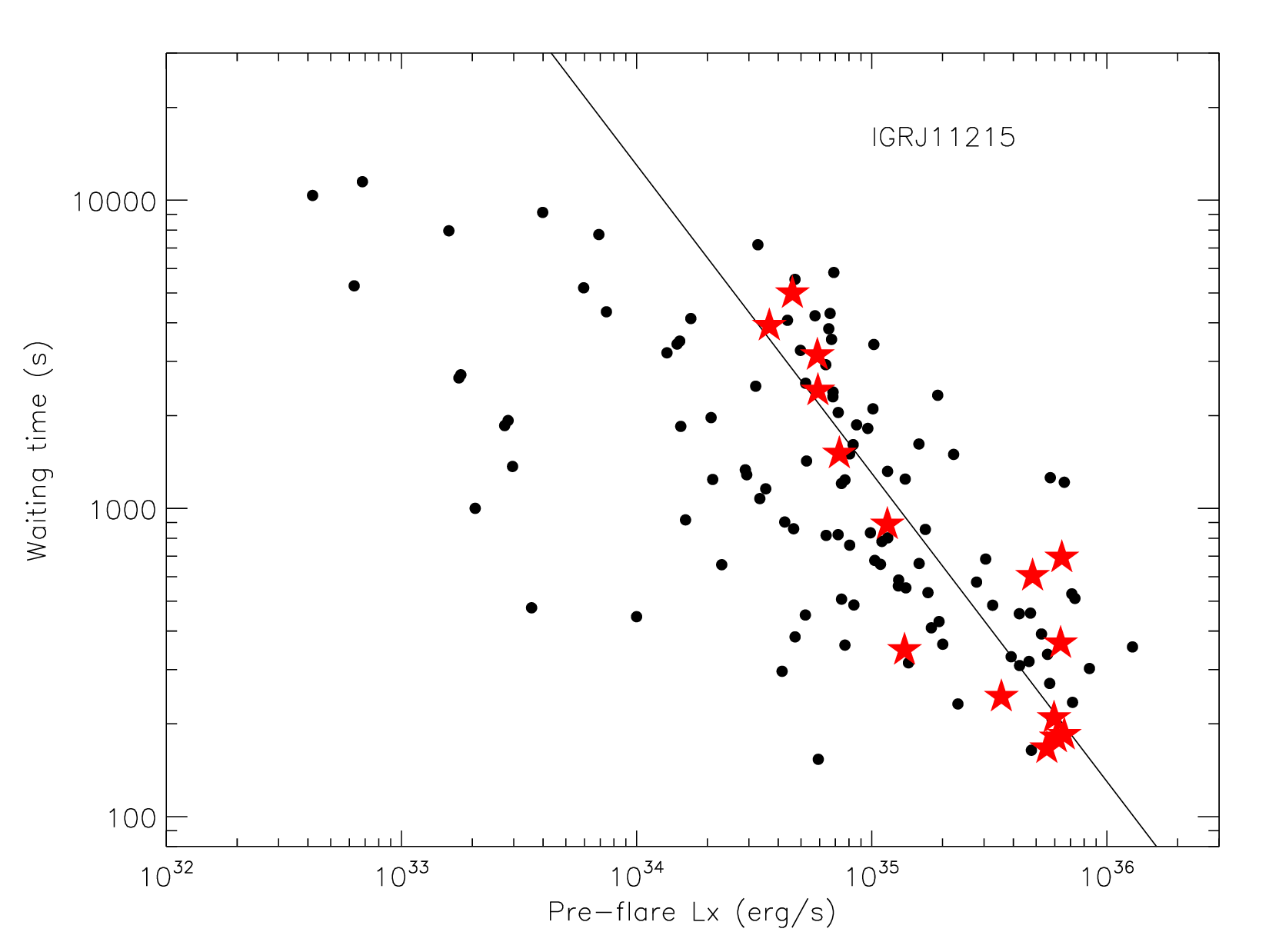} \\
\vspace{-0.4cm}
\includegraphics[scale=0.29,angle=0]{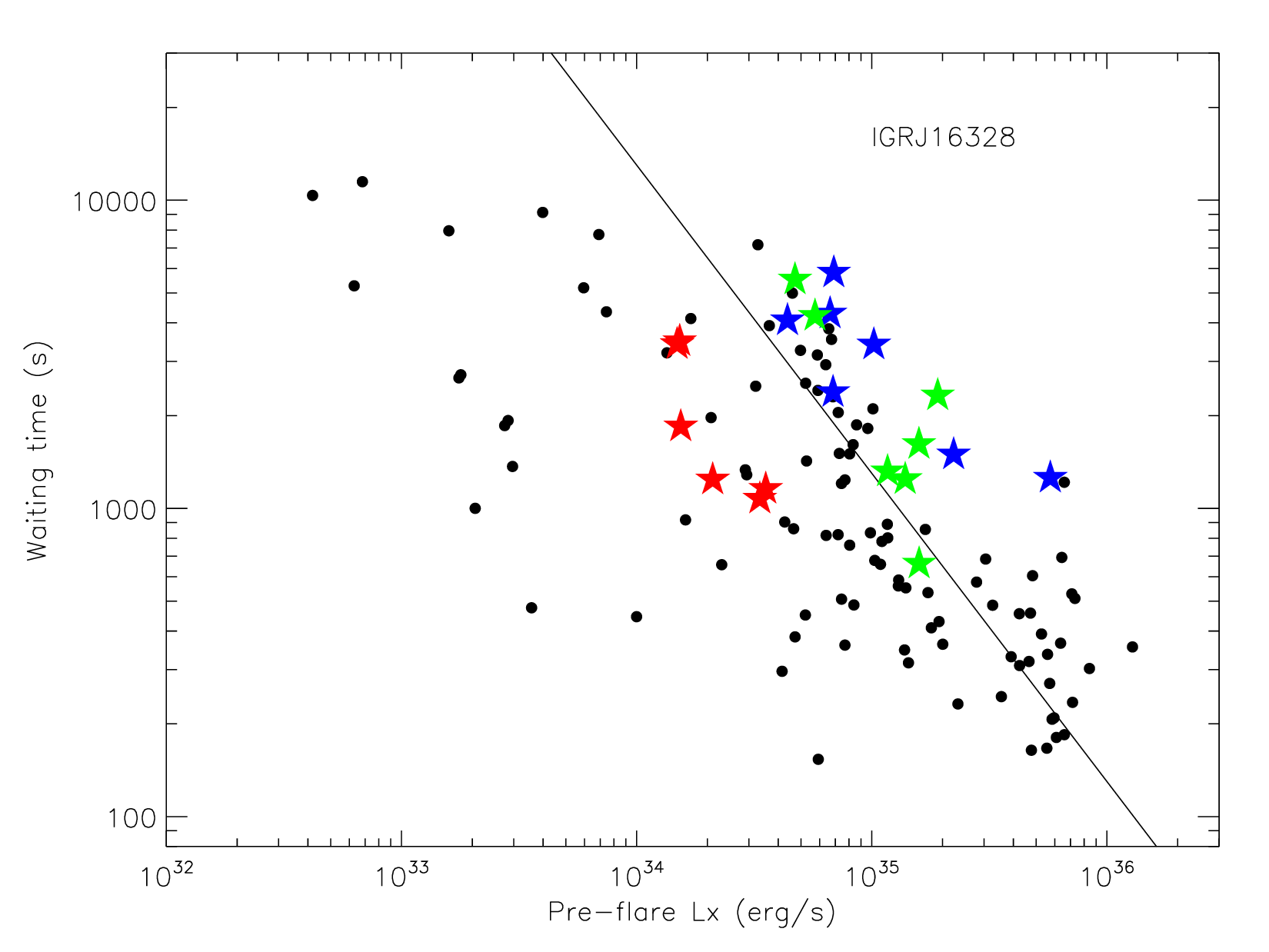} 
\includegraphics[scale=0.29,angle=0]{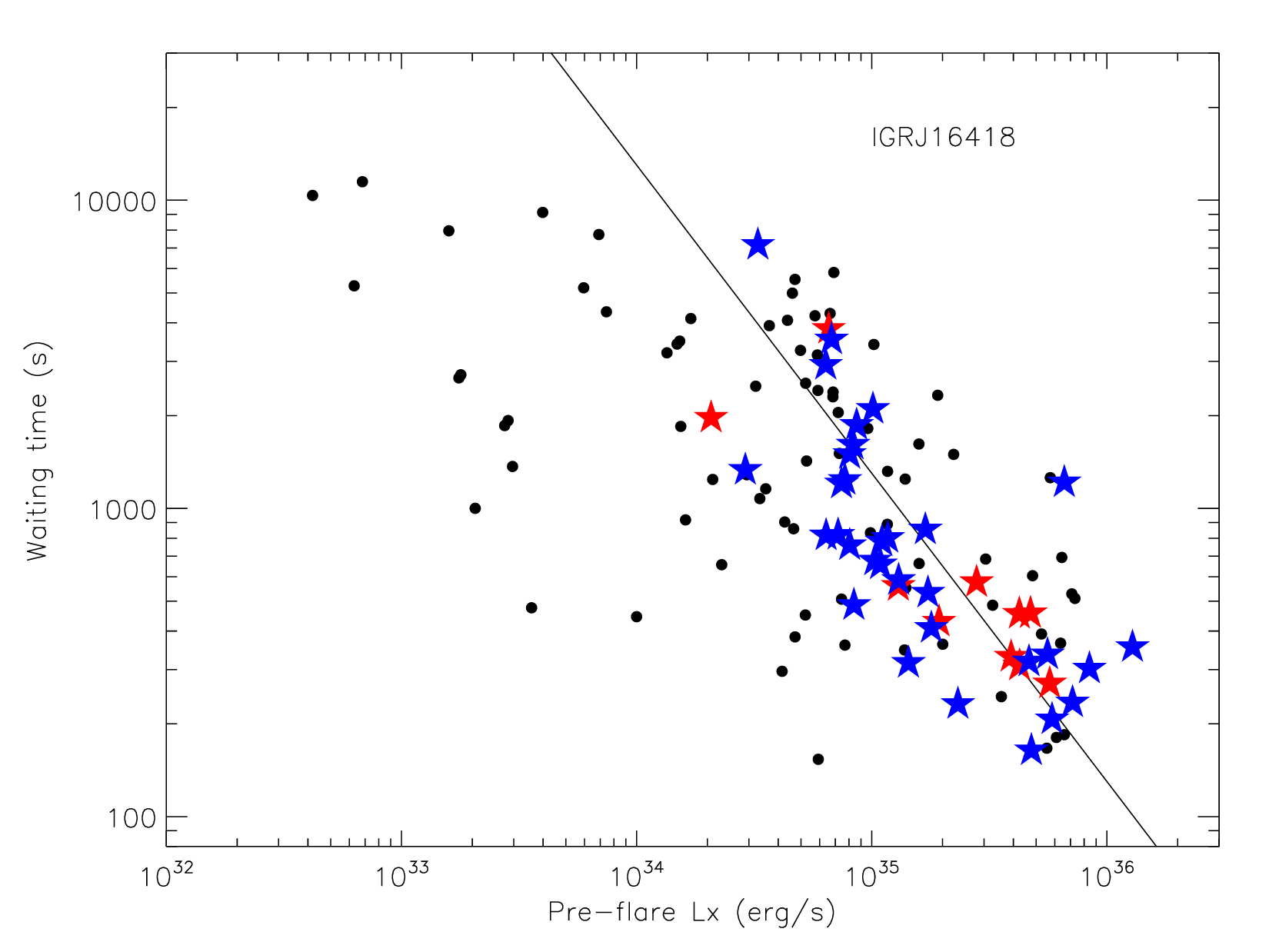} \\
\vspace{-0.4cm}
\includegraphics[scale=0.29,angle=0]{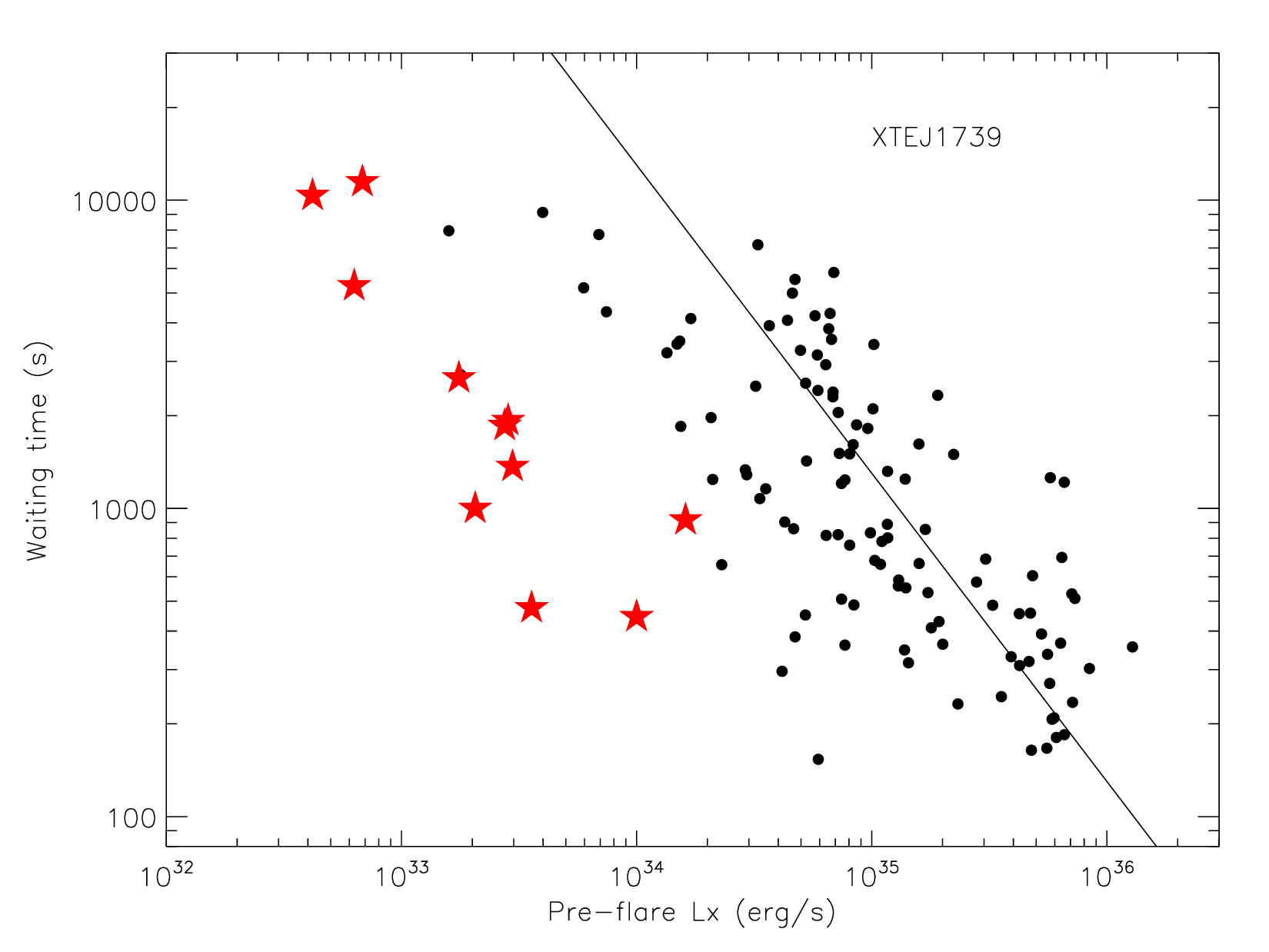} 
\includegraphics[scale=0.29,angle=0]{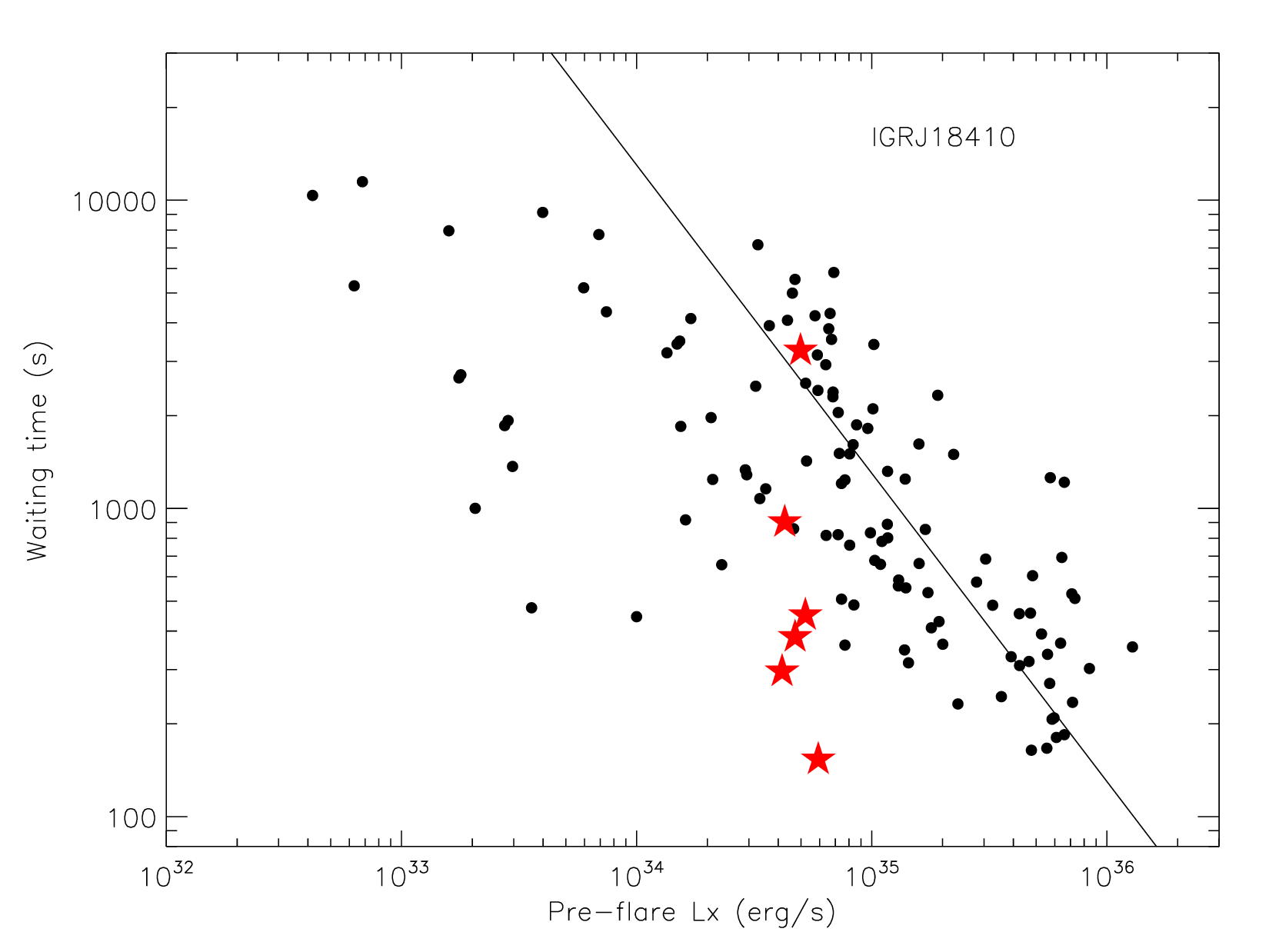} \\
\vspace{-0.4cm}
\includegraphics[scale=0.29,angle=0]{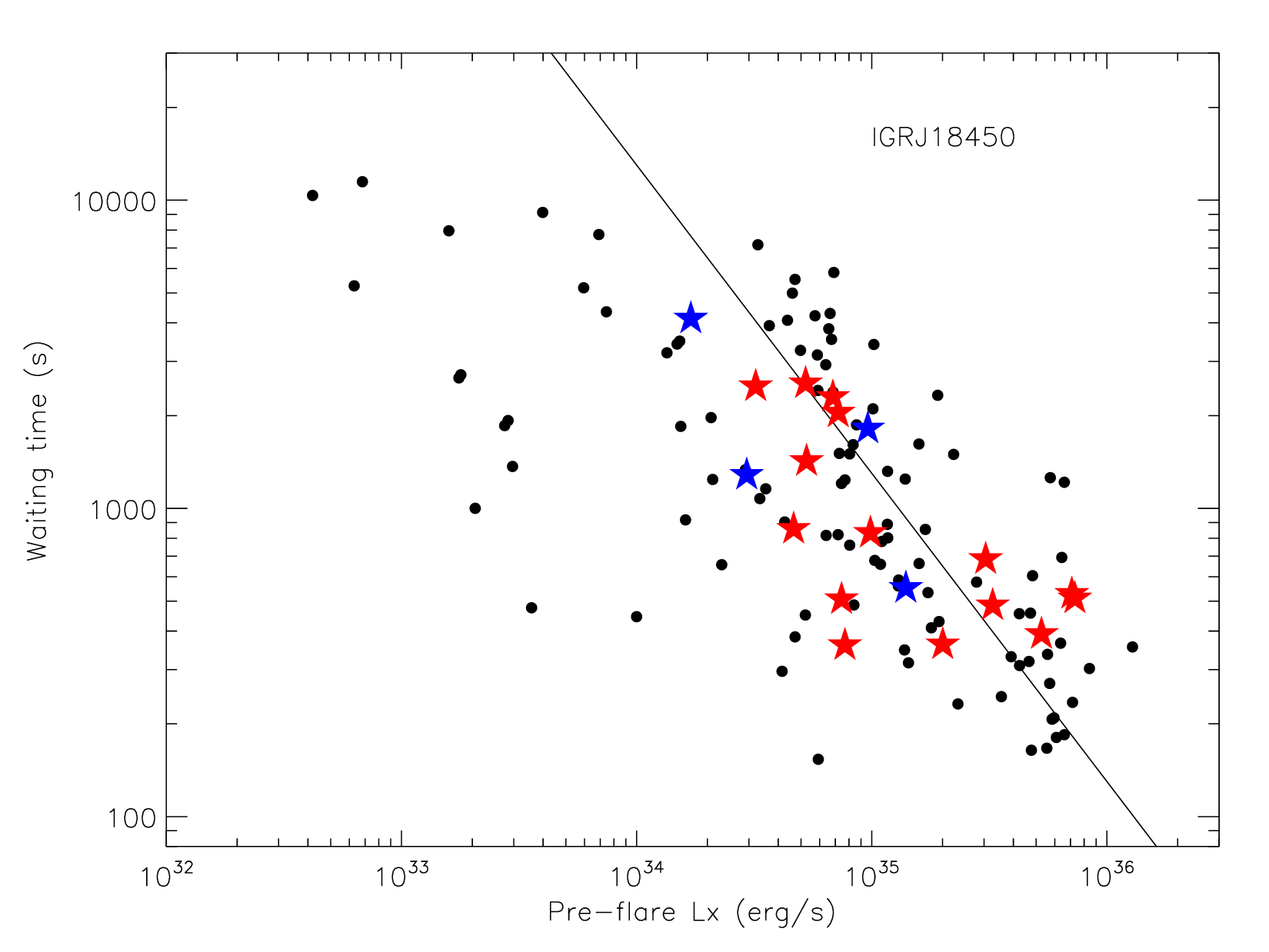} 
\includegraphics[scale=0.29,angle=0]{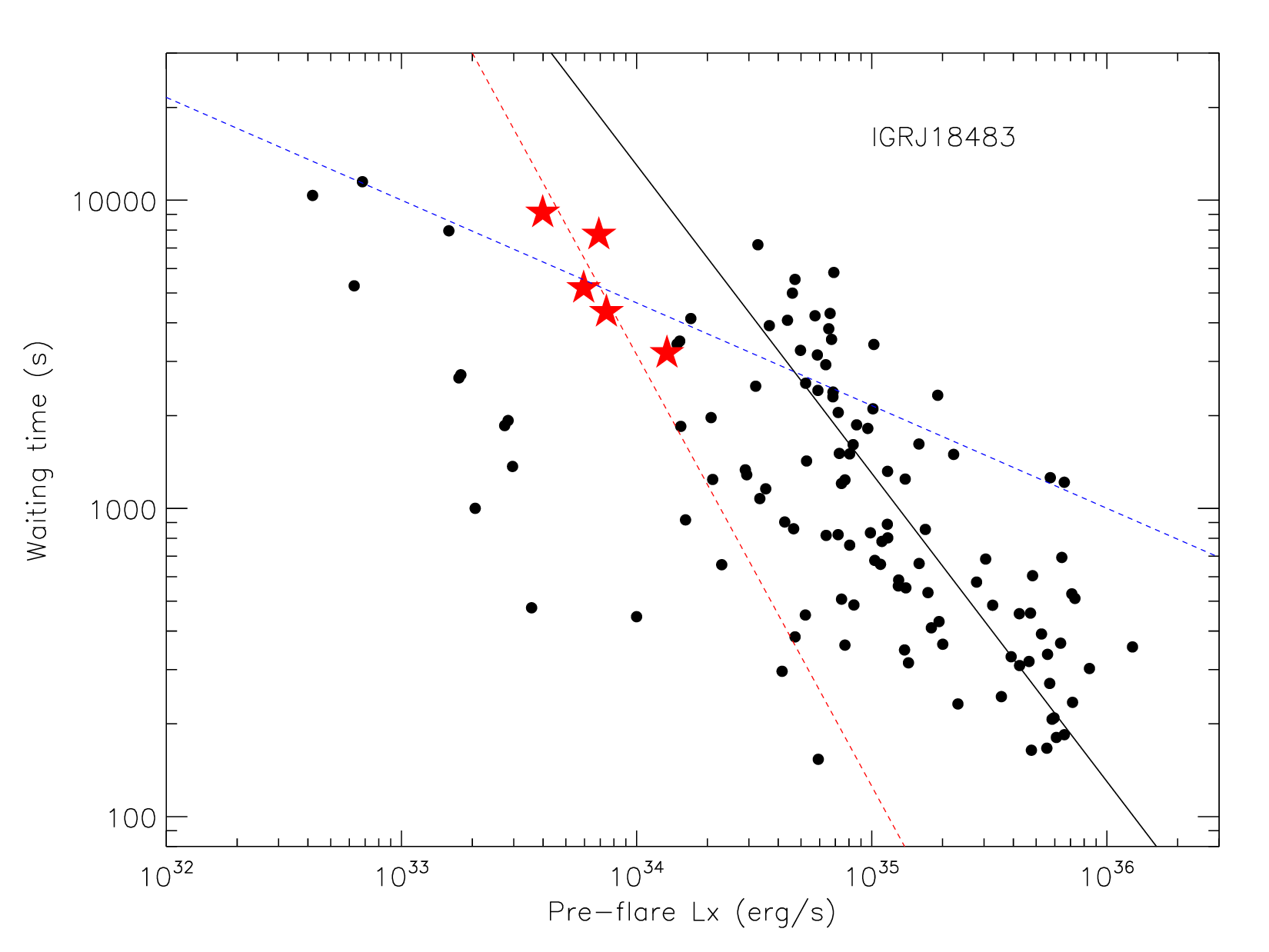} \\
\caption{Flare waiting time against the  pre-flare luminosity. 
The straight line indicates the dependence 
$\Delta T = 130  [\mathrm{s}] \dot M_{16}^{-1}$, where $\dot M_{16}$ is the pre-flare accretion rate. Dashed lines in the last plot mark (with arbitrary normalizations to overlap with IGR~J18483-0311 data points) $\Delta T \propto\dot M_{16}^{-1.4}$ (in red) and $\Delta T \propto\dot M_{16}^{-1/3}$ (in blue) dependences.
Stars mark flares from single sources (different colors mark flares from different observations).
}
\label{fig:wait_lx_quiesc_sources}
\end{figure*}

\begin{figure*}
\includegraphics[scale=0.29,angle=0]{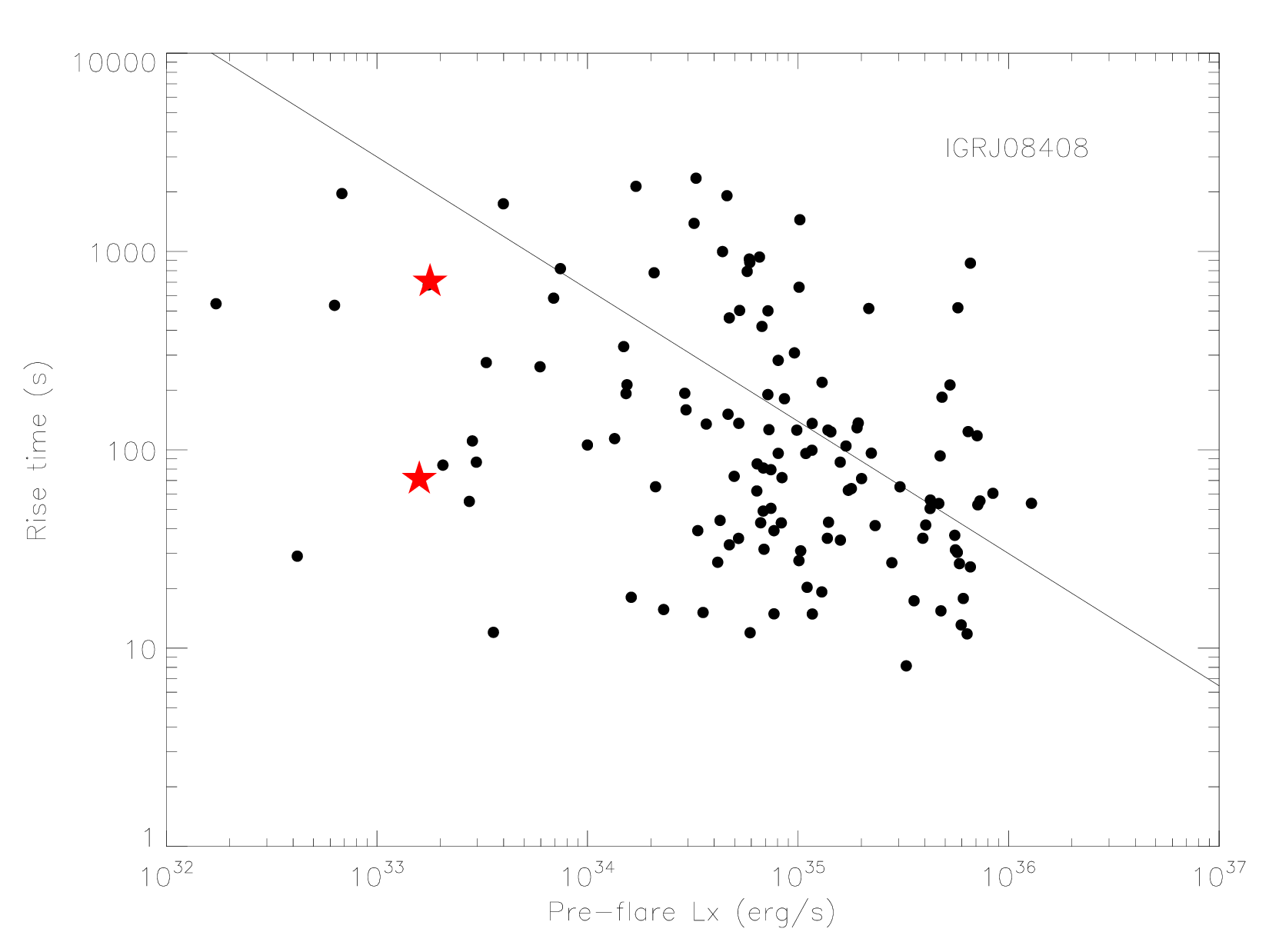} 
\includegraphics[scale=0.29,angle=0]{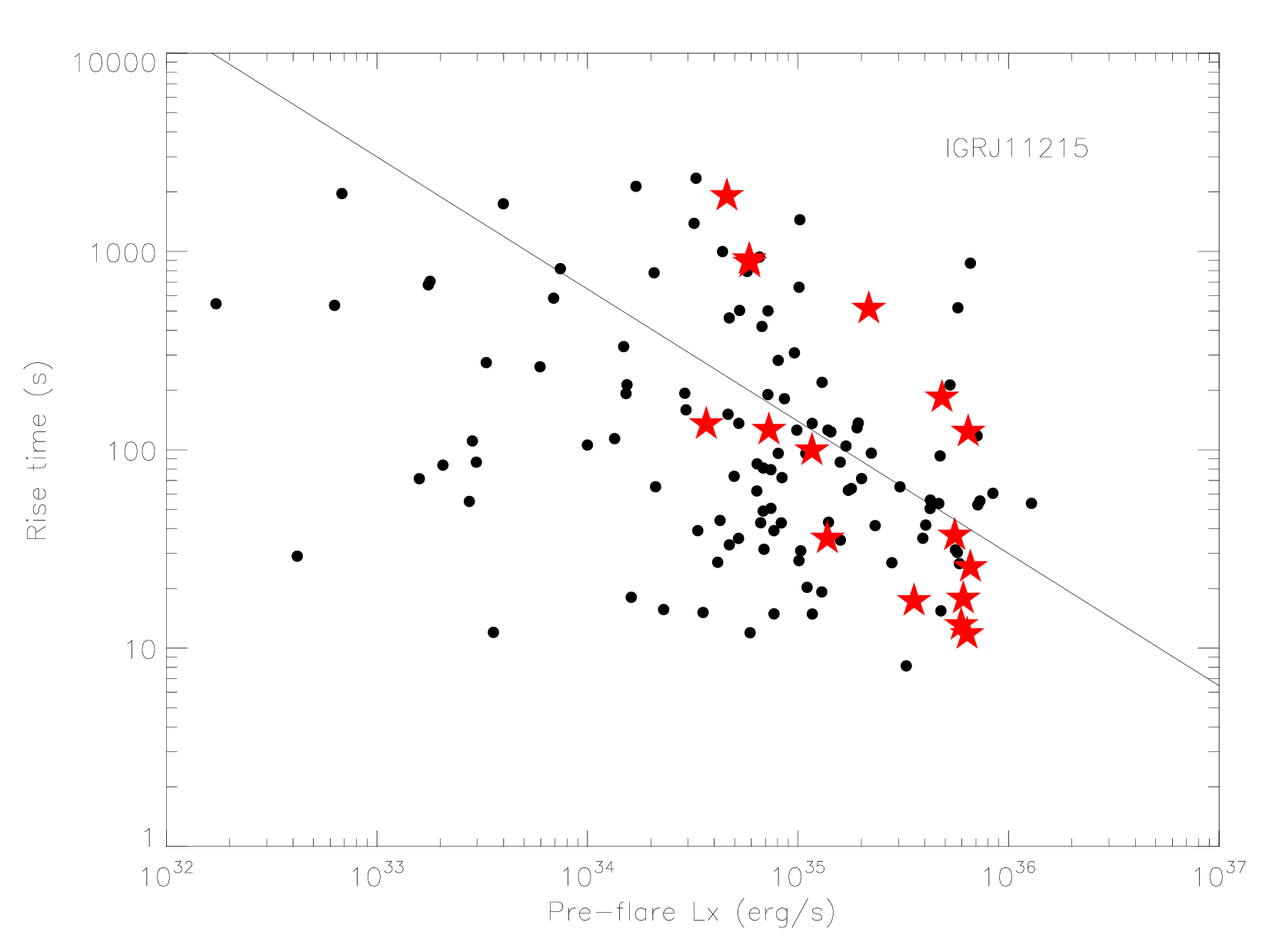} \\
\vspace{-0.4cm}
\includegraphics[scale=0.29,angle=0]{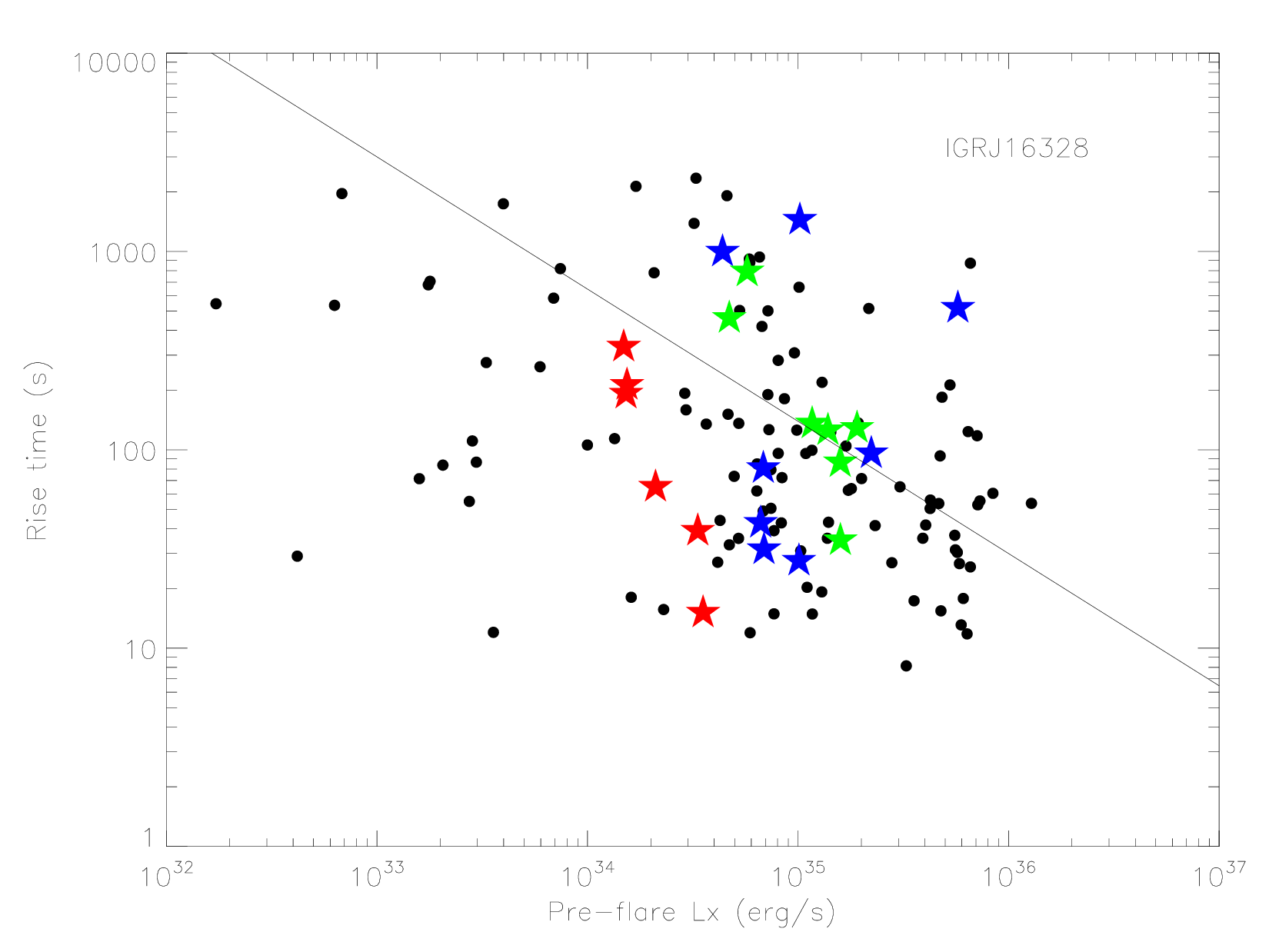} 
\includegraphics[scale=0.29,angle=0]{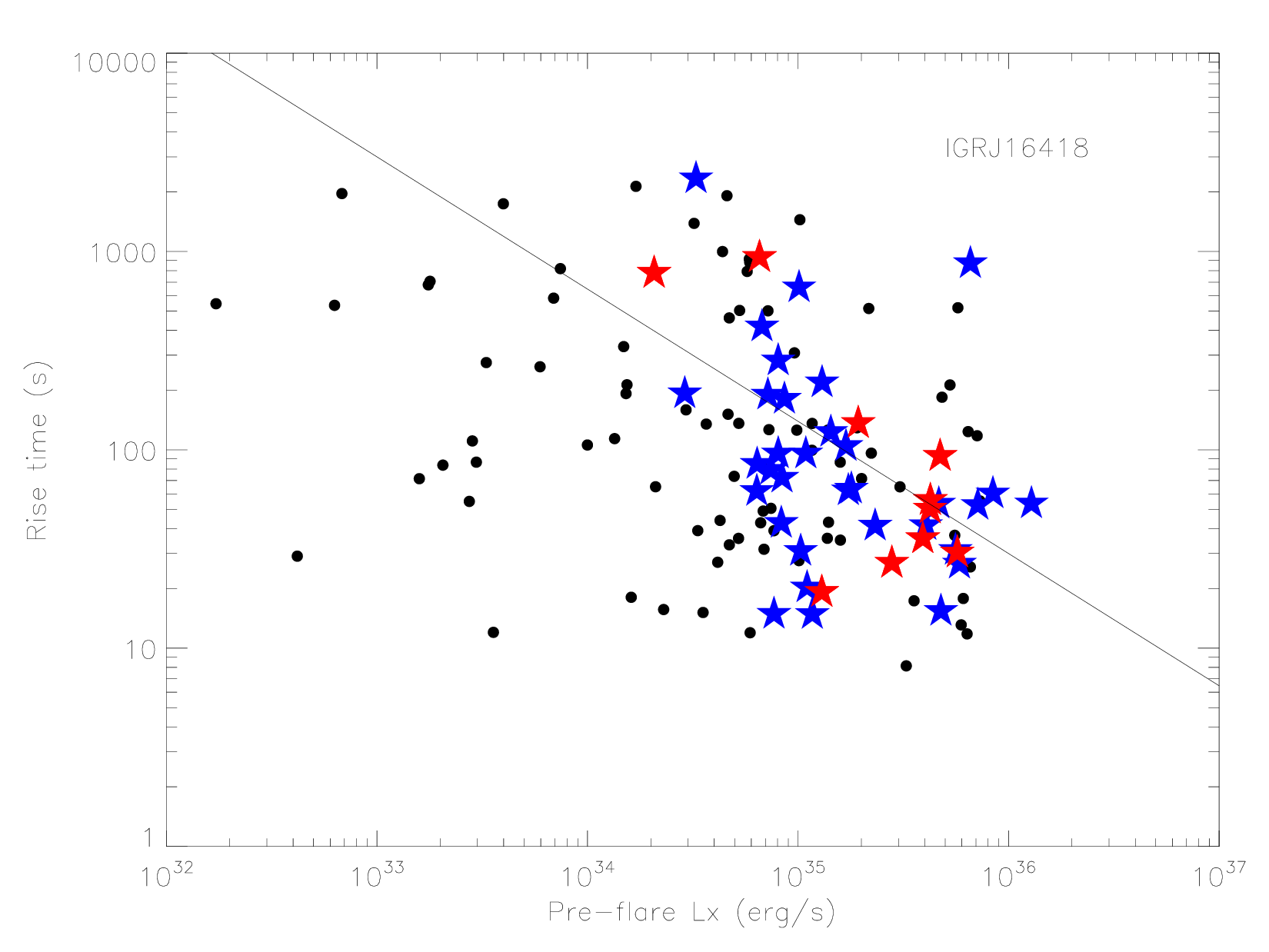} \\
\vspace{-0.4cm}
\includegraphics[scale=0.29,angle=0]{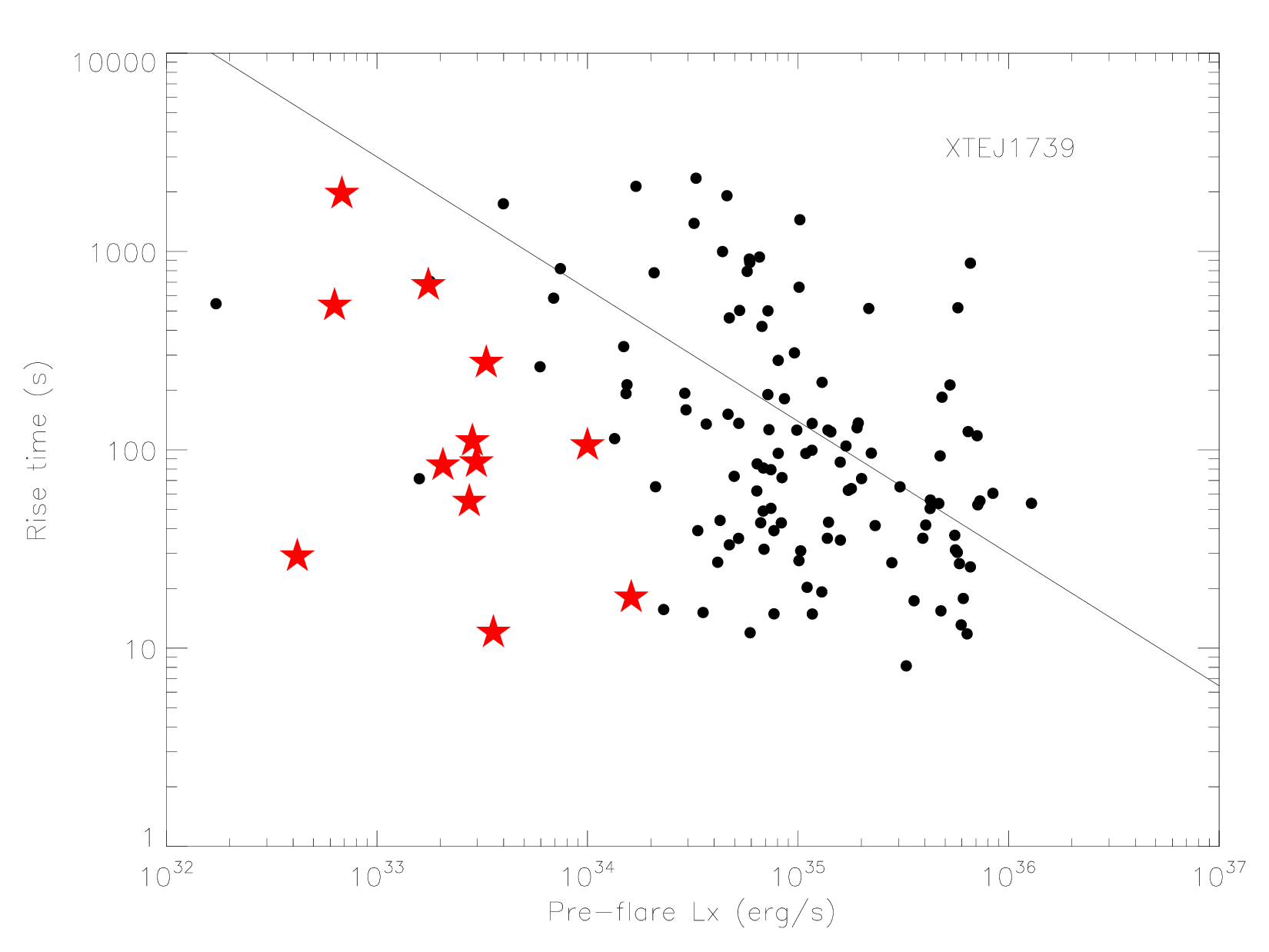} 
\includegraphics[scale=0.29,angle=0]{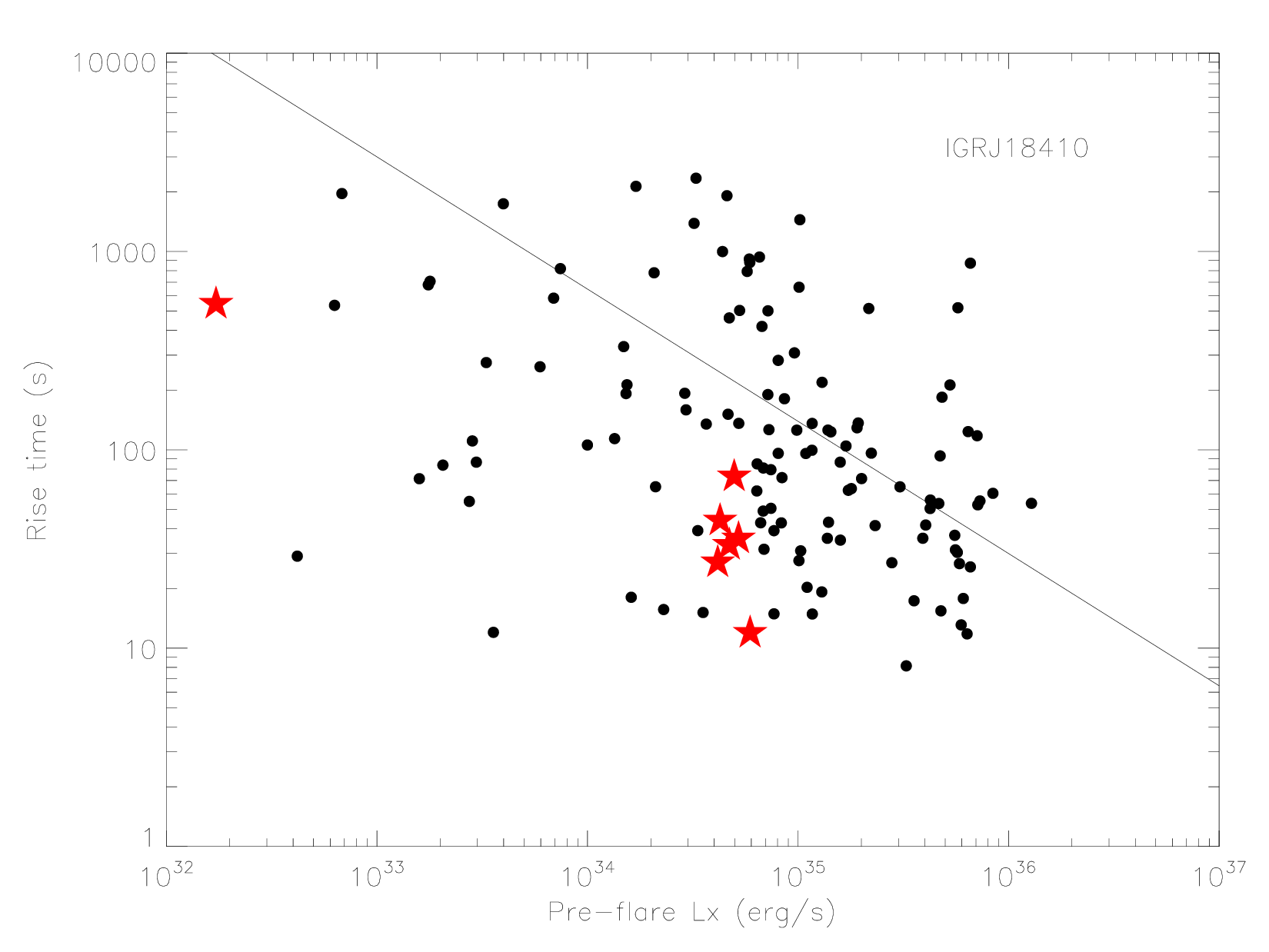} \\
\vspace{-0.4cm}
\includegraphics[scale=0.29,angle=0]{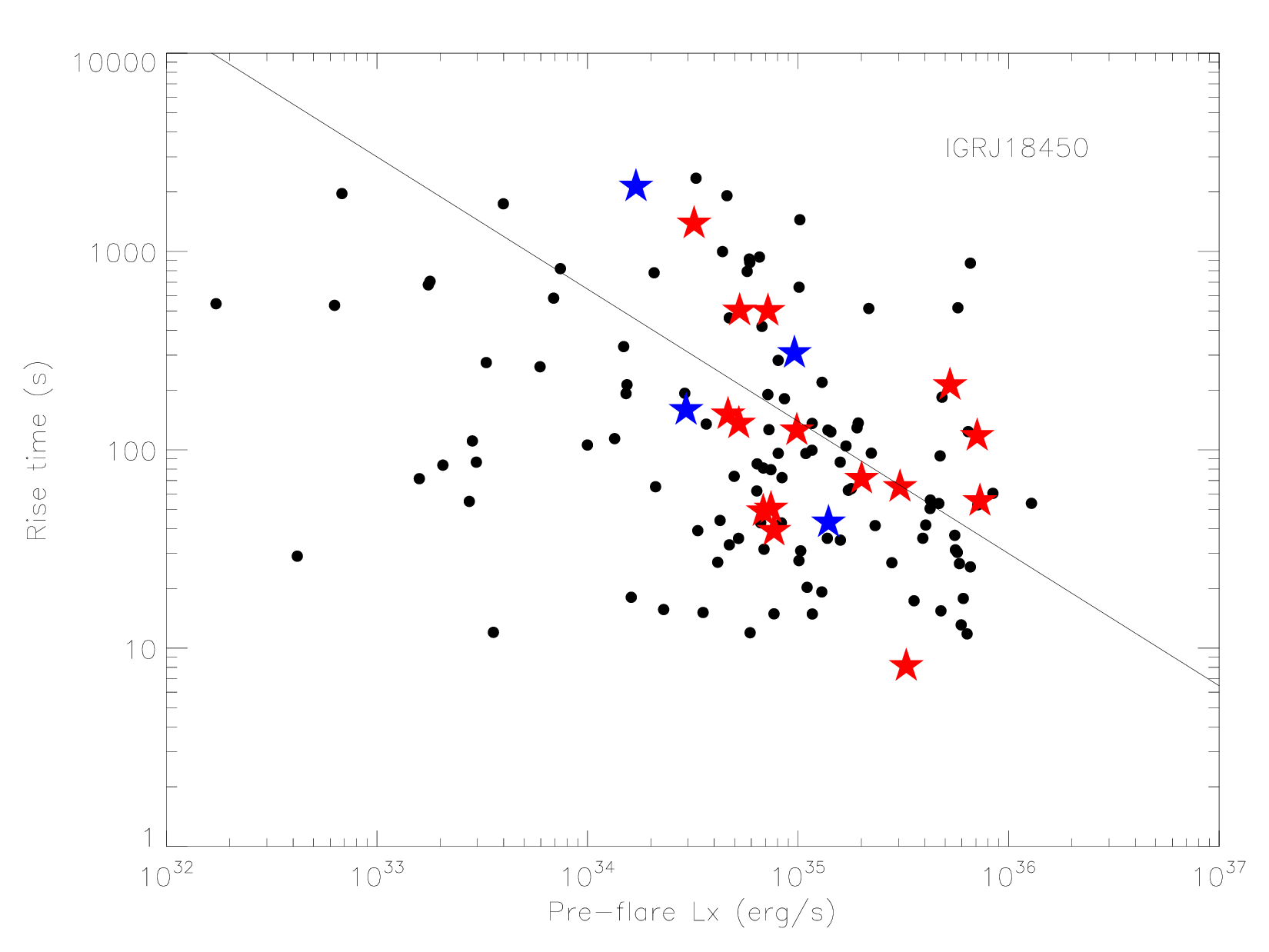} 
\includegraphics[scale=0.29,angle=0]{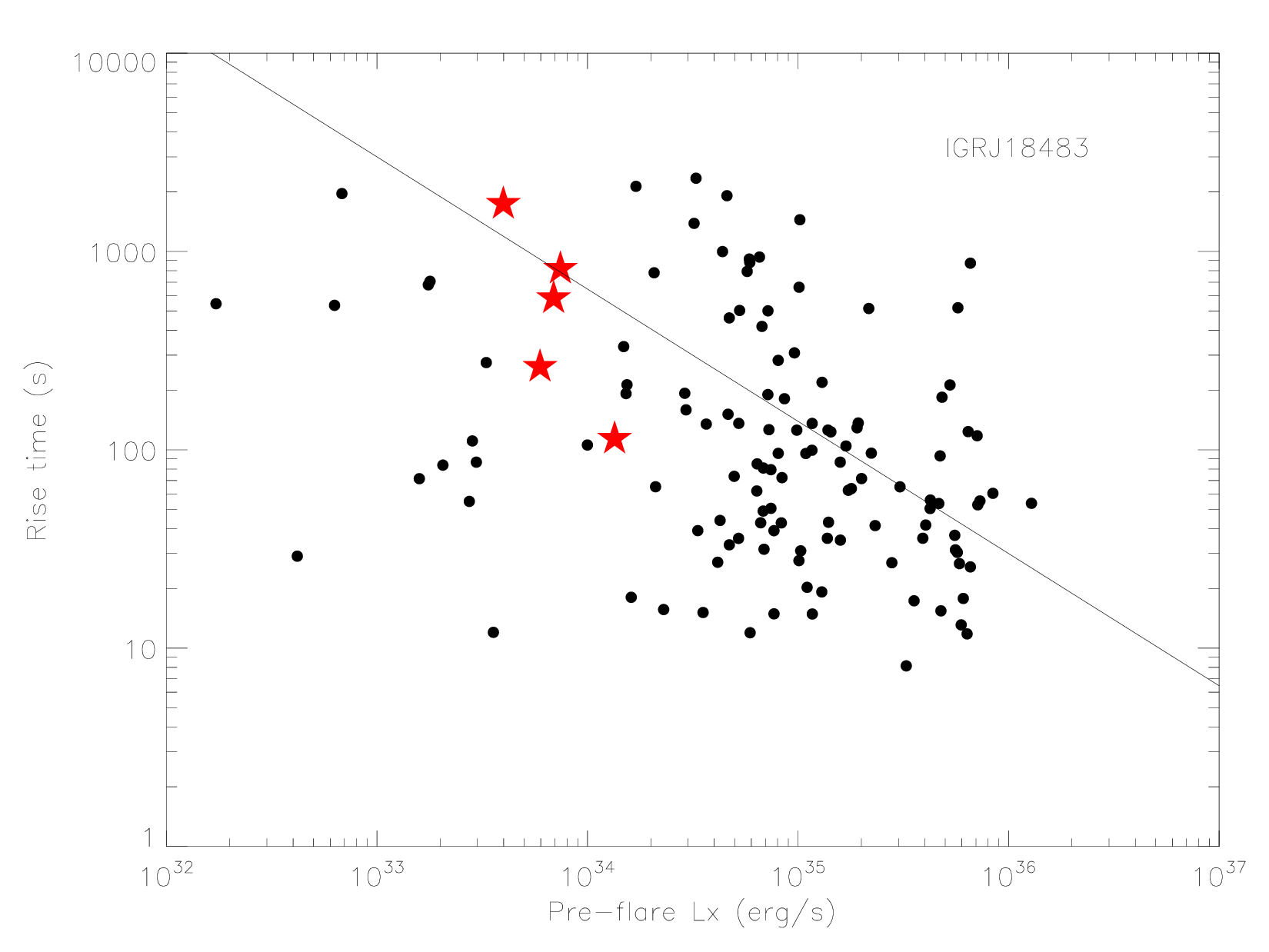} \\
\caption{Rise time to the flare peak vs pre-flare X-ray luminosity. 
Stars mark flares from single sources (different colors indicate flares from different observations).
The solid line marks the theoretical inverse dependence of the rising time from the accretion rate 
in-between the flares ($\delta t_{rise} = 30 \mbox{[s]}\, \dot{M}_{16}^{-2/3}$, see Eq.~\ref{e:rt}). 
}
\label{fig:riset_vs_lx_quiesc_sources}
\end{figure*}

\begin{figure*} 
\includegraphics[scale=0.29,angle=0]{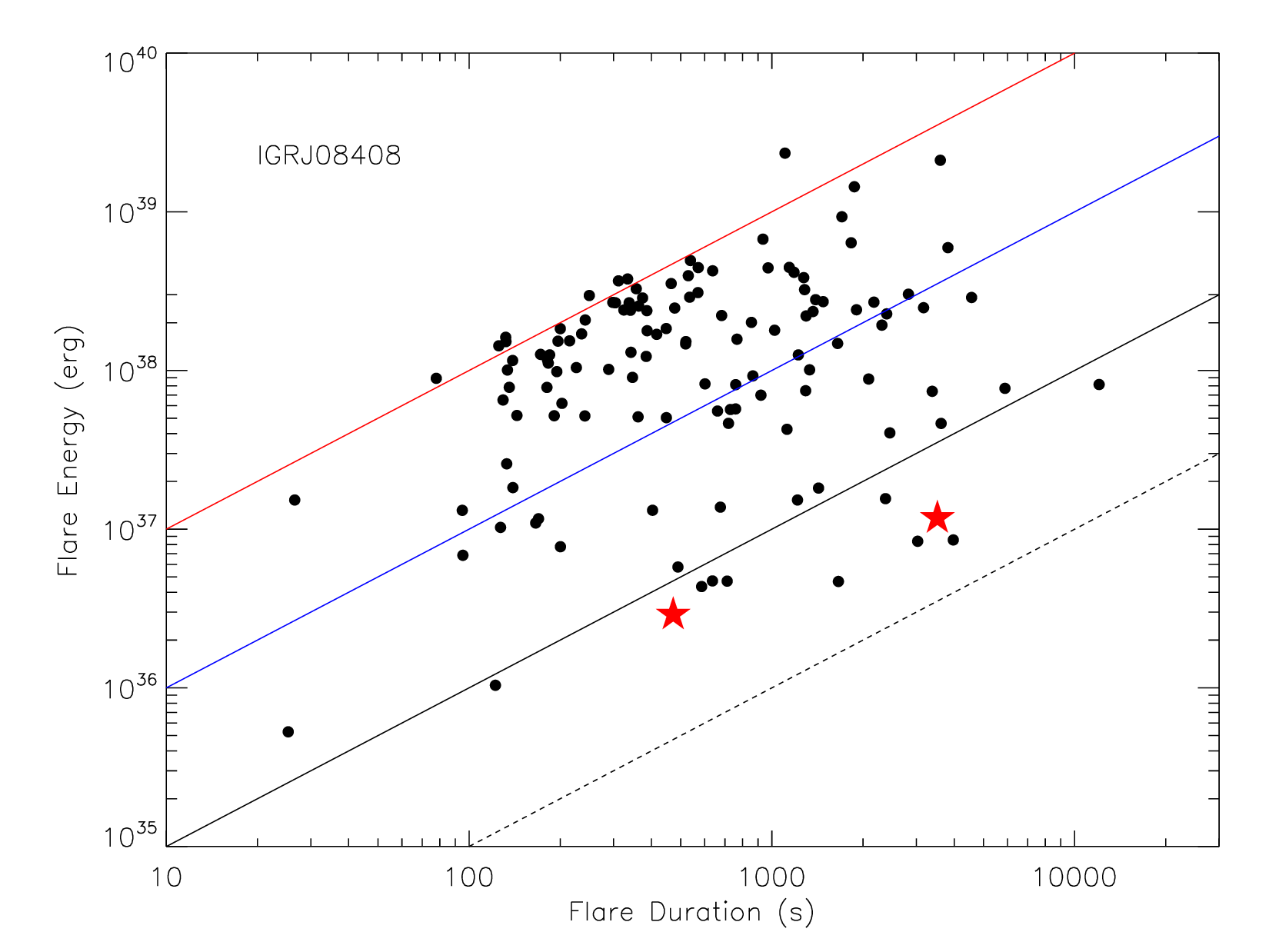} 
\includegraphics[scale=0.29,angle=0]{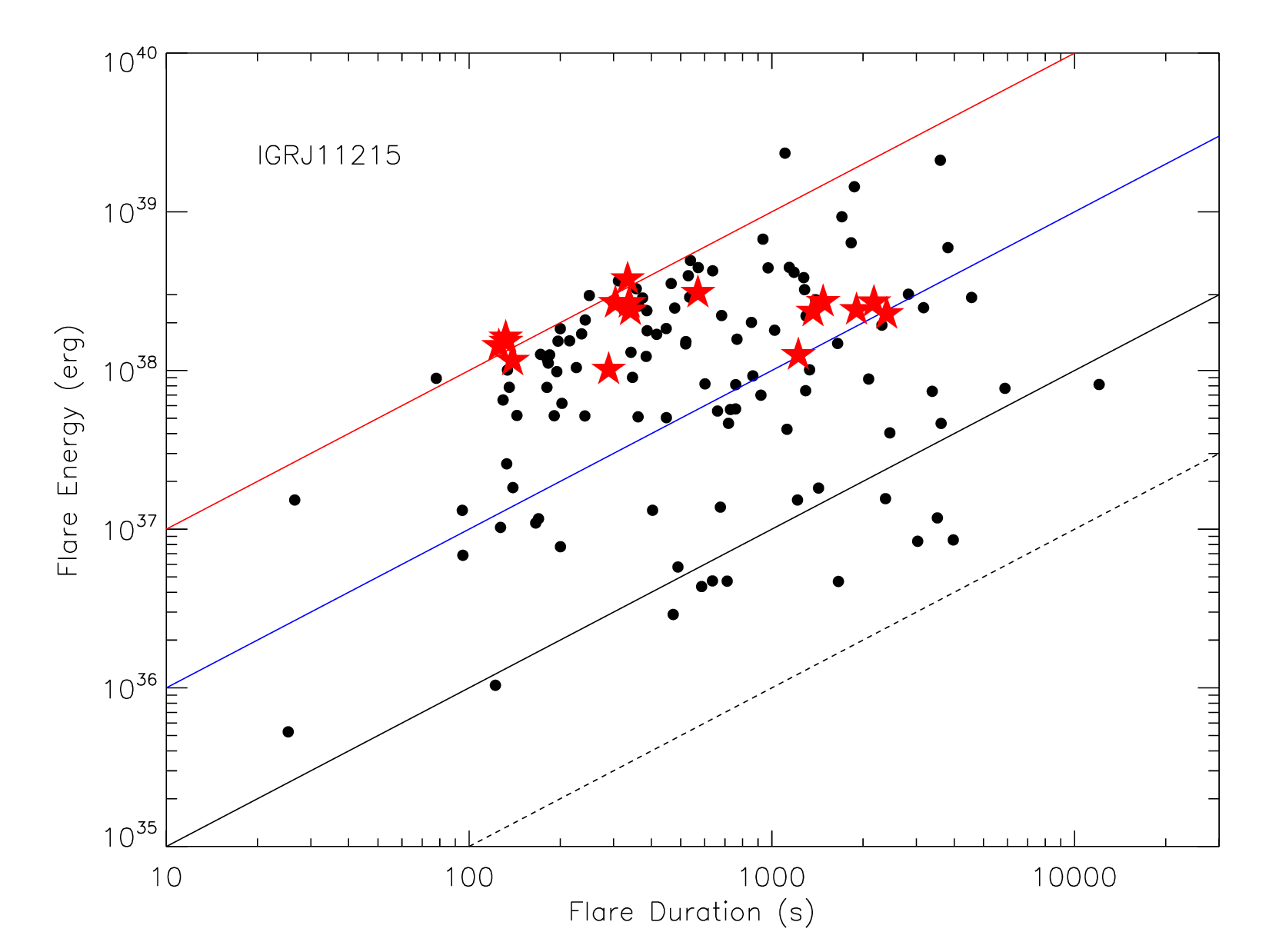} \\
\vspace{-0.4cm}
\includegraphics[scale=0.29,angle=0]{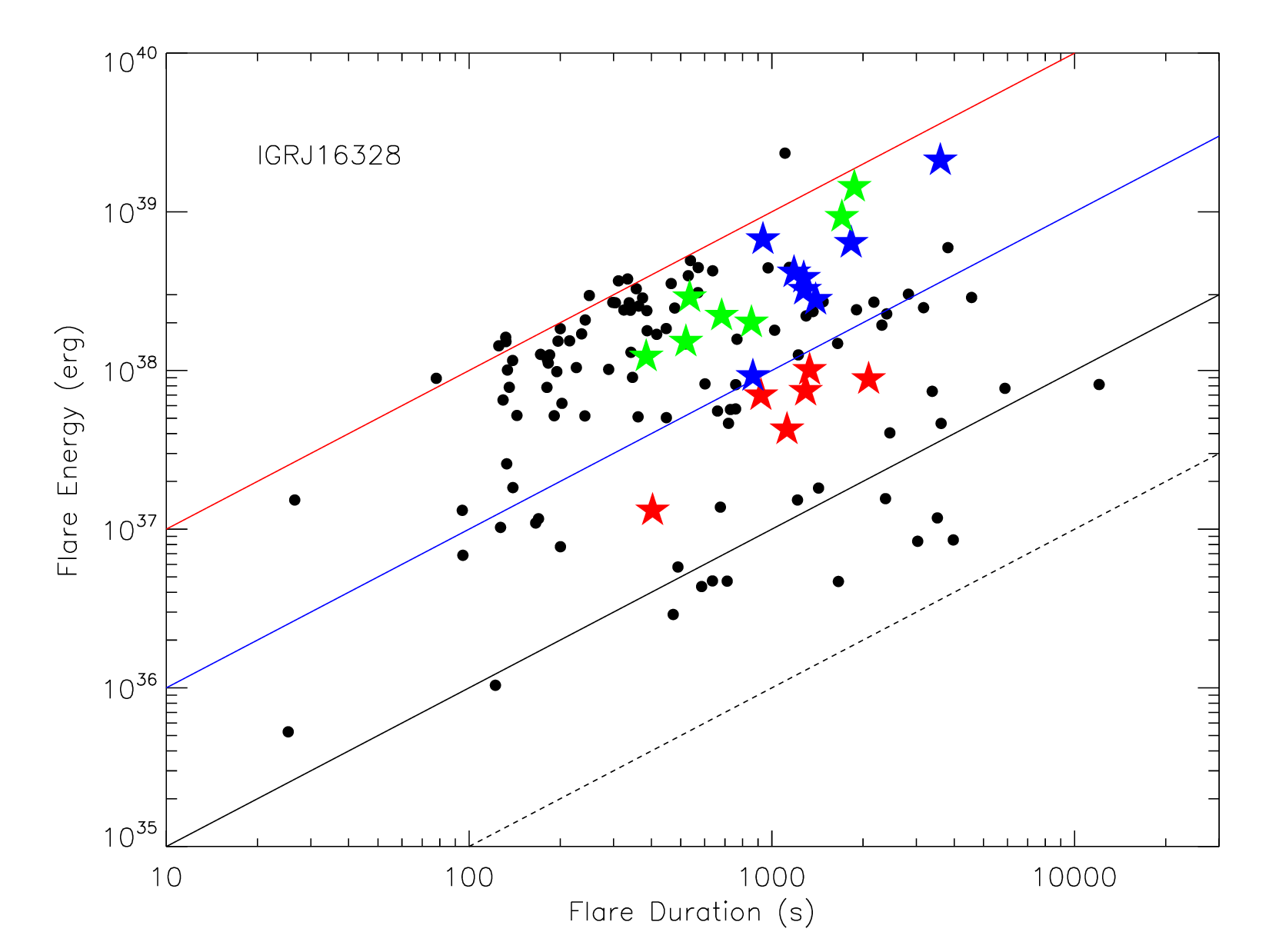} 
\includegraphics[scale=0.29,angle=0]{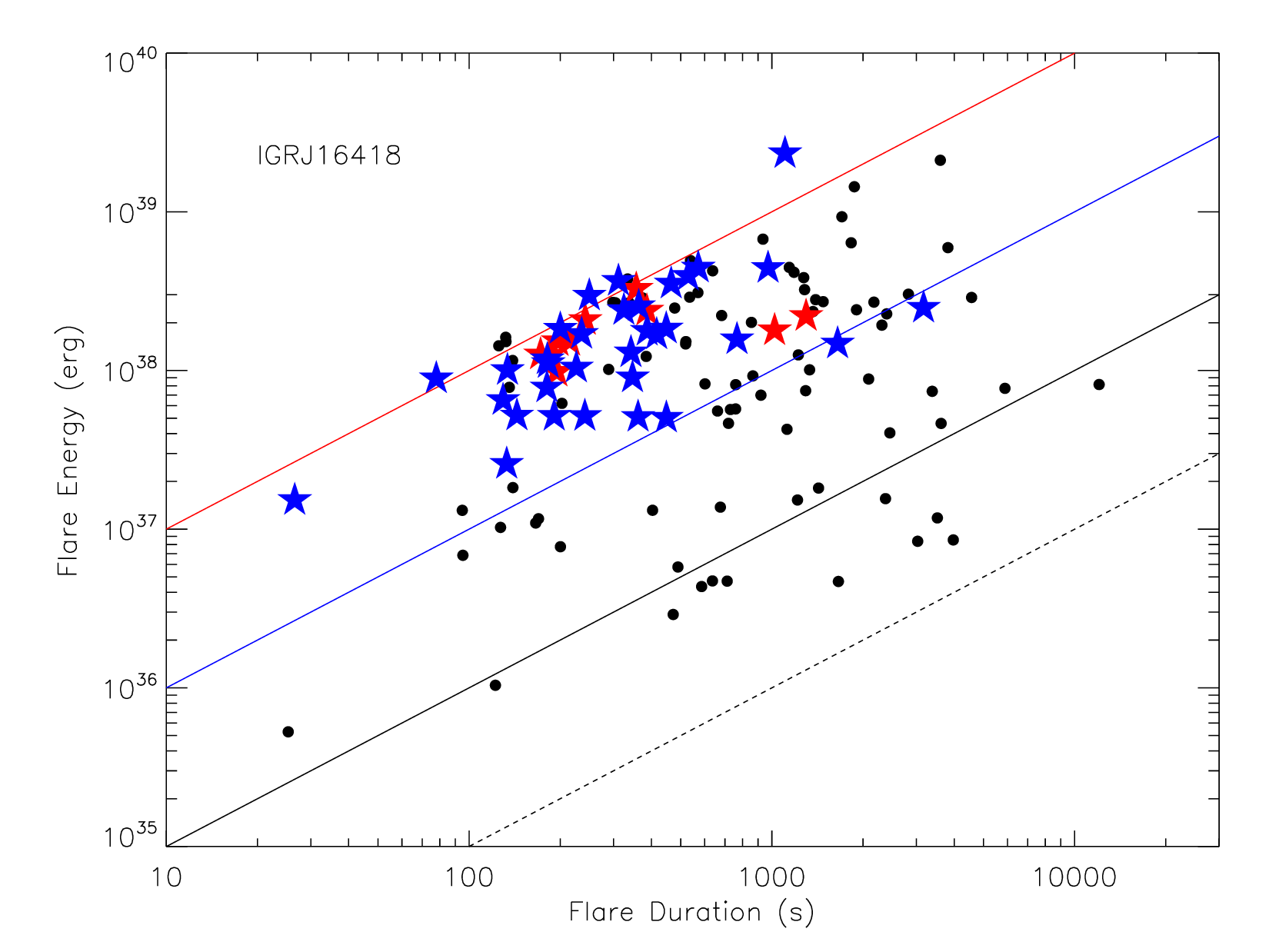}  \\
\vspace{-0.4cm}
\includegraphics[scale=0.29,angle=0]{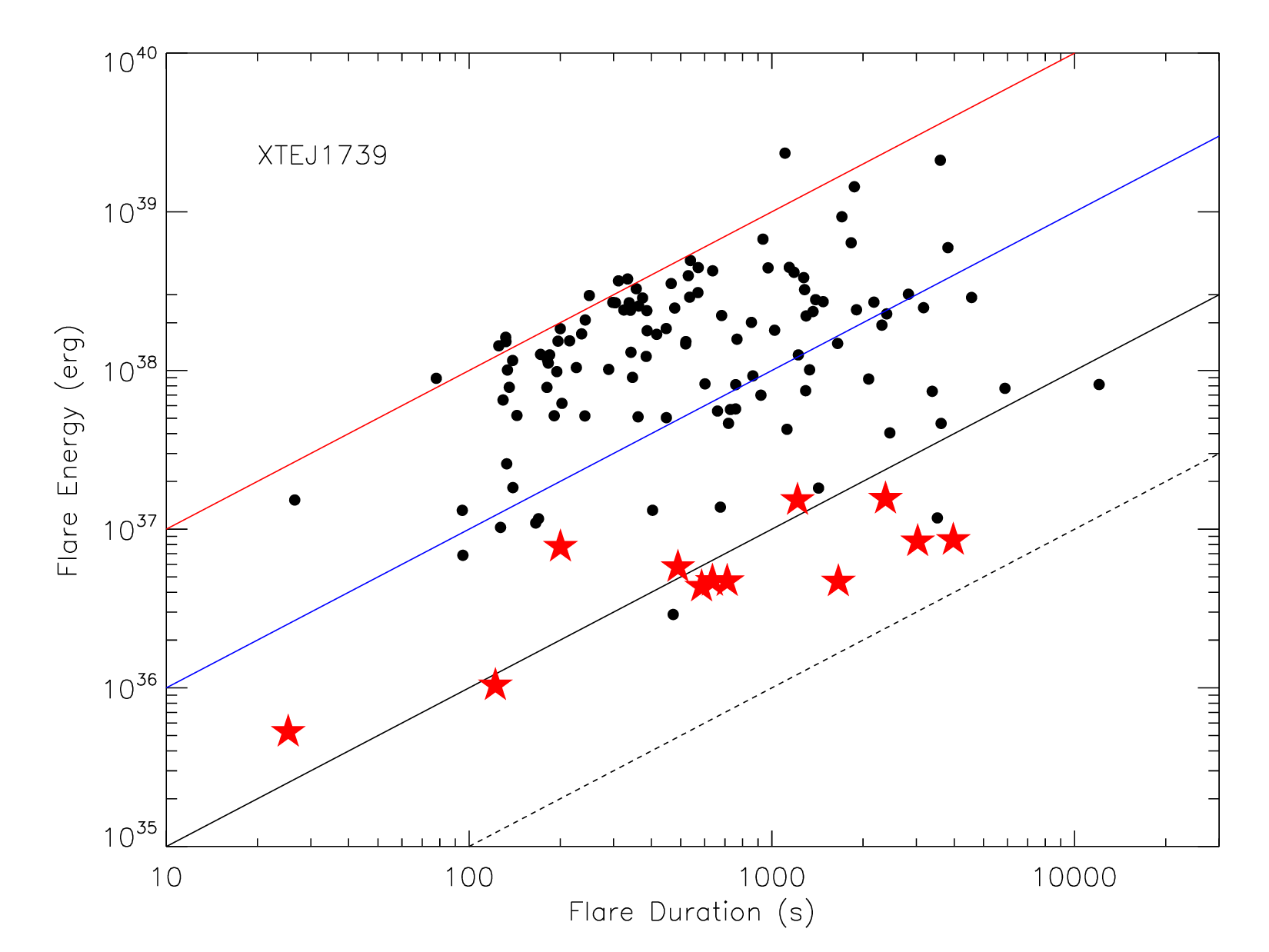} 
\includegraphics[scale=0.29,angle=0]{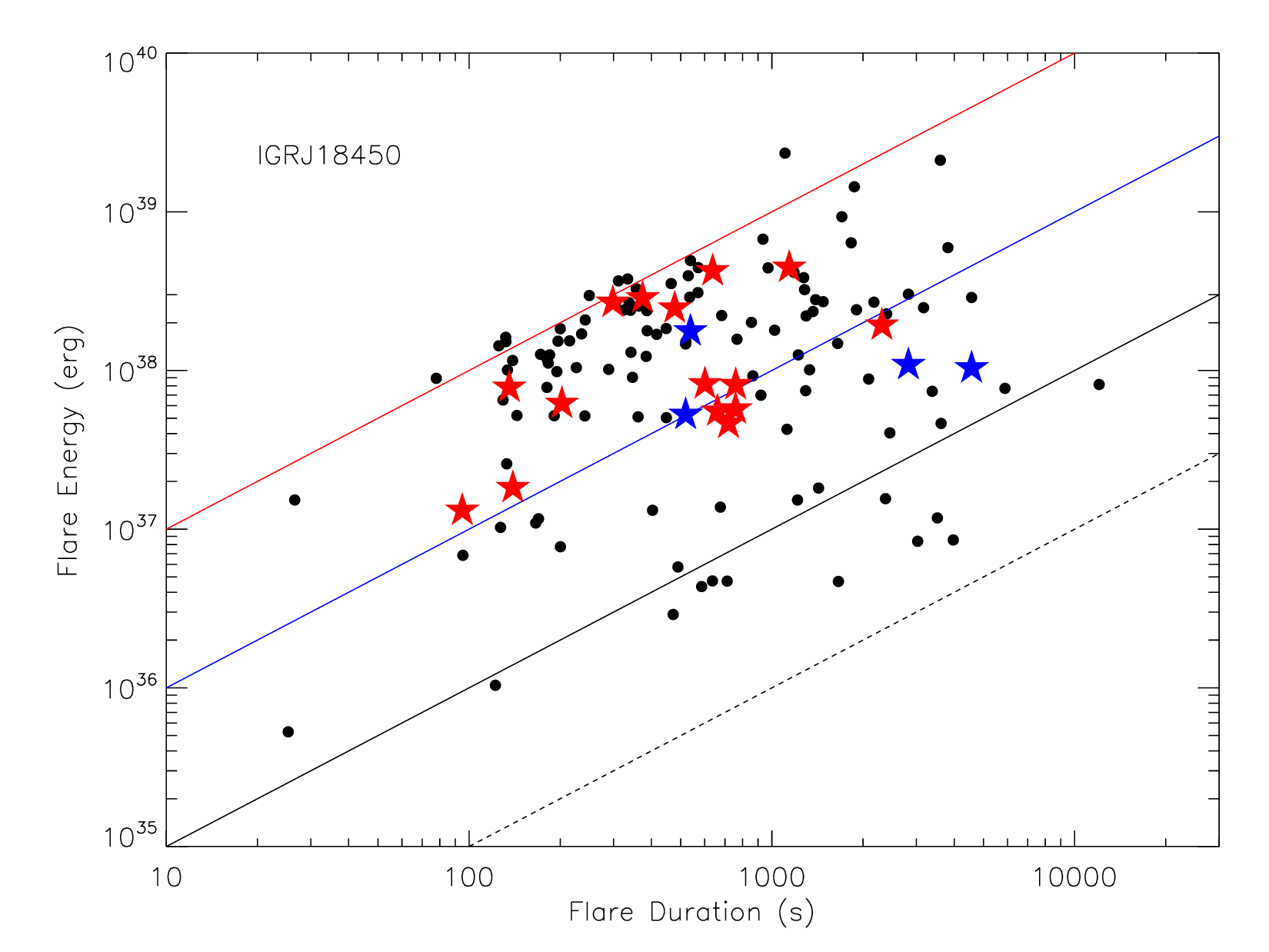} \\
\vspace{-0.4cm}
\includegraphics[scale=0.29,angle=0]{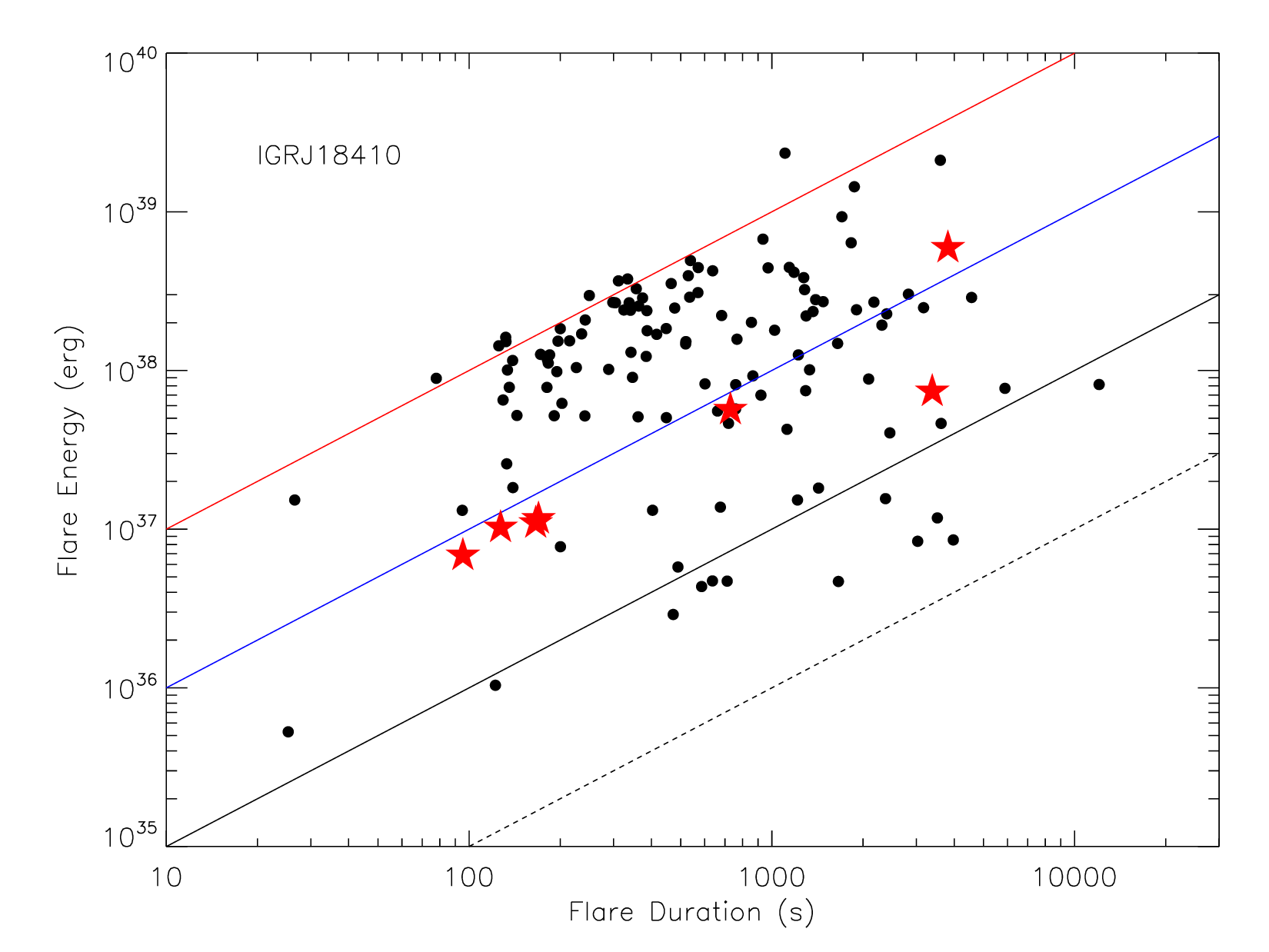}
\includegraphics[scale=0.29,angle=0]{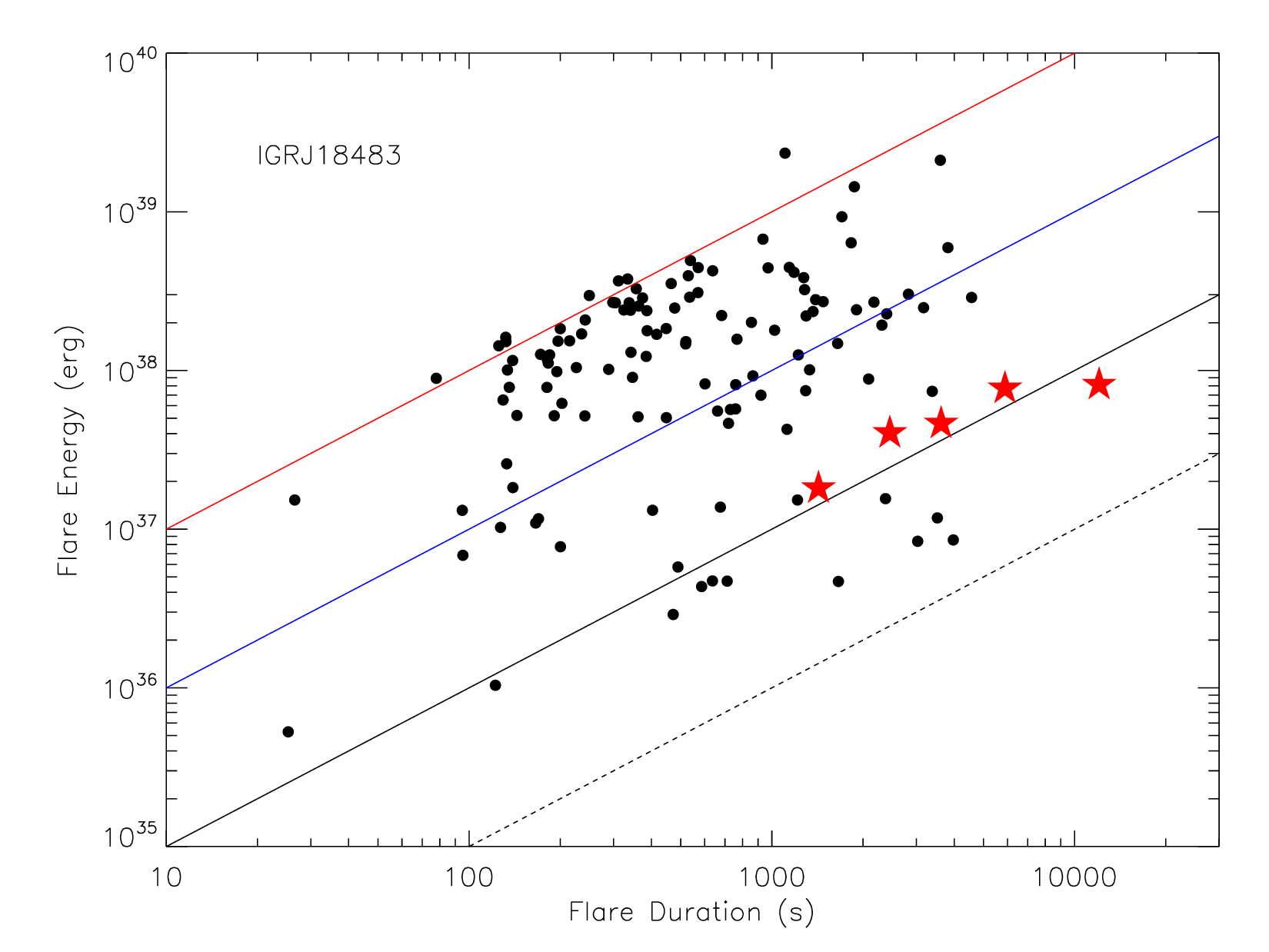} \\
\caption{Energy released in flares vs  flare duration for flares in individual sources. 
The overlaid lines result from the the eq. $\Delta$E = constant $\times$ $\Delta t_{flare}$ 
for different values for the constant, as follows: 
10$^{33}$ (dashed black line), 10$^{34}$ (solid black line), 
10$^{35}$ (dashed blue line), 10$^{36}$ (solid red line). 
}
\label{fig:ene_dur_sources}
\end{figure*}

\begin{figure*} 
\includegraphics[scale=0.29,angle=0]{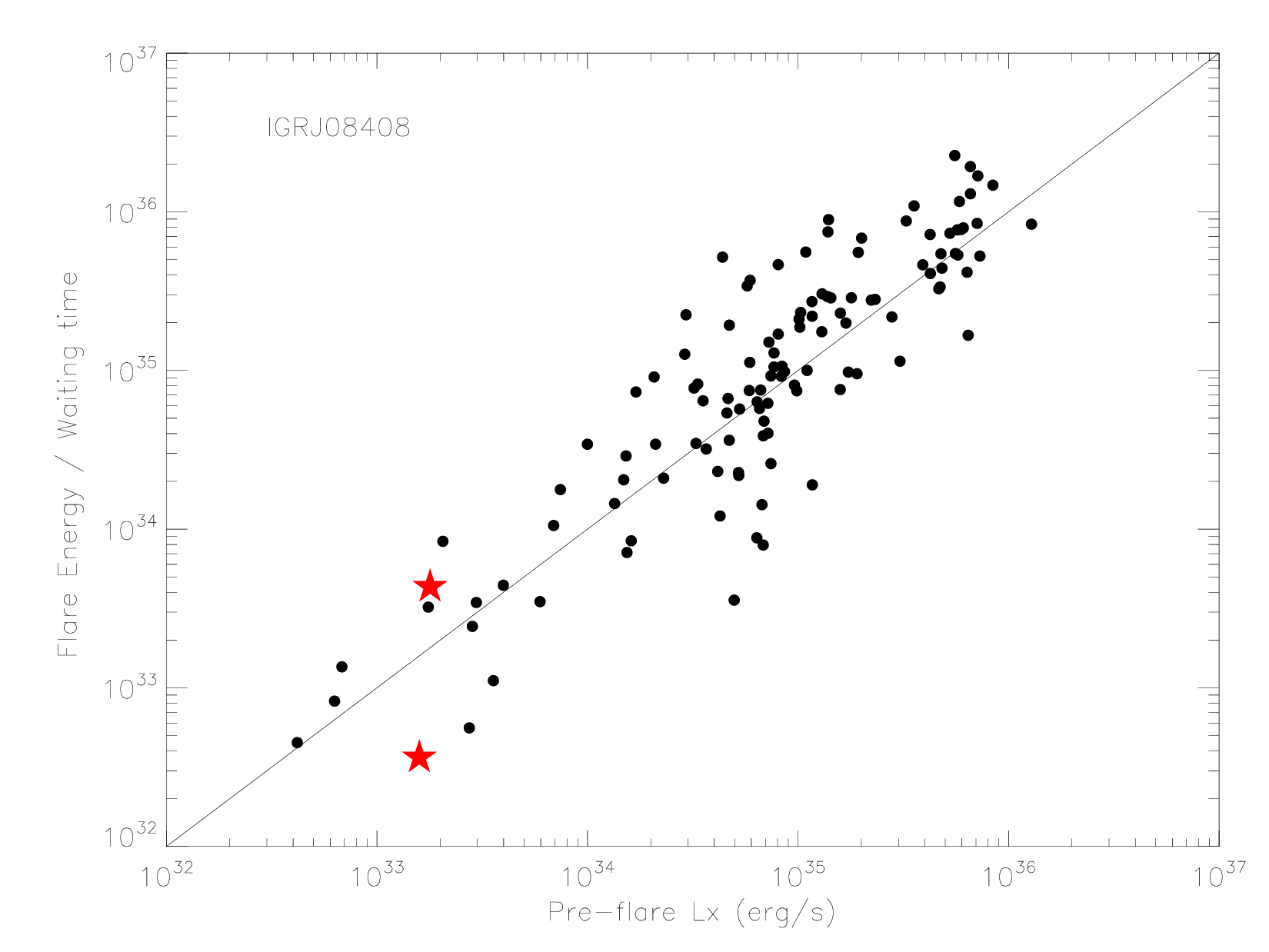}  
\includegraphics[scale=0.29,angle=0]{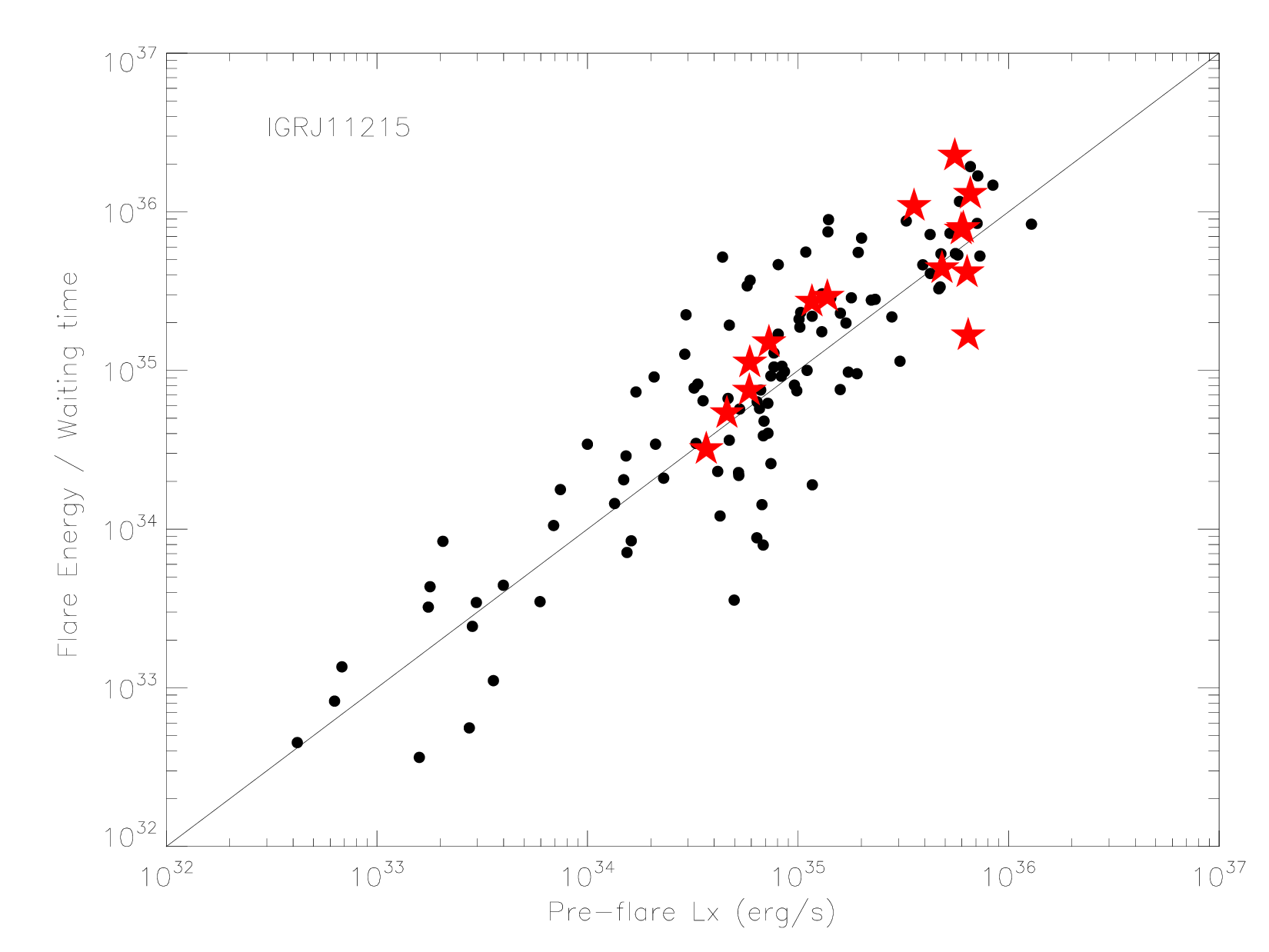} \\
\vspace{-0.4cm}
\includegraphics[scale=0.29,angle=0]{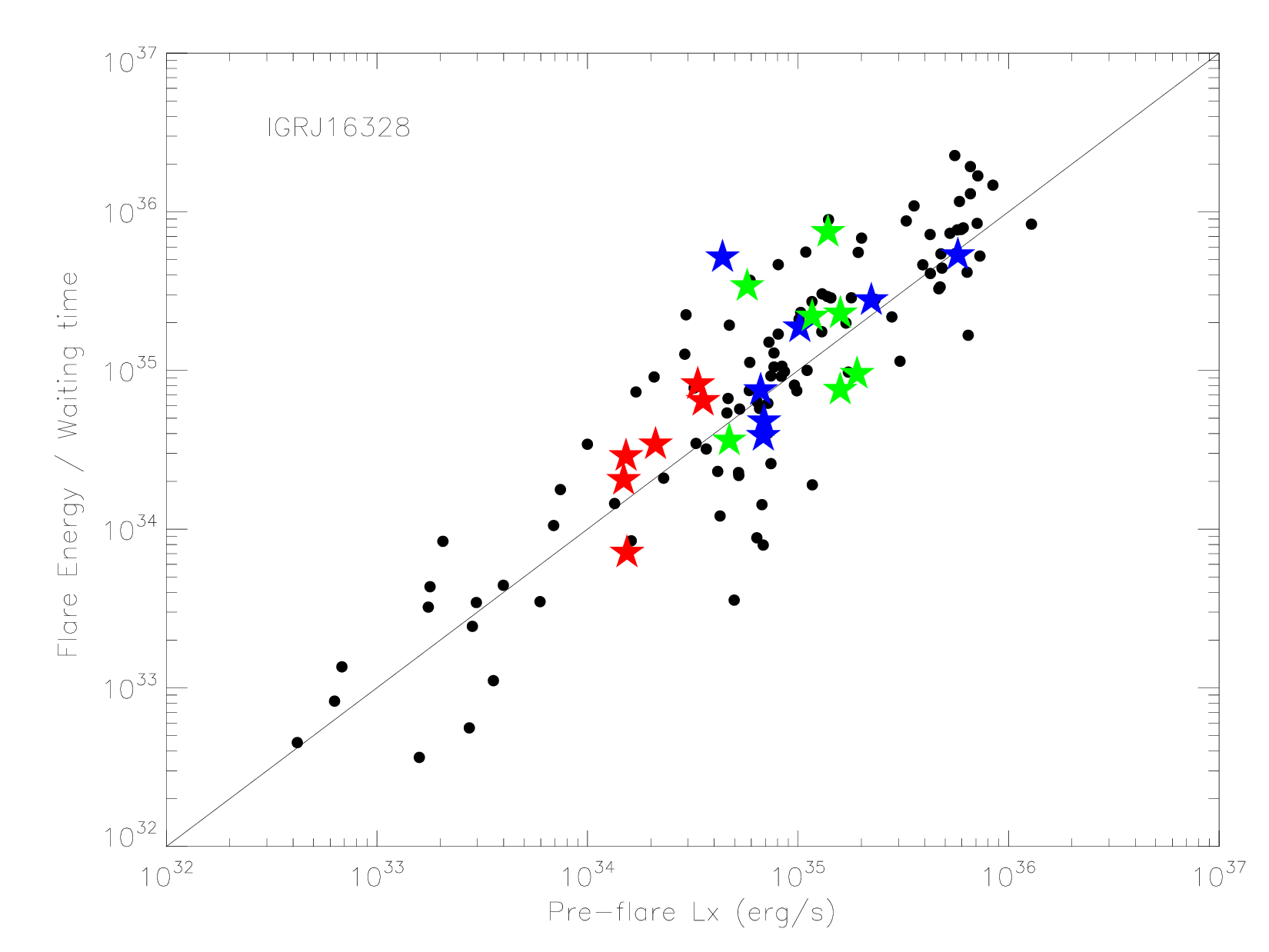}  
\includegraphics[scale=0.29,angle=0]{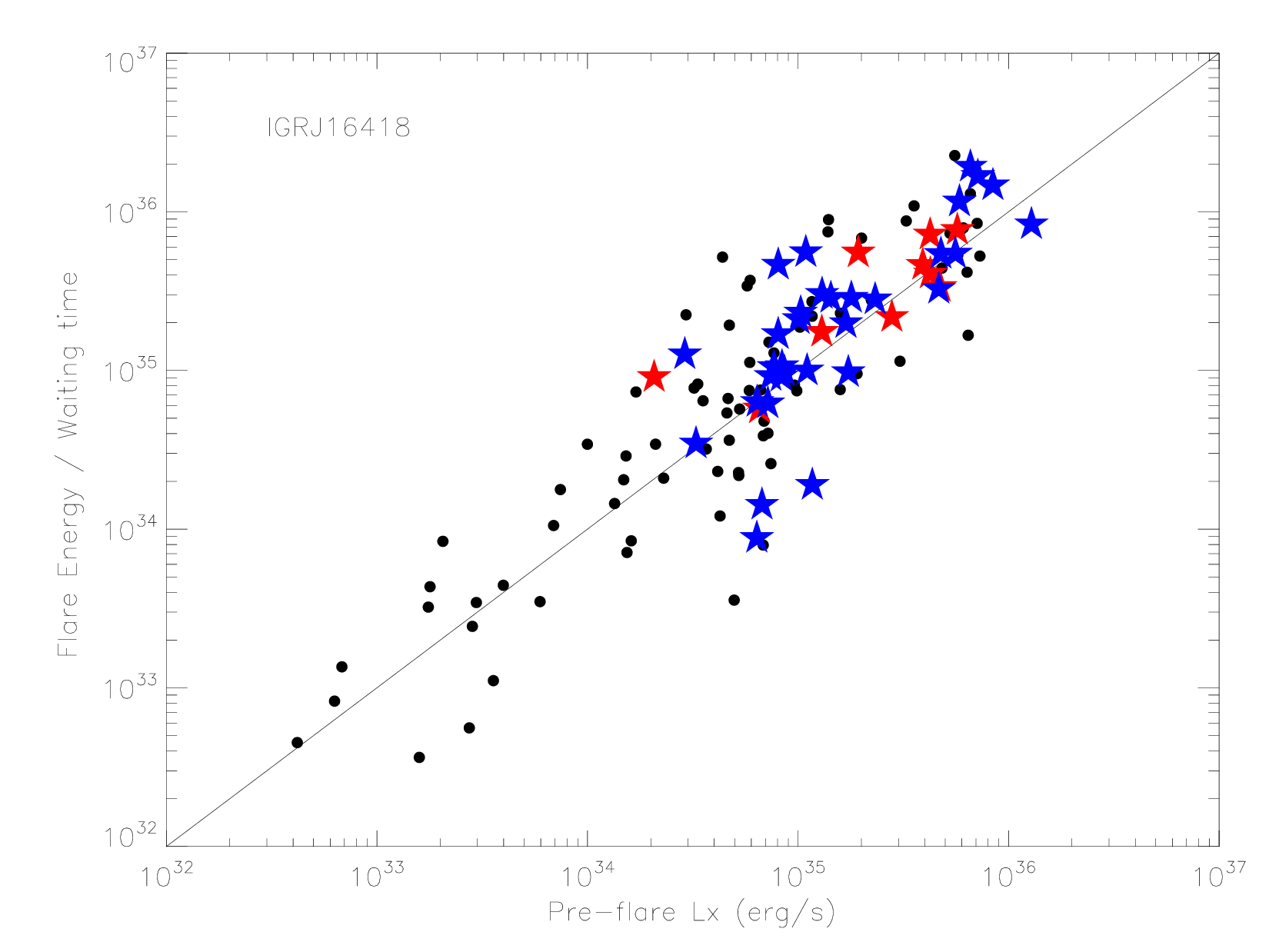} \\
\vspace{-0.4cm}
\includegraphics[scale=0.29,angle=0]{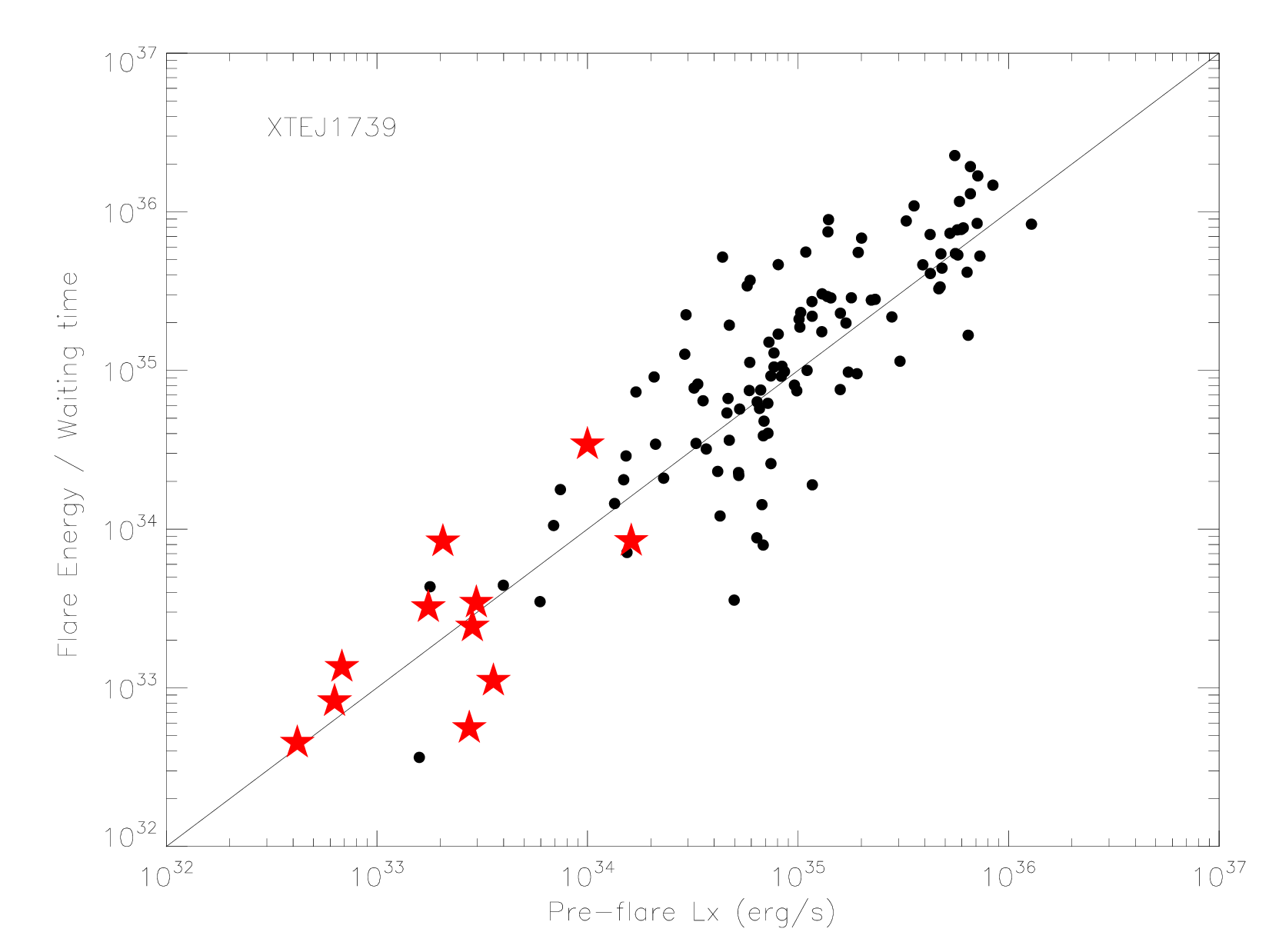}  
\includegraphics[scale=0.29,angle=0]{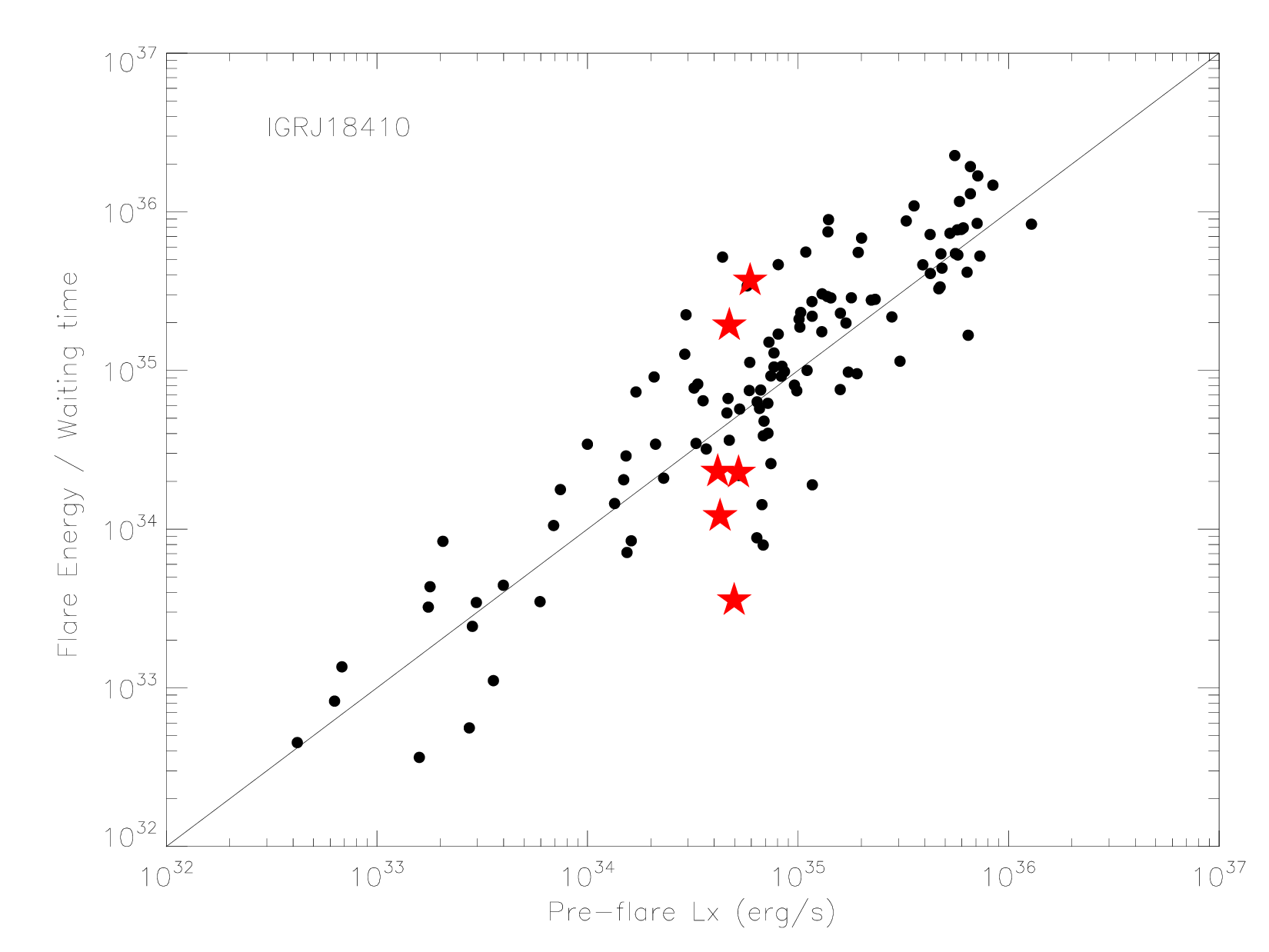} \\ 
\vspace{-0.4cm}
\includegraphics[scale=0.29,angle=0]{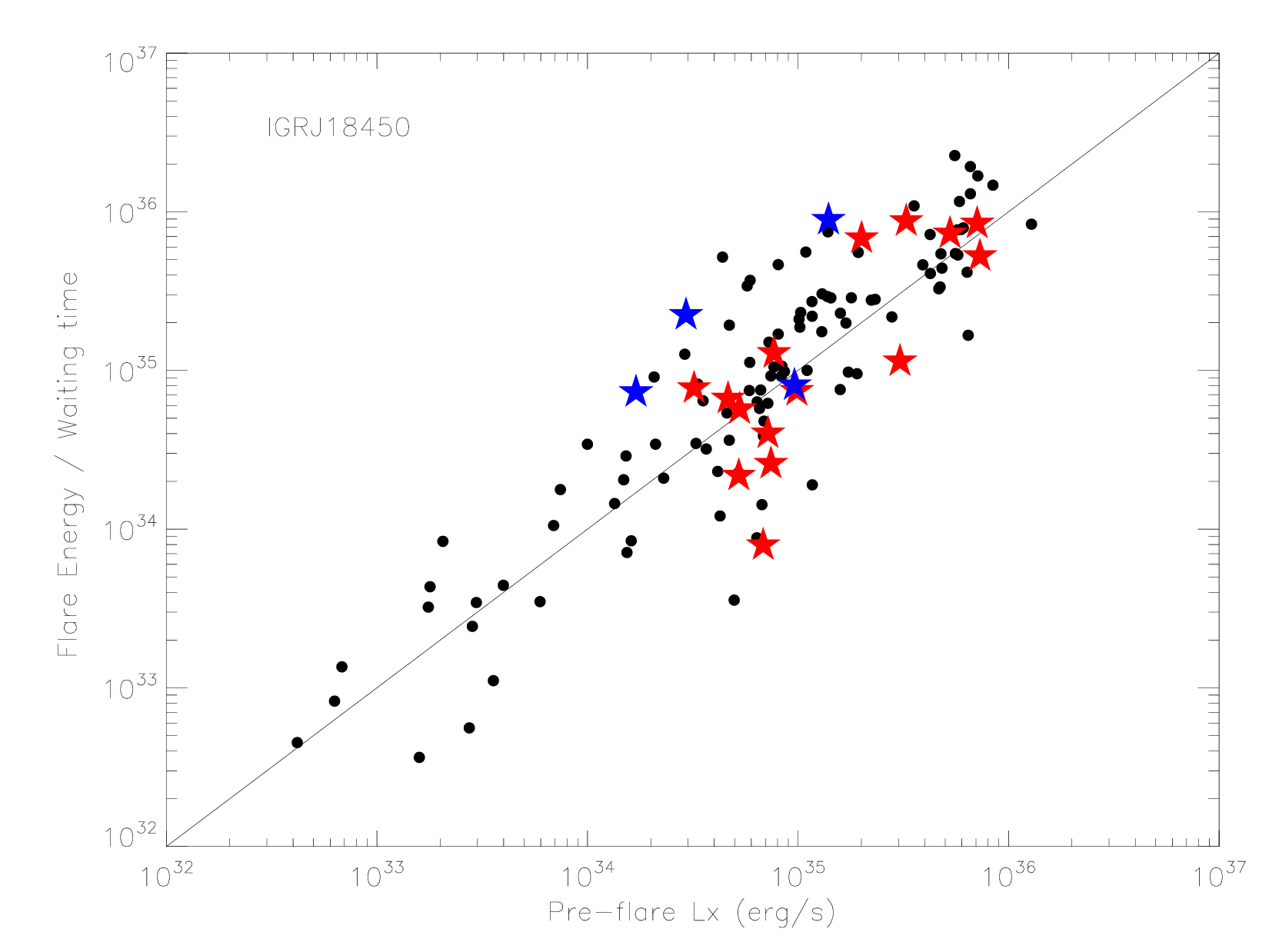} 
\includegraphics[scale=0.29,angle=0]{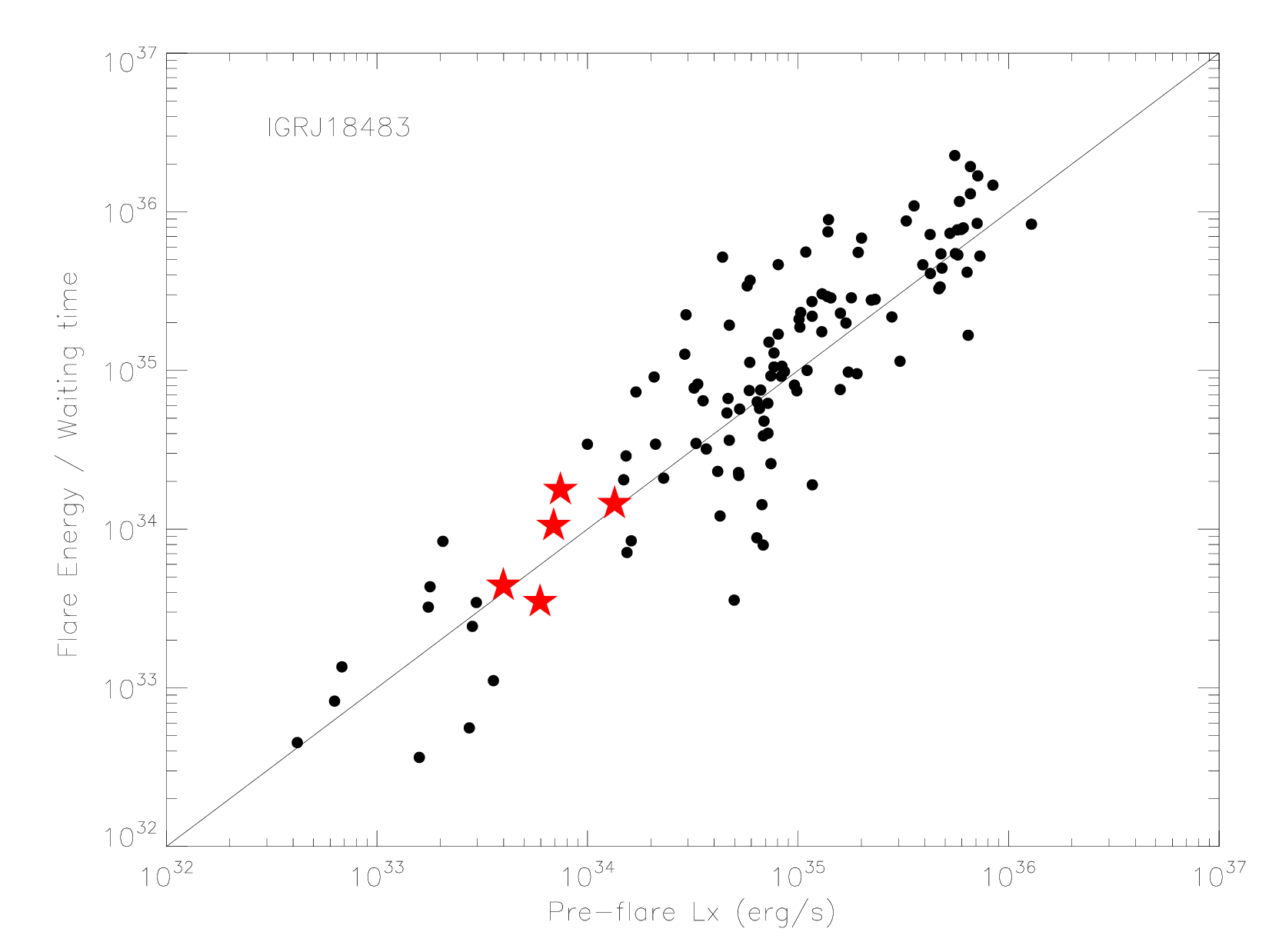} \\
\caption{Ratio of the energy released in flares to the waiting times between consecutive flares, 
plotted against the pre-flare X-ray luminosity, defined as the
luminosity level at the local minimum just before the flare. In color, we show flares from individual sources. 
For a single source, different colors mark flares from different observations.
}
\label{fig:ene_wt_vs_lx_quies_sources}
\end{figure*}

In Table~\ref{tab:flares} we list the properties of the X-ray flares selected from the sample of SFXTs investigated in this work.
Temporal quantities have been rounded to the significant digits. The ($-$) symbol means that the value of the parameters 
could not be determined, according to the definitions assumed in Sect.~\ref{sect:obsdef}.
In Figs.~\ref{fig:lc1} and ~\ref{fig:lc2} we report the EPIC source light curves, segmented in B.b.
In Figs.~\ref{fig:wait_lx_quiesc_sources}, \ref{fig:riset_vs_lx_quiesc_sources}, \ref{fig:ene_dur_sources} and \ref{fig:ene_wt_vs_lx_quies_sources} we highlight the behaviour of flares from single sources.

\begin{table*} 
 \centering
  \caption{Properties of the SFXT flares. The asterisk (*) denotes ``unresolved flares''.
}
  \begin{tabular}{lccrrcrr}
\hline
Flare      &   Peak Luminosity          &    Energy          &  Waiting time $\Delta T$   &  Duration $\Delta$t$_{f}$    &  Average Luminosity     & $\delta$t$_{rise}$ & $\delta$t$_{decay}$ \\
  ID       &  (10$^{34}$~erg~s$^{-1}$)  &  (10$^{37}$ erg)   & (s)                        &     (s)                      & (10$^{34}$~erg~s$^{-1}$)   &   (s)       &   (s)     \\
\hline
\multicolumn{8}{c}{IGR~J08408-4503} \cr
%
       1 &          1.2${\pm         0.9}$  &           $-$                         &        $-$ &        $-$ &           $-$                     &        $-$ &        240 \\ 
       2* &         0.6${\pm         0.5}$  &          0.29${\pm         0.22 }$    &       7900 &        470 &           0.6${\pm         0.5}$  &         70 &         80 \\ 
       3 &          0.9${\pm         0.7}$  &           1.2${\pm          0.4 }$    &       2700 &       3500 &          0.33${\pm         0.13}$ &        700 &       2500 \\ 
       4 &          0.27${\pm       0.20}$  &           $-$                         &      14200 &       $-$  &           $-$                     &        400 &        $-$ \\ 
\hline
\multicolumn{8}{c}{IGR~J11215-5952} \cr
       5 &           83${\pm          24}$   &           31${\pm          7 }$ &        $-$ &        570 &           54${\pm           12}$ &        520 &         15 \\ 
       6*&           79${\pm          23}$   &           27${\pm          8 }$ &        200 &        340 &           79${\pm           23}$ &         17 &        190 \\ 
       7*&           83${\pm          24}$   &           12${\pm          3 }$ &        700 &        140 &           83${\pm           24}$ &        120 &         32 \\ 
       8 &           112${\pm          32}$  &           27${\pm          5 }$ &        600 &        304 &           87${\pm           18}$ &        180 &         19 \\ 
       9*&           114${\pm          33}$  &           14${\pm          4 }$ &        180 &        125 &          114${\pm           33}$ &         18 &         14 \\ 
      10 &           142${\pm          41}$  &           38${\pm          8 }$ &        160 &        330 &          113${\pm           23}$ &         40 &        227 \\ 
      11 &           132${\pm          38}$  &           15${\pm          3 }$ &        360 &        130 &          115${\pm           24}$ &         12 &         60 \\ 
      12*&           123${\pm          35}$  &           16${\pm          5 }$ &        210 &        132 &          123${\pm           35}$ &         13 &         17 \\ 
      13 &           111${\pm          32}$  &           24${\pm          5 }$ &        180 &        341 &           70${\pm           14}$ &         26 &        212 \\ 
      14 &           25${\pm           7}$   &           27${\pm          6 }$ &       2400 &       1480 &         18.4${\pm          3.7}$ &        880 &         70 \\ 
      15 &           22${\pm           6}$   &           24${\pm          5 }$ &        900 &       1900 &         12.7${\pm          2.6}$ &        100 &       1400 \\ 
      16 &           34${\pm          10}$   &           24${\pm          3 }$ &       3100 &       1400 &         17.2${\pm          2.5}$ &        900 &        323 \\ 
      17 &           15${\pm           4}$   &           23${\pm          5 }$ &       1500 &       2390 &          9.5${\pm          2.0}$ &        130 &       1860 \\ 
      18*&           10${\pm           3}$   &           13${\pm          4 }$ &       3900 &       1220 &           10${\pm            3}$ &        130 &        230 \\ 
      19 &           31${\pm          9}$    &           27${\pm          4 }$ &       4900 &       2200 &         12.4${\pm          2.1}$ &       1900 &         50 \\ 
      20*&           35${\pm         10}$    &           10${\pm          3 }$ &        300 &        290 &           35${\pm          10}$  &         35 &         50 \\ 
      21 &           35${\pm         10}$    &           $-$                   &        400 &        $-$ &           $-$                    &         50 &         $-$ \\ 
\hline
\multicolumn{8}{c}{IGR~J16328-4726 (a)} \cr
      22 &           9.1${\pm          0.8}$  &           $-$                   &        $-$ &        $-$ &           $-$                     &        $-$ &         46 \\ 
      23 &           9.7${\pm          0.8}$  &           8.8${\pm       0.5 }$ &       1100 &       2100 &           4.2${\pm          0.2}$ &         39 &       1600 \\ 
      24 &          12.7${\pm          1.1}$  &           7.0${\pm       0.3 }$ &       3400 &        920 &           7.6${\pm          0.4}$ &        331 &        430 \\ 
      25 &           6.1${\pm          0.5}$  &           4.3${\pm       0.3 }$ &       1200 &       1100 &           3.8${\pm          0.2}$ &         70 &        700 \\ 
      26*&           3.3${\pm          0.3}$  &           1.3${\pm       0.1 }$ &       1800 &        410 &           3.3${\pm          0.3}$ &        210 &        210 \\ 
      27 &           9.5${\pm          0.8}$  &          10.1${\pm       0.6 }$ &       3500 &       1330 &           7.6${\pm          0.4}$ &        190 &        650 \\ 
      28 &          15.6${\pm          1.3}$  &           7.5${\pm       0.5 }$ &       1200 &       1290 &           5.8${\pm          0.4}$ &         15 &       1170 \\ 
\hline
\multicolumn{8}{c}{IGR~J16328-4726 (b)} \cr
     29 &            39${\pm          3}$  &           39${\pm          2}$  &        $-$ &       1274 &           30.2${\pm           1.8}$ &         28 &        867 \\ 
     30*&            25${\pm          2}$  &           32${\pm          3}$  &       4300 &       1282 &           25.2${\pm           2.1}$ &         43 &         77 \\ 
     31 &            74${\pm          6}$  &           64${\pm          3}$  &       3400 &       1800 &           34.9${\pm           1.8}$ &       1450 &        230 \\ 
     32 &           102${\pm          9}$  &           67${\pm          4}$  &       1300 &        900 &           71.9${\pm           3.8}$ &        500 &        253 \\ 
     33 &            40${\pm          3}$  &           42${\pm          3}$  &       1500 &       1180 &           35.1${\pm           2.6}$ &        100 &        286 \\ 
     34 &            96${\pm          8}$  &          210${\pm         10}$  &       4100 &       3610 &           58.7${\pm           2.9}$ &        990 &       1410 \\ 
     35 &            30${\pm          2}$  &           28${\pm          2}$  &       5800 &       1390 &           20.0${\pm           1.2}$ &         32 &        870 \\ 
     36 &            19${\pm          2}$  &           9.2${\pm        0.6}$ &       2400 &        900 &           10.7${\pm           0.6}$ &         80 &        700 \\ 
     37 &           147${\pm         12}$  &           $-$                   &       4790 &       $-$  &           $-$                       &        750 &        $-$ \\  
\hline  
\multicolumn{8}{c}{IGR~J16328-4726 (c)} \cr         
     38 &           24${\pm          2}$  &           $-$                  &        $-$ &        $-$ &           $-$                      &        $-$ &        480 \\ 
     39 &           34${\pm          3}$  &           20${\pm         1 }$ &       5500 &        860 &          23.5${\pm       1.5}$     &        460 &         44 \\ 
     40 &           70${\pm          6}$  &           29${\pm         2 }$ &       1300 &        540 &          54.1${\pm       2.9}$     &        136 &        160 \\ 
     41*&           32${\pm          3}$  &           12${\pm         1 }$ &       1600 &        380 &          31.9${\pm       2.7}$     &         90 &         70 \\ 
     42 &           101${\pm         8}$  &           93${\pm         5 }$ &       1200 &       1703 &          54.6${\pm       3.0}$     &        126 &       1009 \\ 
     43 &           121${\pm        10}$  &           144${\pm        6 }$ &       4200 &       1870 &          76.8${\pm       3.3}$     &        790 &        744 \\ 
     44*&           33${\pm          3}$  &           22${\pm         2 }$ &       2300 &        680 &          32.5${\pm       2.7}$     &        130 &         80 \\ 
     45 &           43${\pm          4}$  &         15.2${\pm       0.9 }$ &        700 &        519 &          29.2${\pm       1.8}$     &         35 &        391 \\ 
     46 &           21${\pm          2}$  &          $-$                   &       1200 &        $-$ &           $-$                      &         60 &         $-$ \\ 
\hline
\end{tabular}
\label{tab:flares}
\end{table*} 
     \setcounter{table}{0}
\begin{table*}
 \centering
  \caption{Properties of the SFXT flares. {\it (continued)}
}
  \begin{tabular}{lccrrcrr}
\hline
Flare      &   Peak Luminosity          &    Energy          &  Waiting time $\Delta T$   &  Duration $\Delta$t$_{f}$    &  Average Luminosity     & $\delta$t$_{rise}$ & $\delta$t$_{decay}$ \\
  ID       &  (10$^{34}$~erg~s$^{-1}$)  &  (10$^{37}$ erg)   & (s)                        &     (s)                      & (10$^{34}$~erg~s$^{-1}$)   &   (s)       &   (s)     \\

\hline 
\multicolumn{8}{c}{IGR~J16418-4532 (a)} \cr
      47 &           38${\pm          6}$  &           $-$                   &        $-$ &     $-$    &         $-$                     &        $-$ &        298 \\ 
      48 &           79${\pm          12}$ &           18${\pm          1 }$ &       1970 &       1020 &          17.6${\pm      1.4}$   &        780 &        168 \\ 
      49 &           31${\pm          5}$  &           22${\pm          2 }$ &       3800 &       1290 &          17.0${\pm      1.9}$   &        940 &         60 \\ 
      50*&           51${\pm          8}$  &           10${\pm          2 }$ &        600 &        195 &          51${\pm         8}$    &         19 &         32 \\ 
      51 &           73${\pm          11}$ &           24${\pm          3 }$ &        400 &        390 &          62${\pm         8}$    &        140 &         60 \\ 
      52*&           73${\pm          11}$ &           13${\pm          2 }$ &        300 &        170 &          73${\pm         11}$   &         60 &         80 \\ 
      53*&           72${\pm          11}$ &           15${\pm          2 }$ &        500 &        210 &          72${\pm         11}$   &         90 &         23 \\ 
      54*&           68${\pm          11}$ &           13${\pm          2 }$ &        600 &        180 &          68${\pm         11}$   &         27 &         59 \\ 
      55*&           78${\pm          12}$ &           15${\pm          2 }$ &        300 &        190 &          78${\pm         12}$   &         36 &        150 \\ 
      56 &           107${\pm         16}$ &           21${\pm          2 }$ &        300 &       240 &           86${\pm         9}$    &         31 &        140 \\ 
      57 &           132${\pm         20}$ &           33${\pm          3 }$ &        460 &       360 &           92${\pm         8}$    &         50 &        179 \\ 
      58 &           15${\pm          2}$  &              $-$                 &      1800 &       $-$ &           $-$                    &        160 &        $-$ \\ 
\hline
\multicolumn{8}{c}{IGR~J16418-4532 (b)} \cr
      59*&           75${\pm          12}$  &           10${\pm        2 }$  &        $-$ &        130 &           75${\pm          12}$ &         40 &         70 \\ 
      60*&          115${\pm          18}$  &            9${\pm        1 }$  &        160 &         80 &          115${\pm          18}$ &         15 &         40 \\ 
      61 &          110${\pm          17}$  &           40${\pm        4 }$  &        230 &        530 &           75${\pm           8}$ &         50 &        370 \\ 
      62*&           58${\pm           9}$  &           1.5${\pm      0.2 }$ &        800 &         27 &           58${\pm           9}$ &         15 &         12 \\ 
      63 &           90${\pm          14}$  &           11${\pm         1 }$ &       1210 &        180 &           61${\pm           8}$ &         80 &        120 \\ 
      64 &          405${\pm          62}$  &           234${\pm       15 }$ &       1220 &       1105 &          212${\pm          14}$ &        873 &        172 \\ 
      65 &          180${\pm          28}$  &            30${\pm       3 }$  &        360 &        250 &          119${\pm          12}$ &         50 &        164 \\ 
      66 &           58${\pm           9}$  &           18${\pm        2 }$  &       1900 &        450 &           41${\pm           4}$ &        180 &         60 \\ 
      67 &           68${\pm          10}$  &           18${\pm        2 }$  &        600 &        390 &           46${\pm           5}$ &        220 &          9 \\ 
      68 &           37${\pm           6}$  &           16${\pm        2 }$  &        677 &        770 &           21${\pm           2}$ &         31 &        610 \\ 
      69 &          128${\pm          20}$  &           26${\pm         2 }$ &       1510 &        363 &           70${\pm           6}$ &        282 &         16 \\ 
      70 &          110${\pm          17}$  &           24${\pm         2 }$ &        210 &        324 &           74${\pm           7}$ &         27 &        208 \\ 
      71*&           11${\pm           2}$  &           5.0${\pm       0.8}$ &       3500 &        500 &           11${\pm           2}$ &        400 &        300 \\ 
      72*&           19${\pm           3}$  &           2.6${\pm      0.4 }$ &       2900 &        130 &           19${\pm           3}$ &         60 &         90 \\ 
      73*&           21${\pm           3}$  &           5.2${\pm      0.8 }$ &        500 &        240 &           21${\pm           3}$ &         70 &         57 \\ 
      74*&           14${\pm           2}$  &           5.1${\pm      0.8 }$ &        800 &        400 &           14${\pm           2}$ &        190 &        300 \\ 
      75 &           26${\pm           4}$  &           15${\pm       1 }$   &       1600 &       1600 &          9.0${\pm         0.8}$ &         40 &       1500 \\ 
      76 &           15${\pm           2}$  &           25${\pm       2 }$   &       7200 &       3200 &          7.9${\pm         0.6}$ &       2300 &        390 \\ 
      77 &           64${\pm          10}$  &           17${\pm       2 }$   &       1300 &        420 &           41${\pm           5}$ &        190 &         21 \\ 
      78*&           50${\pm           8}$  &            7${\pm       1 }$   &        200 &        130 &           50${\pm           8}$ &         40 &         11 \\ 
      79 &           37${\pm           6}$  &           5.2${\pm      0.6}$  &        820 &        190 &           27${\pm           3}$ &         80 &         40 \\ 
      80*&           26${\pm           4}$  &           9${\pm        1 }$   &        300 &        350 &           26${\pm           4}$ &        120 &         42 \\ 
      81 &           89${\pm          14}$  &           35${\pm       5 }$   &        800 &        460 &           76${\pm          11}$ &         90 &         32 \\ 
      82 &           78${\pm          12}$  &           10${\pm       1 }$   &        300 &        230 &           46${\pm           6}$ &         50 &        130 \\ 
      83 &           43${\pm           7}$  &           13${\pm       2 }$   &       1200 &        340 &           38${\pm           5}$ &         15 &        110 \\ 
      84 &          152${\pm          23}$  &           37${\pm       5 }$   &        700 &        311 &          118${\pm          16}$ &         96 &         21 \\ 
      85 &          122${\pm          19}$  &           45${\pm       4 }$   &        300 &        570 &           78${\pm           7}$ &         60 &        399 \\ 
      86 &          103${\pm          16}$  &           44${\pm       4 }$   &       2100 &        970 &           46${\pm           4}$ &        660 &        150 \\ 
      87 &           90${\pm          14}$  &           17${\pm       2 }$   &        900 &        240 &           72${\pm           8}$ &        105 &         50 \\ 
      88 &          102${\pm          16}$  &           18${\pm       3 }$   &        300 &        201 &           92${\pm           13}$ &         31 &         32 \\ 
      89*&           43${\pm           7}$  &            8${\pm       1 }$   &        800 &        181 &           43${\pm           7}$ &         21 &         37 \\ 
      90*&           36${\pm           6}$  &           5.2${\pm     0.8}$   &        500 &        140 &           36${\pm           6}$ &         60 &         70 \\ 
      91 &           78${\pm          12}$  &           12${\pm       1 }$   &        410 &        180 &           65${\pm           8}$ &         60 &         40 \\ 
      92 &          102${\pm          16}$  &                 $-$            &        300 &        $-$ &            $-$                  &         18 &         $-$ \\ 
\hline
\end{tabular}
\label{tab:flares}
\end{table*} 
\setcounter{table}{0}
\begin{table*}
 \centering
  \caption{Properties of the SFXT flares. {\it (continued)}
}
   \begin{tabular}{lccrrcrr}
 \hline
Flare      &   Peak Luminosity          &    Energy          &  Waiting time $\Delta T$   &  Duration $\Delta$t$_{f}$  & Average Luminosity  & $\delta$t$_{rise}$ & $\delta$t$_{decay}$ \\
  ID       &  (10$^{34}$~erg~s$^{-1}$)  &  (10$^{37}$ erg)   & (s)                        &     (s)                    & (10$^{34}$~erg~s$^{-1}$)   &   (s)       &   (s)     \\
\hline 
\multicolumn{8}{c}{XTE~J1739-302} \cr
      93 &           1.7${\pm          1.3}$  &          0.58${\pm          0.31 }$ &         $-$ &        490 &          1.2${\pm          0.6}$    &        280 &         12 \\ 
      94 &           0.6${\pm         0.4}$   &          0.84${\pm          0.43 }$ &        1100 &       3100 &          0.28${\pm         0.14}$   &         80 &       2800 \\ 
      95 &           1.7${\pm          1.2}$  &           1.6${\pm          0.7  }$ &       11500 &       2370 &          0.66${\pm         0.29}$   &       1960 &        100 \\ 
      96 &           3.9${\pm        2.9}$    &           1.5${\pm          0.6  }$ &         500 &       1200 &          1.3${\pm          0.5}$    &        110 &       1100 \\ 
      97*&          0.9${\pm         0.6}$    &          0.10${\pm         0.08  }$ &        1860 &        120 &          0.9${\pm          0.6}$    &         50 &         60 \\ 
      98*&          0.7${\pm         0.5}$    &          0.47${\pm          0.35 }$ &        1900 &        710 &          0.7${\pm          0.5}$    &        110 &        120 \\ 
      99*&          0.7${\pm         0.6}$    &          0.47${\pm          0.35 }$ &        1400 &        640 &          0.7${\pm          0.5}$    &         80 &        130 \\ 
     100*&           2.0${\pm          1.6}$  &         0.053${\pm         0.039 }$ &         500 &         25 &          2.1${\pm          1.6}$    &         12 &          8 \\ 
     101 &           0.3${\pm         0.2}$   &          0.86${\pm          0.49 }$ &        3000 &       3900 &          0.22${\pm         0.12}$   &        700 &       1700 \\ 
     102 &           0.6${\pm         0.4}$   &          0.47${\pm          0.25 }$ &       11000 &       1650 &          0.28${\pm         0.15}$   &         29 &       1290 \\ 
     103 &           4.1${\pm          3.1}$  &          0.44${\pm          0.23 }$ &        5300 &        586 &          0.7${\pm          0.4}$    &        535 &         14 \\ 
     104*&           3.9${\pm          2.9}$  &          0.78${\pm          0.57 }$ &         900 &        210 &          3.9${\pm          2.9}$    &         18 &         60 \\ 
     105 &           4.2${\pm          3.1}$  &              $-$                    &         300 &       $-$  &            $-$                      &         42 &      $-$ \\ 
\hline 
\multicolumn{8}{c}{IGR~J17544-2619} \cr
     106 &           3.6${\pm         0.5}$  &           $-$                       &        $-$ &       $-$  &            $-$                       &        $-$ &        110 \\ 
     107 &           6.4${\pm         0.8}$  &      1.38${\pm         0.12 }$      &        700 &        680 &           2.0${\pm          0.2}$    &         16 &        570 \\ 
     108 &           0.8${\pm         0.1}$  &         $-$                         &       5300 &        $-$ &            $-$                       &        400 &        $-$ \\ 
\hline
\multicolumn{8}{c}{IGR~J18410-0535} \cr
     109 &           23  $^{+40} _{-20}$      &                   60 $^{+100} _{-50}$ &        $-$ &       3810 &                     16$^{+38} _{-9}$   &        550 &       2169 \\ 
     110*&            7  $^{+12} _{-6}$       &                   1.2 $^{+2 } _{-1}$  &       3300 &        170 &                      7$^{+11} _{-6}$   &         70 &         33 \\ 
     111*&            7  $^{+12} _{-6}$       &               0.7 $^{+1.2 } _{-0.6}$  &        290 &         90 &                      7$^{+11} _{-6}$   &         27 &         70 \\ 
     112 &            8  $^{+14} _{-7}$       &                 1.0$^{+1.8 } _{-0.9}$ &        450 &        130 &                      8$^{+15} _{-6}$   &         36 &         70 \\ 
     113 &           12  $^{+20} _{-10}$      &                  6$^{+10} _{-5}$      &        150 &        729 &                      8$^{+11} _{-6}$   &         11 &        688 \\ 
     114*&             7 $^{+12} _{-6}$       &              1.1 $^{+1.9} _{-0.9}$    &        910 &        170 &                      7$^{+11} _{-6}$   &         40 &         70 \\ 
     115 &             8 $^{+14} _{-7}$       &           7 $^{+13} _{-6}$            &        380 &       3390 &                 2.2$^{+3.8} _{-1.9}$   &         33 &       3280 \\ 
     116 &          0.7 $^{+1.3} _{-0.6}$     &                                  $-$  &       4500 &        $-$ &           $-$                          &        260 &        $-$ \\ 
\hline   
\multicolumn{8}{c}{IGR~J18450-0435 (a)} \cr
     117 &           51${\pm          16}$  &           $-$                    &         $-$ &        $-$ &           $-$                            &        $-$ &        100 \\ 
     118 &           77${\pm          24}$  &           25${\pm          5 }$  &         400 &        477 &           52${\pm                   10}$ &         72 &        237 \\ 
     119*&           14${\pm           4}$  &           1.3${\pm         0.4 }$ &        510 &         90 &           14${\pm                    4}$ &         50 &        150 \\ 
     120 &           43${\pm          14}$  &           6.2${\pm         1.4 }$ &        830 &        210 &           31${\pm                    7}$ &        126 &         70 \\ 
     121*&           67${\pm          21}$  &           43${\pm          13 }$  &        500 &        637 &           67${\pm                   21}$ &          8 &          7 \\ 
     122 &           68${\pm          21}$  &           7.8${\pm         1.8 }$ &        700 &        130 &           58${\pm                   13}$ &         65 &         50 \\ 
     123 &           84${\pm          26}$  &           29${\pm            6 }$ &        390 &        370 &           77${\pm                   17}$ &        210 &        130 \\ 
     124*&           90${\pm          28}$  &           27${\pm            8 }$ &        500 &        290 &           90${\pm                   28}$ &         60 &         43 \\ 
     125 &           82${\pm          26}$  &           44${\pm            8 }$ &        500 &       1100 &           39${\pm                    7}$ &        120 &        900 \\ 
     126*&           8.4${\pm        2.6}$  &           5.6${\pm         1.7 }$ &       2500 &        660 &           8.4${\pm                 2.6}$ &        140 &         90 \\ 
     127*&           7.5${\pm        2.4}$  &           5.7${\pm         1.8 }$ &        800 &        760 &           7.5${\pm                 2.4}$ &        150 &        250 \\ 
     128 &           13${\pm           4}$  &           8.1${\pm         1.8 }$ &       1400 &        760 &          10.7${\pm                 2.4}$ &        510 &         80 \\ 
     129 &           15${\pm           5}$  &           4.6${\pm         1.3 }$ &        400 &        720 &           6.5${\pm                 1.8}$ &         40 &        680 \\ 
     130 &           15${\pm           5}$  &           19${\pm            4 }$ &       2490 &       2310 &           8.4${\pm                 1.7}$ &       1390 &        830 \\ 
     131*&           13${\pm           4}$  &           1.8${\pm         0.6 }$ &       2310 &        140 &           13${\pm                    4}$ &         50 &         60 \\ 
     132 &           32${\pm          10}$  &           8.2${\pm        1.40 }$ &       2050 &        600 &          13.7${\pm                 2.3}$ &        500 &         43 \\ 
\hline  
\multicolumn{8}{c}{IGR~J18450-0435 (b)} \cr    
     133 &           5.7${\pm      0.6}$   &          $-$                    &        $-$ &        $-$ &           $-$                              &        $-$ &        250 \\ 
     134 &           42${\pm        5}$    &           30${\pm          4 }$ &       4100 &       2800 &          10.7${\pm        1.6}$            &       2100 &        500 \\ 
     135 &           39${\pm        4}$    &           15${\pm          3 }$ &       1800 &        520 &           28${\pm          6}$             &        310 &         43 \\ 
     136 &           161${\pm       18}$   &           49${\pm          9 }$ &        600 &        540 &           92${\pm          17}$            &         43 &        380 \\ 
     137 &           23${\pm         3}$   &           29${\pm          7 }$ &       1300 &       4600 &           6.3${\pm         1.4}$           &        160 &       4100 \\ 
     138 &           165${\pm       19}$   &           $-$                   &       7200 &        $-$ &           $-$                              &        400 &        $-$ \\ 
\hline  
\multicolumn{8}{c}{IGR~J18483-0311 } \cr     
     139 &           4${\pm         1}$      &         $-$                      &        $-$ &        $-$ &           $-$                      &        $-$ &       4100 \\ 
     140 &           3.1${\pm         0.9}$  &           4.0${\pm        0.7 }$ &       9100 &       2450 &          1.65${\pm          0.30}$ &       1740 &         90 \\ 
     141 &           2.9${\pm         0.8}$  &           4.6${\pm        1.1 }$ &       3200 &       3620 &          1.28${\pm          0.31}$ &        110 &       3340 \\ 
     142*&           1.27${\pm       0.36}$  &           1.8${\pm        0.5 }$ &       5200 &       1400 &          1.27${\pm          0.36}$ &        260 &        500 \\ 
     143 &           4${\pm         1}$      &           7.7${\pm        1.4 }$ &       4300 &       5900 &          1.31${\pm          0.24}$ &        820 &       4800 \\ 
     144 &           1.2${\pm         0.3}$  &           8.2${\pm        1.4 }$ &       8000 &      12100 &          0.68${\pm          0.11}$ &        600 &       9900 \\ 
\hline
\end{tabular}
\label{tab:flares}
\end{table*}

\end{appendix}

\bibliographystyle{mn2e} 
\bibliographystyle{mnras}

\begin{thebibliography}{}

\bibitem[\protect\citeauthoryear{{Arons} \& {Lea}}{{Arons} \&
  {Lea}}{1976}]{1976ApJ...207..914A}
{Arons} J.,  {Lea} S.~M.,  1976, \apj, 207, 914

\bibitem[\protect\citeauthoryear{{Bozzo}, {Bernardini}, {Ferrigno}, {Falanga},
  {Romano} \& {Oskinova}}{{Bozzo} et~al.}{2017}]{Bozzo2017}
{Bozzo} E.,  {Bernardini} F.,  {Ferrigno} C.,  {Falanga} M.,  {Romano} P.,
  {Oskinova} L.,  2017, \aap, 608, A128

\bibitem[\protect\citeauthoryear{{Bozzo}, {Falanga} \& {Stella}}{{Bozzo}
  et~al.}{2008}]{2008ApJ...683.1031B}
{Bozzo} E.,  {Falanga} M.,    {Stella} L.,  2008, \apj, 683, 1031

\bibitem[\protect\citeauthoryear{{Burnard}, {Arons} \& {Lea}}{{Burnard}
  et~al.}{1983}]{1983ApJ...266..175B}
{Burnard} D.~J.,  {Arons} J.,    {Lea} S.~M.,  1983, \apj, 266, 175

\bibitem[\protect\citeauthoryear{{Carlyle} \& {Hillier}}{{Carlyle} \&
  {Hillier}}{2017}]{2017A&A...605A.101C}
{Carlyle} J.,  {Hillier} A.,  2017, \aap, 605, A101

\bibitem[\protect\citeauthoryear{{De Luca}, {Salvaterra}, {Tiengo},
  {D'Agostino}, {Watson}, {Haberl} \& {Wilms}}{{De Luca}
  et~al.}{2017}]{Deluca2017}
{De Luca} A.,  {Salvaterra} R.,  {Tiengo} A.,  {D'Agostino} D.,  {Watson} M.,
  {Haberl} F.,    {Wilms} J.,  2017, in {Ness} J.-U.,  {Migliari} S.,  eds, The
  X-ray Universe 2017 {EXTraS: Exploring the X-ray Transient and variable Sky}.
p.~65

\bibitem[\protect\citeauthoryear{{De Luca}, {Salvaterra}, {Tiengo},
  {D'Agostino}, {Watson}, {Haberl} \& {Wilms}}{{De Luca}
  et~al.}{2016}]{deluca16}
{De Luca} A.,  {Salvaterra} R.,  {Tiengo} A.,  {D'Agostino} D.,  {Watson}
  M.~G.,  {Haberl} F.,    {Wilms} J.,  2016, The Universe of Digital Sky
  Surveys, 42, 291

\bibitem[\protect\citeauthoryear{{Elsner} \& {Lamb}}{{Elsner} \&
  {Lamb}}{1977}]{1977ApJ...215..897E}
{Elsner} R.~F.,  {Lamb} F.~K.,  1977, \apj, 215, 897

\bibitem[\protect\citeauthoryear{{Elsner} \& {Lamb}}{{Elsner} \&
  {Lamb}}{1984}]{1984ApJ...278..326E}
{Elsner} R.~F.,  {Lamb} F.~K.,  1984, \apj, 278, 326

\bibitem[\protect\citeauthoryear{{Gim{\'e}nez-Garc{\'{\i}}a}, {Torrej{\'o}n},
  {Eikmann}, {Mart{\'{\i}}nez-N{\'u}{\~n}ez}, {Oskinova}, {Rodes-Roca} \&
  {Bernab{\'e}u}}{{Gim{\'e}nez-Garc{\'{\i}}a} et~al.}{2015}]{Gimenez2015}
{Gim{\'e}nez-Garc{\'{\i}}a} A.,  {Torrej{\'o}n} J.~M.,  {Eikmann} W.,
  {Mart{\'{\i}}nez-N{\'u}{\~n}ez} S.,  {Oskinova} L.~M.,  {Rodes-Roca} J.~J.,
   {Bernab{\'e}u} G.,  2015, \aap, 576, A108

\bibitem[\protect\citeauthoryear{{Grebenev} \& {Sunyaev}}{{Grebenev} \&
  {Sunyaev}}{2007}]{2007AstL...33..149G}
{Grebenev} S.~A.,  {Sunyaev} R.~A.,  2007, Astronomy Letters, 33, 149

\bibitem[\protect\citeauthoryear{{Illarionov} \& {Sunyaev}}{{Illarionov} \&
  {Sunyaev}}{1975}]{1975A&A....39..185I}
{Illarionov} A.~F.,  {Sunyaev} R.~A.,  1975, \aap, 39, 185

\bibitem[\protect\citeauthoryear{{Kulkarni} \& {Romanova}}{{Kulkarni} \&
  {Romanova}}{2008}]{2008MNRAS.386..673K}
{Kulkarni} A.~K.,  {Romanova} M.~M.,  2008, \mnras, 386, 673

\bibitem[\protect\citeauthoryear{{Lii}, {Romanova}, {Ustyugova}, {Koldoba} \&
  {Lovelace}}{{Lii} et~al.}{2014}]{2014MNRAS.441...86L}
{Lii} P.~S.,  {Romanova} M.~M.,  {Ustyugova} G.~V.,  {Koldoba} A.~V.,
  {Lovelace} R.~V.~E.,  2014, \mnras, 441, 86

\bibitem[\protect\citeauthoryear{{Marelli}, {Salvetti}, {Gastaldello},
  {Ghizzardi}, {Molendi}, {Luca}, {Moretti}, {Rossetti} \& {Tiengo}}{{Marelli}
  et~al.}{2017}]{marelli17}
{Marelli} M.,  {Salvetti} D.,  {Gastaldello} F.,  {Ghizzardi} S.,  {Molendi}
  S.,  {Luca} A.~D.,  {Moretti} A.,  {Rossetti} M.,    {Tiengo} A.,  2017,
  Experimental Astronomy, 44, 297

\bibitem[\protect\citeauthoryear{{Mart{\'{\i}}nez-N{\'u}{\~n}ez}, {Kretschmar},
  {Bozzo}, {Oskinova}, {Puls}, {Sidoli}, {Sundqvist} \&
  {Blay}}{{Mart{\'{\i}}nez-N{\'u}{\~n}ez} et~al.}{2017}]{Martinez-Nunez2017}
{Mart{\'{\i}}nez-N{\'u}{\~n}ez} S.,  {Kretschmar} P.,  {Bozzo} E.,  {Oskinova}
  L.~M.,  {Puls} J.,  {Sidoli} L.,  {Sundqvist} J.~O.,    {Blay} P.,  2017,
  \ssr, 212, 59

\bibitem[\protect\citeauthoryear{{Negueruela}, {Smith}, {Harrison} \&
  {Torrej{\'o}n}}{{Negueruela} et~al.}{2006}]{Negueruela2006}
{Negueruela} I.,  {Smith} D.~M.,  {Harrison} T.~E.,    {Torrej{\'o}n} J.~M.,
  2006, \apj, 638, 982

\bibitem[\protect\citeauthoryear{{Plesset} \& {Whipple}}{{Plesset} \&
  {Whipple}}{1974}]{1974PhFl...17....1P}
{Plesset} M.~S.,  {Whipple} C.~G.,  1974, Physics of Fluids, 17, 1

\bibitem[\protect\citeauthoryear{{Postnov}, {Oskinova} \&
  {Torrej{\'o}n}}{{Postnov} et~al.}{2017}]{2017MNRAS.465L.119P}
{Postnov} K.,  {Oskinova} L.,    {Torrej{\'o}n} J.~M.,  2017, \mnras, 465, L119

\bibitem[\protect\citeauthoryear{{Postnov}, {Staubert}, {Santangelo},
  {Klochkov}, {Kretschmar} \& {Caballero}}{{Postnov}
  et~al.}{2008}]{2008A&A...480L..21P}
{Postnov} K.,  {Staubert} R.,  {Santangelo} A.,  {Klochkov} D.,  {Kretschmar}
  P.,    {Caballero} I.,  2008, \aap, 480, L21

\bibitem[\protect\citeauthoryear{{Pradhan}, {Bozzo} \& {Paul}}{{Pradhan}
  et~al.}{2018}]{Pradhan2018}
{Pradhan} P.,  {Bozzo} E.,    {Paul} B.,  2018, \aap, 610, A50

\bibitem[\protect\citeauthoryear{{Rosen}, {Webb}, {Watson}, {Ballet}, {Barret},
  {Braito}, {Carrera} \& {Ceballos}}{{Rosen} et~al.}{2016}]{Rosen2016}
{Rosen} S.~R.,  {Webb} N.~A.,  {Watson} M.~G.,  {Ballet} J.,  {Barret} D.,
  {Braito} V.,  {Carrera} F.~J.,    {Ceballos} M.~T.,  2016, \aap, 590, A1

\bibitem[\protect\citeauthoryear{{Scargle}}{{Scargle}}{1998}]{scargle98}
{Scargle} J.~D.,  1998, \apj, 504, 405

\bibitem[\protect\citeauthoryear{{Scargle}, {Norris}, {Jackson} \&
  {Chiang}}{{Scargle} et~al.}{2013}]{scargle13}
{Scargle} J.~D.,  {Norris} J.~P.,  {Jackson} B.,    {Chiang} J.,  2013, \apj,
  764, 167

\bibitem[\protect\citeauthoryear{{Sguera}, {Barlow}, {Bird}, {Clark}, {Dean},
  {Hill}, {Moran}, {Shaw}, {Willis}, {Bazzano}, {Ubertini} \&
  {Malizia}}{{Sguera} et~al.}{2005}]{Sguera2005}
{Sguera} V.,  {Barlow} E.~J.,  {Bird} A.~J.,  {Clark} D.~J.,  {Dean} A.~J.,
  {Hill} A.~B.,  {Moran} L.,  {Shaw} S.~E.,  {Willis} D.~R.,  {Bazzano} A.,
  {Ubertini} P.,    {Malizia} A.,  2005, \aap, 444, 221

\bibitem[\protect\citeauthoryear{{Sguera}, {Bazzano}, {Bird}, {Dean},
  {Ubertini}, {Barlow}, {Bassani}, {Clark}, {Hill}, {Malizia}, {Molina} \&
  {Stephen}}{{Sguera} et~al.}{2006}]{Sguera2006}
{Sguera} V.,  {Bazzano} A.,  {Bird} A.~J.,  {Dean} A.~J.,  {Ubertini} P.,
  {Barlow} E.~J.,  {Bassani} L.,  {Clark} D.~J.,  {Hill} A.~B.,  {Malizia} A.,
  {Molina} M.,    {Stephen} J.~B.,  2006, \apj, 646, 452

\bibitem[\protect\citeauthoryear{{Shakura}, {Postnov} \&
  {Hjalmarsdotter}}{{Shakura} et~al.}{2013}]{2013MNRAS.428..670S}
{Shakura} N.,  {Postnov} K.,    {Hjalmarsdotter} L.,  2013, \mnras, 428, 670

\bibitem[\protect\citeauthoryear{{Shakura}, {Postnov}, {Kochetkova} \&
  {Hjalmarsdotter}}{{Shakura} et~al.}{2012}]{Shakura2012}
{Shakura} N.,  {Postnov} K.,  {Kochetkova} A.,    {Hjalmarsdotter} L.,  2012,
  \mnras, 420, 216

\bibitem[\protect\citeauthoryear{{Shakura}, {Postnov}, {Kochetkova} \&
  {Hjalmarsdotter}}{{Shakura} et~al.}{2018}]{2018ASSL..454..331S}
{Shakura} N.,  {Postnov} K.,  {Kochetkova} A.,    {Hjalmarsdotter} L.,  2018,
  in {Shakura} N.,  ed., Astrophysics and Space Science Library Vol.~454 of
  Astrophysics and Space Science Library, {Quasi-Spherical Subsonic Accretion
  onto Magnetized Neutron Stars}.
p.~331

\bibitem[\protect\citeauthoryear{{Shakura}, {Postnov}, {Sidoli} \&
  {Paizis}}{{Shakura} et~al.}{2014}]{Shakura2014}
{Shakura} N.,  {Postnov} K.,  {Sidoli} L.,    {Paizis} A.,  2014, \mnras, 442,
  2325

\bibitem[\protect\citeauthoryear{{Shakura}}{{Shakura}}{1973}]{1973SvA....16..756S}
{Shakura} N.~I.,  1973, Sov. Astron., 16, 756

\bibitem[\protect\citeauthoryear{{Sidoli}}{{Sidoli}}{2017}]{Sidoli2017review}
{Sidoli} L.,  2017, in Proceedings of the XII Multifrequency Behaviour of High
  Energy Cosmic Sources Workshop. 12-17 June, 2017 Palermo, Italy (MULTIF2017)
  Online at https://pos.sissa.it/cgi-bin/reader/conf.cgi?confid=306, id.52
  (arXiv:1710.03943) {Supergiant Fast X-ray Transients - A short review}.
p.~52

\bibitem[\protect\citeauthoryear{{Sidoli} \& {Paizis}}{{Sidoli} \&
  {Paizis}}{2018}]{SP2018}
{Sidoli} L.,  {Paizis} A.,  2018, \mnras, 481, 2779

\bibitem[\protect\citeauthoryear{{Sidoli}, {Romano}, {Mangano}, {Pellizzoni},
  {Kennea}, {Cusumano}, {Vercellone}, {Paizis}, {Burrows} \&
  {Gehrels}}{{Sidoli} et~al.}{2008}]{Sidoli2008:sfxts_paperI}
{Sidoli} L.,  {Romano} P.,  {Mangano} V.,  {Pellizzoni} A.,  {Kennea} J.~A.,
  {Cusumano} G.,  {Vercellone} S.,  {Paizis} A.,  {Burrows} D.~N.,    {Gehrels}
  N.,  2008, \apj, 687, 1230

\bibitem[\protect\citeauthoryear{{Sidoli}, {Romano}, {Mereghetti}, {Paizis},
  {Vercellone}, {Mangano} \& {G{\"o}tz}}{{Sidoli} et~al.}{2007}]{Sidoli2007}
{Sidoli} L.,  {Romano} P.,  {Mereghetti} S.,  {Paizis} A.,  {Vercellone} S.,
  {Mangano} V.,    {G{\"o}tz} D.,  2007, \aap, 476, 1307

\bibitem[\protect\citeauthoryear{{Syunyaev} \& {Shakura}}{{Syunyaev} \&
  {Shakura}}{1977}]{1977SvAL....3..138S}
{Syunyaev} R.~A.,  {Shakura} N.~I.,  1977, Soviet Astronomy Letters, 3, 138

\bibitem[\protect\citeauthoryear{{Tsygankov}, {Lutovinov}, {Doroshenko},
  {Mushtukov}, {Suleimanov} \& {Poutanen}}{{Tsygankov}
  et~al.}{2016}]{2016A&A...593A..16T}
{Tsygankov} S.~S.,  {Lutovinov} A.~A.,  {Doroshenko} V.,  {Mushtukov} A.~A.,
  {Suleimanov} V.,    {Poutanen} J.,  2016, \aap, 593, A16

\bibitem[\protect\citeauthoryear{{Walter}, {Lutovinov}, {Bozzo} \&
  {Tsygankov}}{{Walter} et~al.}{2015}]{Walter2015}
{Walter} R.,  {Lutovinov} A.~A.,  {Bozzo} E.,    {Tsygankov} S.~S.,  2015,
  \aapr, 23, 2

\end{thebibliography}

\bsp

\label{lastpage}

\end{document}